\documentclass[a4paper,10pt]{article}
\usepackage{authblk}
\usepackage{amscd}
\usepackage{subfigure}
\usepackage{color}
\usepackage{amssymb,latexsym,amsmath,amsfonts}
\usepackage{graphicx,epstopdf}
\usepackage[a4paper,top=5cm,bottom=2.5cm,left=2.5cm,right=2.5cm]{geometry}

\newcommand{\bm}{\boldmath}

\newcommand{\m}{\mbox}

\newcommand{\ubm}{\unboldmath}

\newcommand{\bu}{\m{\bm$u$\ubm}}

\begin{document}

\title{Complex singularities and PDEs}

\date{}

\author[1]{R.E.Caflisch\thanks{rcaflisch@ipam.ucla.edu}}
\author[2]{F. Gargano\thanks{francesco.gargano@unipa.it}}
\author[3]{M. Sammartino\thanks{marcomarialuigi.sammartino@unipa.it}}
\author[3]{V. Sciacca\thanks{vincenzo.sciacca@unipa.it}}
\affil[1]{Mathematics Department, University of California at Los Angeles, Los Angeles, CA 90036, United States}
\affil[2]{Department of Mathematical Modelling - DEIM , University of Palermo,  90123 Palermo, Italy}
\affil[3]{Department of Mathematics , University of Palermo,  90123 Palermo, Italy}
\renewcommand\Authands{ and }

\maketitle

\begin{abstract}
In this paper we give a review on the computational methods used to  characterize the complex singularities developed by some relevant PDEs.
We begin by reviewing the  singularity tracking method based on the analysis of the Fourier spectrum.
We then introduce  other methods generally used to detect the hidden singularities.
In particular we show some applications of the Pad\'e approximation, of the Kida method, and of
Borel-Polya method.
We apply these techniques to the study of the singularity formation of some nonlinear dispersive and dissipative one dimensional PDE of the 2D Prandtl equation, of
the 2D KP equation, and to Navier-Stokes equation for high Reynolds number incompressible flows
in the case of  interaction with  rigid boundaries.

\end{abstract}

\section{Introduction}\label{sec:0}

Many nonlinear partial differential equations (PDE)  develop finite time singularities which signal the limit of applicability of a PDE as mathematical
model and often has  physical significance.
Therefore there is considerable interest in methods that can give indications whether a singularity is forming,  where it is forming, and on its nature.
In this review paper we shall focus on methods based on analytic continuation in the complex domain  of
numerical solutions of PDEs derived through spectral discretization.

The study in the complex plane of the analytic structure of the solutions of nonlinear PDE
has in fact revealed to be a powerful method for the understanding of the process of singularity formation.
The main idea behind  the singularity tracking method \cite{SSF83} is to consider
the analytic continuation of a function in the independent variable and to detect the width of
the analyticity strip, i.e. the distance from the real domain to the nearest complex singularity.
The width of the analyticity strip can vary with time and if a singularity reaches the real domain
then the solution loses analyticity and becomes singular.
The width of the analyticity strip can alternatively be bounded
away from zero, or tend to zero asymptotically with time, in which case the solution develops
increasingly small scales while remaining smooth.

The numerical implementation of this ideas typically involves high resolution (spectral) numerical computation of a time-evolution problem, while the location and other properties of the nearest complex
singularity are determined from asymptotic behavior of the Fourier transform of the numerical solution.
Indeed, the asymptotic properties of the Fourier transform for an analytic function of a single variable with isolated pole or branch
point singularities at complex locations is determinate by the Laplace asymptotic formula, see \cite{CKP66}
or \cite{Henrici}: more details on this and on the determination of the asymptotic behavior of the spectrum
will be given in the next Section.

During the last three decades these ideas have been used extensively, particularly in the
analysis of the (possible) singular behavior of flows and of  PDEs arising in fluid dynamics.
Examples include: the study of  interface flow problems and of the singularity formation for vortex sheet equation \cite{MOO78,MOO79,CO89,CBT99,Kr86a,Sh92,BCS93,PS98};
the investigation of the complex singularity formation for incompressible Euler flow; \cite{Caf93,FMB03,PMFB06,MBF05,CB05};
the analysis of the singularity formation for Prandtl solution and its connection to the separation
phenomena \cite{COW83,DLSS06,GSS09,GSS11,GSSC14};
as well  the analysis of the singularity formation for Camassa-Holm and Degasperi-Procesi equations \cite{DLSS06,CGS12},
nonlinear Schr\"{o}dinger equation \cite{RM15}, KdV \cite{KR13}, and others \cite{KR14,KR15}.

When a PDE develops a finite time singularity the  tracking of complex singularity gives valuable information on the time of blow up of the solution, on the spatial location and the algebraic character of the singularity.
However, the method, as explained in \cite{SSF83} is able to characterize only the  singularity closest
to the real axis.
In some cases it is also important to analyze other singularities in the complex plain.
For example the analysis of the  hidden complex singularities  in \cite{GSSC14} has revealed how the
separation phenomena for the Navier Stokes equations is not related to the singularity of Prandtl equation.

To detect the hidden one can use the  Pad\'e approximants.
The advantage of the Pad\'{e} approximation method is that it allows one to continue the
function even beyond the radius of convergence (or strip of analyticity).
Pad\'{e} approximants also have been used in the
analysis of complex singularities
of various ordinary and partial differential equations, see \cite{COW83,GSSC14,Gu89}.
A filtering  method used to analyze hidden singularity was introduced by Kida \cite{Ki86}, while  the method
of Borel-Polya-Van der Hoeven has been proposed in \cite{PF07}.
We will discuss these methodologies in Section 3.

Another important implementation of the singularity tracking method is its extension to functions of several variables.
If the singularity of a function of several variables occurs along a single variable, a simple way to analyze the complex singularities
is the application of the method of singularity tracking to this variable, see \cite{GSS09,DLSS06}. In the more general case, it is possible to extend the singularity tracking method and detect complex singularity surface for a function of several variables from the full multidimensional Fourier transform. The main idea of this generalization consist to
consider the analytic continuation in one variable and
detect the complex singularity surface as a function of the other real variables. This analysis is based on the asymptotic properties of the multidimensional Fourier transform and in particular on the fact that the parameters, which characterize the singularity, are determined by the decay of
the Fourier spectrum along or near a distinguished direction in wavenumber, which is the direction with the lowest decay rate.
For example, one can see these applications in \cite{Caf93,CS04,SC09,PMFB06,GSSC14,MCSV13}.

The goal of the present paper is to give a brief review of some
of the most recent advances in the singularity tracking method and to present some applications
of interest in the field of fluid  dynamics.

The plan of the paper is the following.
In Section 2 we present the method and we consider as an application, for the one dimensional case,
the classical study of the singularity formation for Burgers equation.
In Section 3 we present the methods to detect the hidden complex singularities:
in  Section 3.1 we discuss  the Pad\'e approximants theory, while  the Kida method and  the
Borel-Polya-Van der Hoeven method are introduced in Section 3.2 and 3.3 respectively. These techniques are applied in Section 4 to several PDES.
In Section 4.1 we investigate the complex singularities for the Kortweg de Vries equation
and how they are related to the  rapid oscillatory behavior of the solutions in the regime of small dispersion. In Section 4.2 we analyze the complex singularities of the wall shear of the Navier Stokes and Prandtl equations and their relation with the mechanisms of unsteady separation phenomena. Finally in Section 5 we consider the singularity tracking method to analyze the complex singularities
manifold for solutions of two dimensional PDEs; the applications presented are  wall bounded flows at
high Reynolds number.

%%%%%%%%%%%%%%%%%%%%%%%%%%%%%%%%%%%%%%%%%%%%%%%%%%%%%%%%%%%%%%%
\section{Singularity tracking method}\label{sec:1}

The complex singularity tracking method is based on the relationship between
the asymptotic properties of the Fourier spectrum and the radius of analyticity of a real function.

Suppose that $u(z)$ is a real function analytic in the strip of the complex plane
$\left\{z \in \mathbb{C}: \left| z \right| < \delta \right\}$.
We suppose that the  singularity closest to the real axis has
complex location $z^{*} = x^{*} + i \delta$, and that $u(z) \sim \left(z - z^{*} \right)^{\alpha}$.
Using a steepest descent argument it is possible to give the asymptotic (in $k$) behavior of the spectrum of $u(z)$:
\begin{equation}\label{laplace}
\hat{u}_k \sim C \left|k\right|^{-(1+\alpha)} \exp{\left(-\delta \left|k\right|\right)} \exp{(i k x^{*})},
\end{equation}
where with $\hat{u}_k$ we have denoted the Fourier coefficients.

Estimating the rate of exponential decay of the
spectrum of the function $u$ one gets the distance of the singularity from the real axis, the $\delta$ in \eqref{laplace}.
If one estimates the rate of algebraic decay (the $\alpha$ in \eqref{laplace}) one can characterize the singularity, and
moreover the oscillatory behavior of the spectrum (the $x^{*}$ in \eqref{laplace}) gives the location of the singularity.
If the spectrum of the function $u$ has not exponential
decay, this means that the width of the strip of analyticity is zero and $u$ has some kind of blow up.
The  estimate $\delta$, $\alpha$ and $x^{*}$ of \eqref{laplace} requires the use some fitting techniques
(least square fitting for example), and in practical applications reveals to be a delicate matter.

We consider now a one dimensional evolutionary PDE
\begin{equation}\label{pde}
u_t=F\left(u,u_x,u_{xx},...\right),
\end{equation}
where $u=u(x,t)$, $x\in\left[0,2\pi\right]$ is the spatial variable and $t>0$ the time variable, and where with $u_t$ and $u_x$, $u_{xx},\dots$ we denote the various partial derivatives of $u$ with respect to $t$ and $x$ at different orders.

We discretize using Fourier-Galerkin spectral method and we transform the one dimensional PDE in a system of N ODEs
\begin{equation}\label{ode}
\frac{d \hat{u}_k}{d t}=G_k \left(\hat{\bu} \right), \qquad k=-N,\dots, N \quad,
\end{equation}
where $\hat{\bu}=\left( \hat{u}_k \right)_{ \left|k\right| \leq N}$ and $N$ is the order of the numerical
truncation of the discrete Fourier series expansion
\begin{equation}
u_N(x_j,t)=\sum_{k=-N}^{N} \hat{u}_k (t) \exp{\left( i k x_j \right)},
\end{equation}
with $x_j=2\pi j/N$, $j=1,\dots,N$.

Giving the initial condition to system \eqref{ode} and solving numerically the ODE
system \eqref{ode} one can determine the time evolution of the Fourier spectrum $u_k(t)$. Studying the  asymptotic behavior of the spectrum for large $k$ gives the time evolution
of the complex singularity, i.e. the path in the complex plane $(\delta(t), x^{*}(t))$ and
the algebraic characterization $\alpha(t)$.

%%%%%%%%%%%%%%%%%%%%%%%%%%%%%%%%%%%%%%%%%%%%%%%%%%%%%%%%%%%%%%%
\subsection{Application to 1-dimensional PDEs}\label{subsec:1.1}

Burgers equation is a good case study where to test the above ideas and in \cite{SSF83}
the authors studied the shock formation process for the following problem:
\begin{equation}\label{burgers}
u_t+u u_x=0, \qquad x \in \left[0,2\pi\right],
\end{equation}
with initial condition
\begin{equation}\label{IC-burgers}
u(x,0)=\sin(x),
\end{equation}
and periodic boundary conditions.

From the classical results on the existence of classical solutions for conservation law \eqref{burgers} by the method of characteristic, the solution develops a singularity (a blow up on $u_x$) at time
\begin{equation}\label{t_s-burgers}
t_s=\frac{1}{-\inf_{x \in \left[ 0 , 2 \pi \right]} u_x(x,0)}.
\end{equation}

\begin{figure}
\includegraphics[width=12.cm]{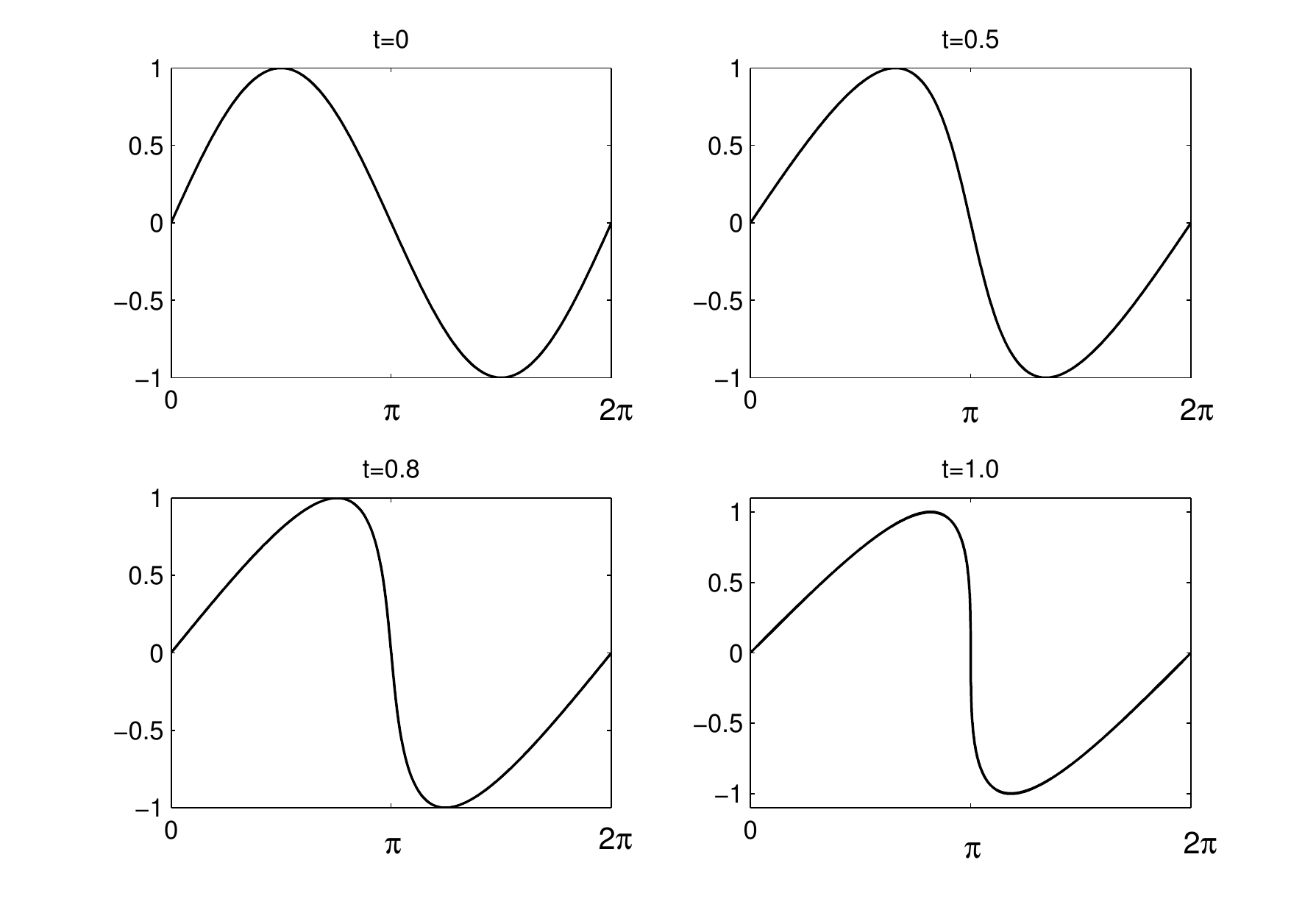}
  \caption{The behavior in time of the numerical solution of
  Burgers equation with initial condition \eqref{IC-burgers}. At time $t_S =1$
  one can see a singularity as a blow up of the first derivative.}
\label{burgers_1}
\end{figure}

The dynamics of the $k$-th Fourier mode of $u$ is described by the following
ODE
\begin{equation}\label{numode}
\frac{d \hat{u}_k }{d t}= - \widehat{(uu_x)}_k
\end{equation}
To solve the above ODE system one can compute efficiently the nonlinear term
through a pseudo spectral procedure, see e.g. \cite{Boy00,CHQZ06}; advancing in time
can be achieved, e.g., using  a 4th order explicit Runge-Kutta method.
The process of progressive steepening of the wave until the blow-up of the spatial derivative
can be observed in Fig.\ref{burgers_1}.

\begin{figure}
\includegraphics[width=12.cm]{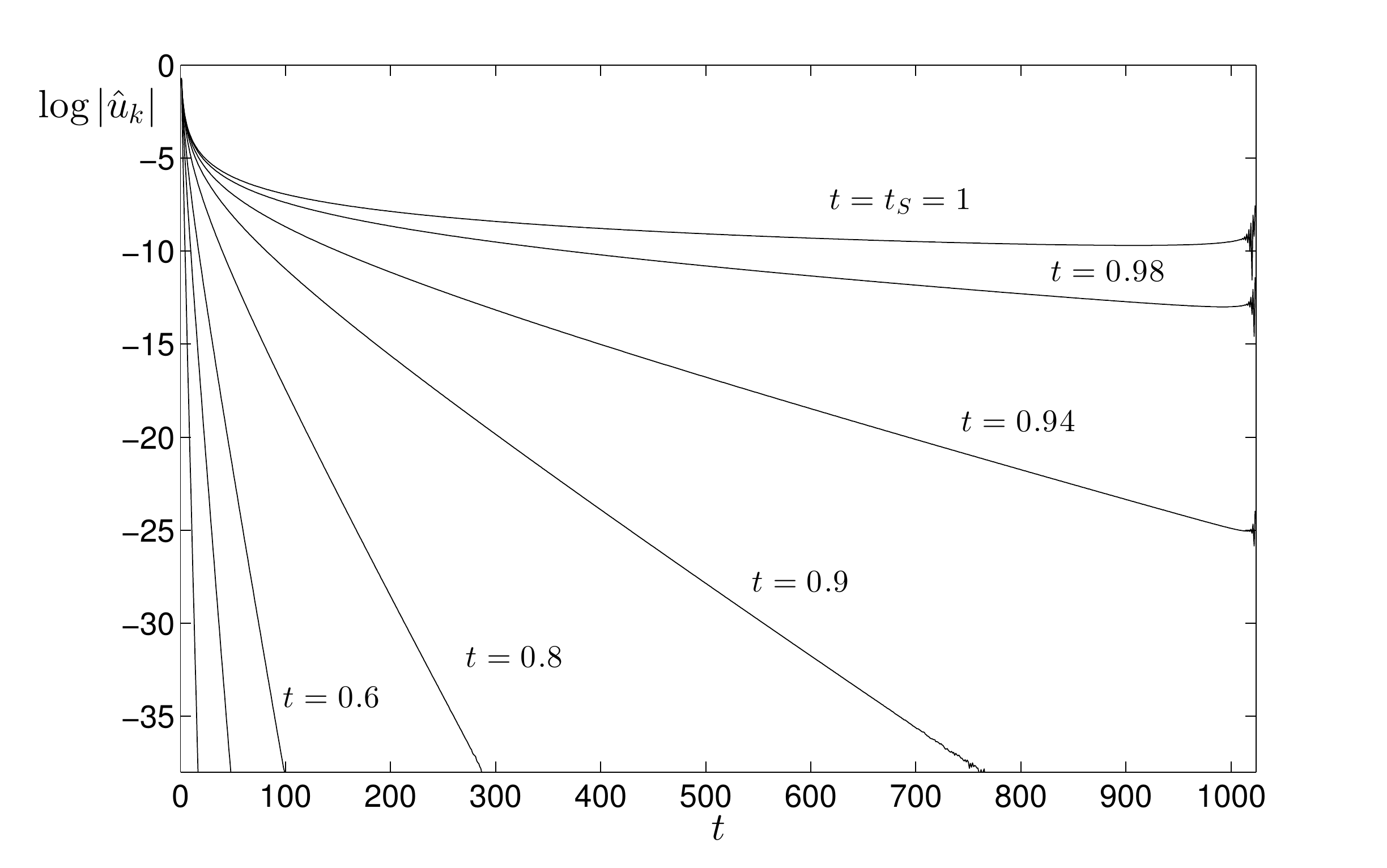}
  \caption{The behavior of the spectrum in time of the Burgers
  equation numerical solution with initial datum \eqref{IC-burgers}, starting at time
  $t = 0.6$ up to singularity time $t_S = 1$.}
\label{burgers_spectrum}
\end{figure}

In Fig.\ref{burgers_spectrum} it is shown the behavior  in time of the spectrum of the solution.
starting at time $t = 0.6$ up to singularity time $t_S = 1$.

\begin{figure}[h]
  \resizebox{13cm}{8cm}{\includegraphics{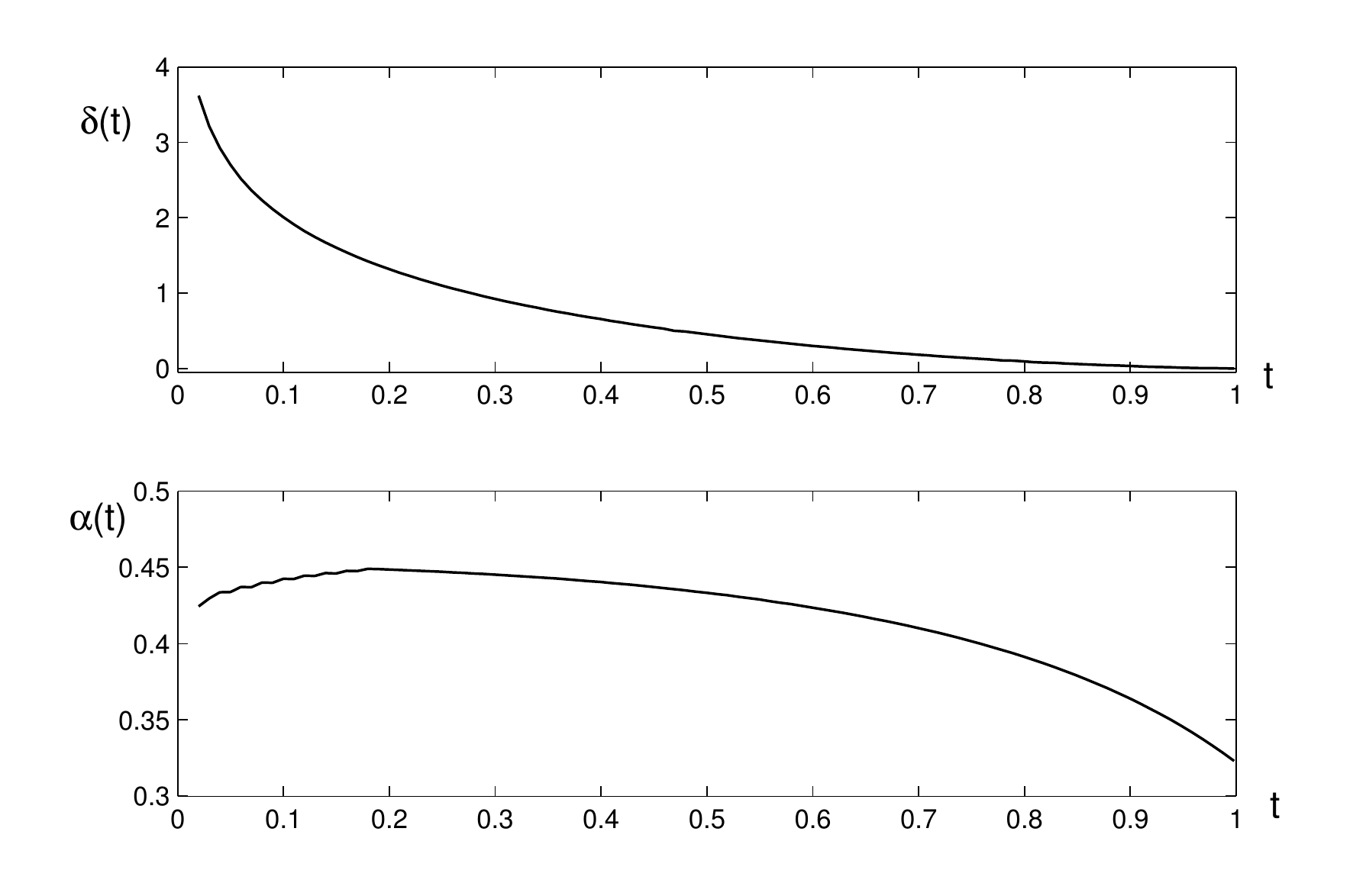}}
  \caption{The tracking singularity method applied to Burgers
  equation numerical solution with initial datum \eqref{IC-burgers}. In the top, the
  behavior in time of the width of the analyticity strip $\delta$. In the
  bottom, the behavior in time of the algebraic character $\alpha$. The
  singularity time is $t_S = 1$. The singularity is of cubic--root type.}
  \label{burgers_fitt}
\end{figure}

\begin{figure}[h]
  \resizebox{13cm}{8cm}{\includegraphics{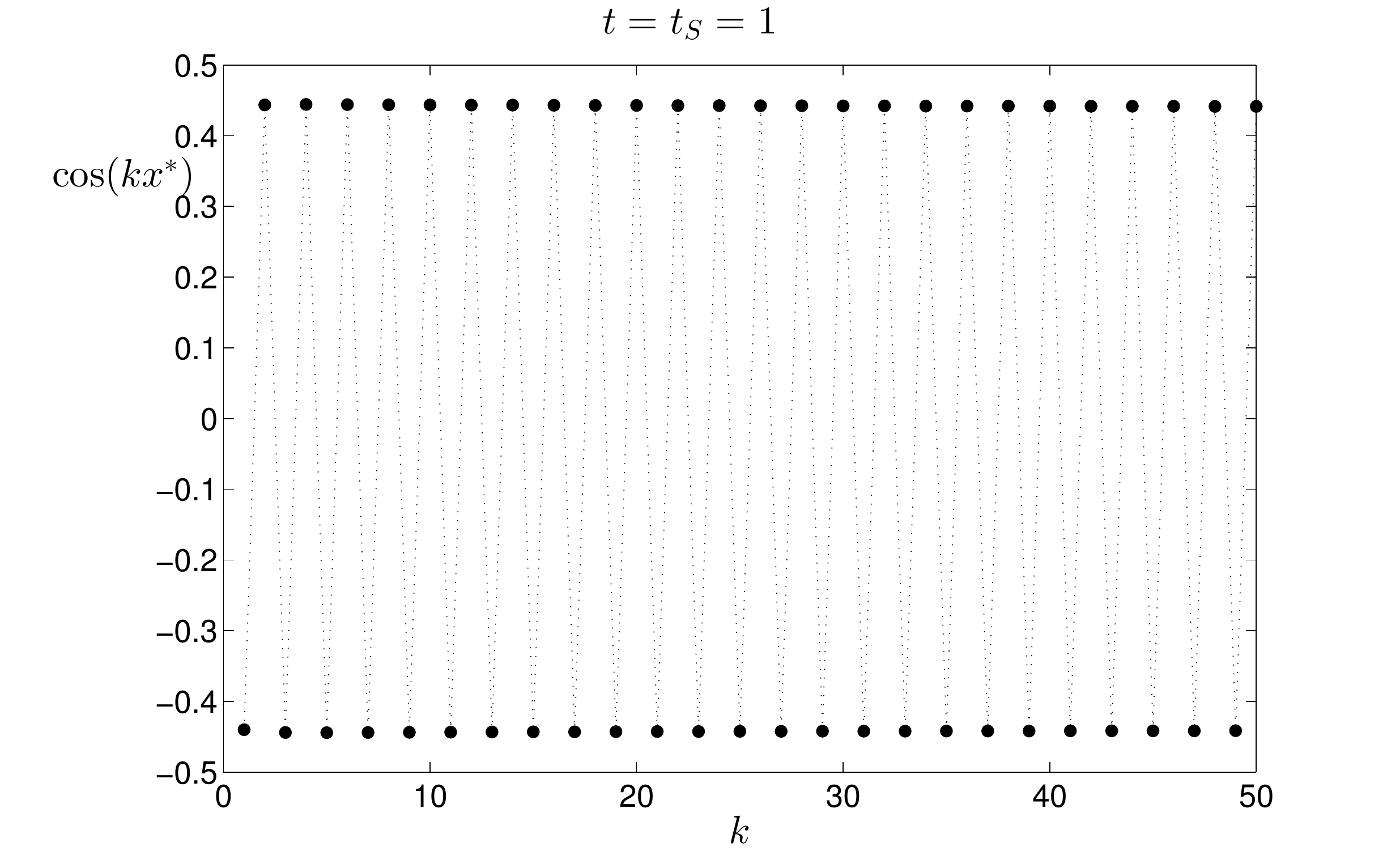}}
  \caption{The tracking singularity method applied to Burgers
  equation numerical solution with initial datum \eqref{IC-burgers}. The
  behavior in $k$ of the term $\cos(k x^{*})$ in formula \eqref{laplace_real}, at the
  singularity time $t_S = 1$}
  \label{burgers_fitt_real}
\end{figure}

For symmetry reasons one has that a complex singularity comes with its complex conjugate.
Therefore the asymptotic formula \eqref{laplace} in this case becomes:
\begin{equation}\label{laplace_real}
\hat{u}_k \sim C \left|k\right|^{-(1+\alpha)} \exp{\left(-\delta \left|k\right|\right)} \cos{(k x^{*})}.
\end{equation}
In this case a simple least square fitting procedure applied to the numerical data of the spectrum
is able to give the values of the parameters $\delta$, $x^*$ and $\alpha$ of \eqref{laplace_real}.
The results  are shown in Fig.\ref{burgers_fitt}.
The critical time $t_S$ where $\delta (t_S) = 0$ is the singularity time $t_S= 1$ and the singularity
algebraic character $\alpha (t_S)$ is of cubic--root type.
One can also notice from fig.\ref{burgers_fitt_real} that the location of the singularity is in $x^{*}=\pi$

The above numerical results are in perfect agreement with the findings of Fournier and Frisch
obtained in \cite{FF83} using asymptotic analysis techniques in the study of the Fourier-Lagrange modes.
In fact they showed that the solution of the above problem has two complex conjugate singularities of square--root type located in $\pi \pm i\delta(t)$ which collide, at time $t=t_S=1$, at $x^{*}=\pi$ and $\delta=0$ to form a real cube-roots singularity.

Other  fitting procedure can certainly be used, see \cite{Boy09} for a discussions on the issues
related to the fitting procedures. Here we mention that
in \cite{KR13} the fitting procedure was based on the  minimizing the $L^\infty$ norm
\begin{equation}
\Delta = \left\| \log| \hat{u}_k| - (C-(1+\alpha) \log|k| -\delta k) \right\|_\infty.
\end{equation}

\begin{figure}[h]
  \resizebox{13cm}{8cm}{\includegraphics{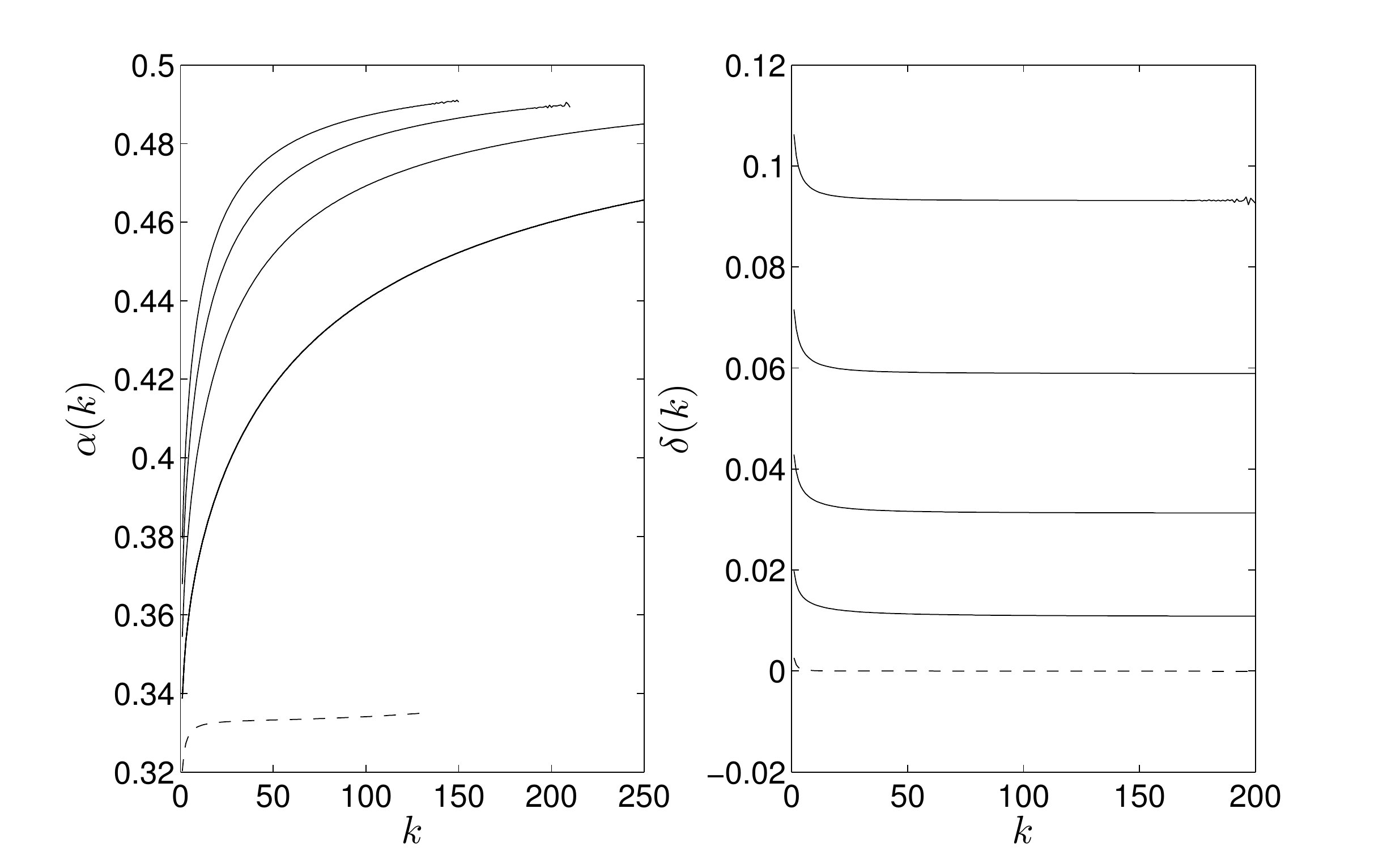}}
  \caption{The sliding fitting procedure of length $3$ for the
  Burgers equation with initial datum \eqref{IC-burgers}. On the right
  (left) figure the $\alpha (k)$ ($\delta (k)$) behaviors w.r.t. $k$, starting
  at time $t = 0.8$ up to singularity time $t_S = 1$ with increments of
  $0.05$ (the dashed lines correspond to the singularity time).}
  \label{burgers_sliding}
\end{figure}

A different analysis of the singularity formation can be performed with
the so called sliding--fitting technique with length $3$, see \cite{Caf93,DLSS06,PMFB06,Sh92} for details.
This procedure consists in searching the values of $C$, $\alpha$ and $\delta$ of the
asymptotic formula \eqref{laplace} locally to each $k$ mode, using
only the $(k - 1)$-th, $k$--th and $(k + 1)$-th modes of the spectrum.
The formulas are:
\begin{eqnarray}
  \alpha (k) & = & \frac{\log \left( \frac{\hat{u}_{k - 1}  \hat{u}_{k +
  1}}{\hat{u}^2_k} \right)}{\log \left( \frac{k^2}{\left( k - 1 \right)
  \left( k + 1 \right)} \right)}, \label{sliding_alpha}\\
  \delta (k) & = & \left[ \log \left( \frac{\hat{u}_k}{\hat{u}_{k + 1}}
  \right) + \alpha \log \left( \frac{k}{k + 1} \right) \right],\label{sliding_delta} \\
  \log C (k) & = & \log | \hat{u}_k | + \alpha (k) \log \left( k \right) + k
  \delta (k) \label{sliding_C}.
\end{eqnarray}
These values depend on $k$, and the asymptotics can be computed with an
extrapolation process using the epsilon algorithm of Wynn, see  \cite{Caf93,PMFB06}.
The results of the sliding fitting procedure with length
$3$, are shown in fig.\ref{burgers_sliding} for the Burgers equation.

Another techniques is the Van der Hoeven asymptotic interpolation method, recently discussed in \cite{PF07}.
An important feature of the
asymptotic interpolation method is that it uses the determination of subleading terms to improve the accuracy
on leading order terms. We now explain this method and we refer to \cite{PF07,GSSC14} for applications of this technique.

Suppose the following asymptotic expansion on the spectrum holds:
\begin{equation}\label{laplace_full}
\hat{u}_k \sim C \frac{1}{k^{\alpha+1}} e^{-\delta k} \left( 1+ \frac{\gamma_1}{k}+ \frac{\gamma_2}{k^2}+\frac{\gamma_3}{k^3}+
 O(k^{-4}) \right).
\end{equation}
We apply successively the following six transformations
to identify the parameters $C$, $\alpha$, $\delta$, $\gamma_1$, $\gamma_2$, $\gamma_3$:
\begin{eqnarray}\label{vdhoven}
&&\hat{u}_k \rightarrow SR(\hat{u}_k)=\hat{u}^{(1)}_k \rightarrow -D(\hat{u}^{(1)}_k)=\hat{u}^{(2)}_k \rightarrow I(\hat{u}^{(2)}_k)=\hat{u}^{(3)}_k \nonumber \\
&&\rightarrow D(\hat{u}^{(3)}_k)=\hat{u}^{(4)}_k \rightarrow D(\hat{u}^{(4)}_k)=\hat{u}^{(5)}_k \rightarrow D(\hat{u}^{(5)}_k)=\hat{u}^{(6)}_k,
\end{eqnarray}
where
 \begin{eqnarray}
 I &:& U_k \rightarrow \frac{1}{U_k} \nonumber \\
 R &:& U_k \rightarrow \frac{U_k}{U_{k-1}} \nonumber \\
 SR &:& U_k \rightarrow \frac{U_k U_{k-2}}{U_{k-1}^2} \nonumber \\
 D &:& U_k \rightarrow U_k - U_{k-1}.  \nonumber
 \end{eqnarray}
The last term $\hat{u}^{(6)}_k = 3/(\alpha+1)$ is a constant which is easy to identify.
Inverting the chain \eqref{vdhoven}, one can find the values of the parameters in \eqref{laplace_full}.
This is the sixth stage procedure and in Table \ref{burgers_table} we give the results for the Burgers equation. As explained in \cite{PF07}, it is possible to consider other parameters in the asymptotic formula \eqref{laplace_full} which can be identified using more stages.

The asymptotic interpolation method is best computed using high digit precision computation.
In  \cite{PF07} the authors perform a six stages procedure with 80-digit precision
(and the 13-stage procedure with 120-digit precision)  obtaining data with accuracy of the order of $10^{-7}$.
Here we have performed the six stage procedure  in double precision and we note that
the coefficients $C$, $\alpha$ and $\delta$ have accuracy of the order of $10^{-4}$, while the coefficients $\gamma_1$, $\gamma_2$ and $\gamma_3$ have worst accuracy.

\begin{scriptsize}
\begin{table}[ht]
\begin{center}
\begin{tabular}{|l|l|l|l|l|l|}
\hline
& $t=0.8$ &  $t=0.85$ & $t=0.9$ & $t=0.95$ & $t=t_S=1$      \\
\hline
 $C$ &  $\;\; 0.70713615783 $ & $\;\;0.61442246421$ &  $\;\;0.51993097752$ &  $\;\;  0.43513434016$ & $\;\;\;0.47175434493 $\\
 \hline
 $\alpha$ & $\;\;0.49965144521  $ &  $\;\;0.49948664577 $ &  $\;\;0.49612632902 $ &  $ \;\;0.49624364601 $ & $ \;\;\;0.33296382179 $    \\
 \hline
 $\delta$ & $\;\;0.09606920975 $ & $\;\;0.05363644145$ &   $\;\;0.01544724997 $ & $\;\;0.00331071154 $ &  $\; \pm 0.00000355044 $    \\
\hline
 $\gamma_1$ &  $-1.47272447227 $ &  $-1.8634299638 $ & $-2.4808780250 $ & $-5.63365314047 $ & $ \;-0.66772591307 $    \\
\hline
 $\gamma_2$ &  $\;\;1.96722910228 $ &  $\;\;6.19701488066 $ & $  13.37505045755 $ & $ 89.95119911733 $  & $\;-0.12956190944  $ \\
\hline
 $\gamma_3$ & $13.10231991191  $ &  $34.55734201601 $ & $58.35449497230$    & $ 81.24726572859  $ & $ -83.44565528569  $ \\
 \hline
\end{tabular}
\vskip0.5cm
\caption{The results at different times for Burgers equation of the Van der Hoeven asymptotic interpolation method applied to \eqref{laplace_full}, using the six stage procedure \eqref{vdhoven}. \label{burgers_table}}
\end{center}
\end{table}
\end{scriptsize}

Formulas \eqref{sliding_alpha}-\eqref{sliding_C} for the sliding fitting , or the Van der Hoeven asymptotic interpolation method, are useful when the spectrum does not have oscillatory behavior, like in the case of Burgers equation considered.

In other case when $x^{*}\neq \pi$, one can fit $L + 1$ modes of the spectrum
starting from a certain $\hat{u}_k$ , i.e. the set $\left\{ \hat{u}_k , . . . , \hat{u}_{k+L} \right\} $, using again a least square fitting method. This way one determines certain $\delta_L(k)$ and
$\alpha_L(k)$ that, if relatively independent from $L$ and $k$ (at least in the central part of the spectrum), give
reliable information on the location and on the nature of the singularity. This technique is applied in \cite{CGS12,DLSS06} for the investigation on the blow up for the Camassa-Holm and the $b$-family equations.

%%%%%%%%%%%%%%%%%%%%%%%%%%%%%%%%%%%%%%%%%%%%%%%%%%%%%%%%%%%%%%%
\section{Secondary complex singularities}\label{sec:2}

As we have seen in the previous section, the method of complex singularity tracking  is useful
to study the blow up phenomena of an evolutionary PDE.
However the method gives information only to the  complex singularity
closest to the real axis.
Often it is also important to analyze the other singularities in the complex plain.
In this section we present some techniques that can reveal the presence of complex singularities
that, being more distant from the real axis, are hidden by  the main (the closest to the real axis)
complex singularity.

Application of the various methods presented here will be given in Section 4.

%%%%%%%%%%%%%%%%%%%%%%%%%%%%%%%%%%%%%%%%%%%%%%%%%%%%%%%%%%%%%%%
\subsection{Pad\'{e} method}\label{subsec:2.1}

In this section we recall the Pad\'{e} approximations.

Suppose there is a
complex
function $f(z)$ expressed by a power series $f(z)=\sum_{k=0}^{\infty}f_iz^k$,
the Pad\'{e} approximant $P_{L/M}$
is a rational function approximating $f$, such that
\begin{equation}
f(z)\approx\frac{\sum_{i=0}^{L}a_i z^i}{1+\sum_{j=1}^{M}b_j z^j}=P_{L/M}(z),
\label{pade}
\end{equation}
with the property that
\begin{equation}\label{pade_property}
f(z) - P_{L/M} (z) = O (z^{L+M+1} ),
\end{equation}
where $L+1$ and $M$ are the number of coefficients in the numerator and
denominator respectively.

The $M$ unknown denominator coefficients $b_j, j=1\ldots,M$ and the $L+1$ unknown, $a_i, i=0,\ldots,L$ are
 determined uniquely by \eqref{pade_property} equating coefficients of equal powers of $z$ between
 $(\sum_{i=0}^{\infty}f_iz^i) (1+\sum_{j=1}^{M}b_j z^j)$  and $\sum_{i=0}^{L}a_i
 z^i$,
 setting the coefficients of order greater than $L$ equal to zero, and $b_0=1$
 by definition.
 The following  set of $M$ linear equations must be solved
 \begin{equation}
 \begin{array}{c}
 b_Mf_{L-M+1}+b_{M-1}+\ldots+b_0f_{L+1}=0,\\
 \vdots \\
 b_Mf_L+b_{M-1}f_{L+1}+\ldots+b_0f_{L+M}=0.
 \end{array}
 \label{padematrix}
 \end{equation}
 Then the $L+1$ unknown numerator coefficients $a_i, i=0,\ldots,L$
 follow from $(\sum_{i=0}^{\infty}f_iz^i) (1+\sum_{j=1}^{M}b_j
 z^j)=\sum_{i=0}^{L}a_i z^i$ by equating coefficients of equal powers of $z$
 less then or equal to $L$.

It is possible to use Pad\'{e} approximants for Fourier series \cite{BGM96,Wei03}. Consider
\begin{equation}
u(x) \approx \sum_{k=-N}^{N} \hat{u}_k e^{i k x},
\end{equation}
an approximate solution to a PDE at a specific time t.
If we denote by $z_{+} = e^{ix}$ and $z_{-} = e^{-ix}$, the Fourier series on the right may be
expressed as the sum of two power series in the complex variables $z_{+}$ and $z_{-}$:
\begin{equation}
u(x) \approx \sum_{k=0}^{N} \hat{u}_k z_{+}+\sum_{k=0}^{N} \hat{u}_k z_{-}-\hat{u}_0.
\end{equation}

Both power series on the right may now be converted to Pad\'{e} approximants
\begin{equation}
 u(z) \approx P_{L/M}(z_{+})+Q_{L/M}(z_{-}) - \hat{u}_0 ,
\end{equation}
with $L+M+1=N$.

 The advantage of the Pad\'{e} approximation method is that it allows one to continue the
 function $f$ even beyond the radius of convergence (or strip of analyticity),    although  convergence issues can arise
 near branch points or branch cuts.
 The disadvantage of the Pad\'{e} approximation method is that not all of the singularities
 represented by a general $P_{L/M}$ are singularities of the function being
 approximated.

 In fact, there are several examples (see, for example, \cite{BGM96}) for which some defects or
 spurious singularities can appear.
 However, these defects can in principle be detected as the spurious singularities generally manifest
 as a pole very close to the zeros of $P_{L/M}$.  Moreover these spurious singularities
 have a transient nature   as they usually disappear by changing the degrees of the
 Pad\'{e} approximation. Another issue is represented by the fact that the linear system \eqref{padematrix} is close to being
singular (ill-conditioned), particularly when one seeks  a high degree Pad\'e approximant.
To overcome these problems one possibility is to compute function values of $P_{L/M} (z)$ for a given $z$. This technique  may be done efficiently using Wynn's epsilon algorithm, and we refer to \cite{BGM96}  for the details.
A different possibility instead is to use high numerical precision computation:
in this paper we shall focus on this technique.

If the singularities are poles it is easy to locate and characterize the singularities.
If we have the explicit expression of $P_{L/M}$ (found by solving the linear system \eqref{padematrix}) one can  simply compute the roots of the denominator of $P_{L/M}$.
Moreover, the algebraic character of the poles is the algebraic multiplicity of the roots.
On the other hand, if one has computed numerically the values of Pad\'e approximants  on given points $z$
 it is possible to find the poles searching the maximum of the following function:
\begin{equation}
 f (z) = \log |u(z)|.
\end{equation}
 Instead, to compute the order of the pole, it is possible to use the argument principle  \cite{Wei03}:
\begin{equation}
\frac{1}{2\pi i} \int_C \frac{u^{'} (z)}{u(z)} dz = Z(u) - P (u),
\end{equation}
where $C$ is a closed curve (for computational reasons it is possible to choose $C$ as the circle centered in the poles location) and where $Z(u)$ is the number of zeros and $P (u)$ is the number of poles (counting multiplicity) of $u$ inside $C$.
As it is known, if $u$ is analytic and nonzero at each point of a simple closed positively
oriented contour $C$, and inside $C$ the only singularities of $u$ are poles, if $C$ is chosen very close to the poles then $Z(u)=0$ and $P(u)$ determines the algebraic order of the pole singularity. If the singularity is an algebraic branch points or other type of singularity like logarithmic
 branch points or essential singularities, the
singularity appears as a string of poles and zeros located along the branch cut (see \cite{BGM96,Wei03}).

 Pad\'{e} approximants also have been used in the
 analysis of complex singularities
 of various ordinary differential equations (see \cite{COW83,GSSC14,Gu89}).
 The theoretical and practical issues related to
 Pad\'{e}-based methods are so numerous that it is
 impossible to cite them all here, and the reader is referred to \cite{BGM96} and \cite{Gu89} for a good
 discussion of this topic.

\begin{figure}
\includegraphics[width=13.5cm]{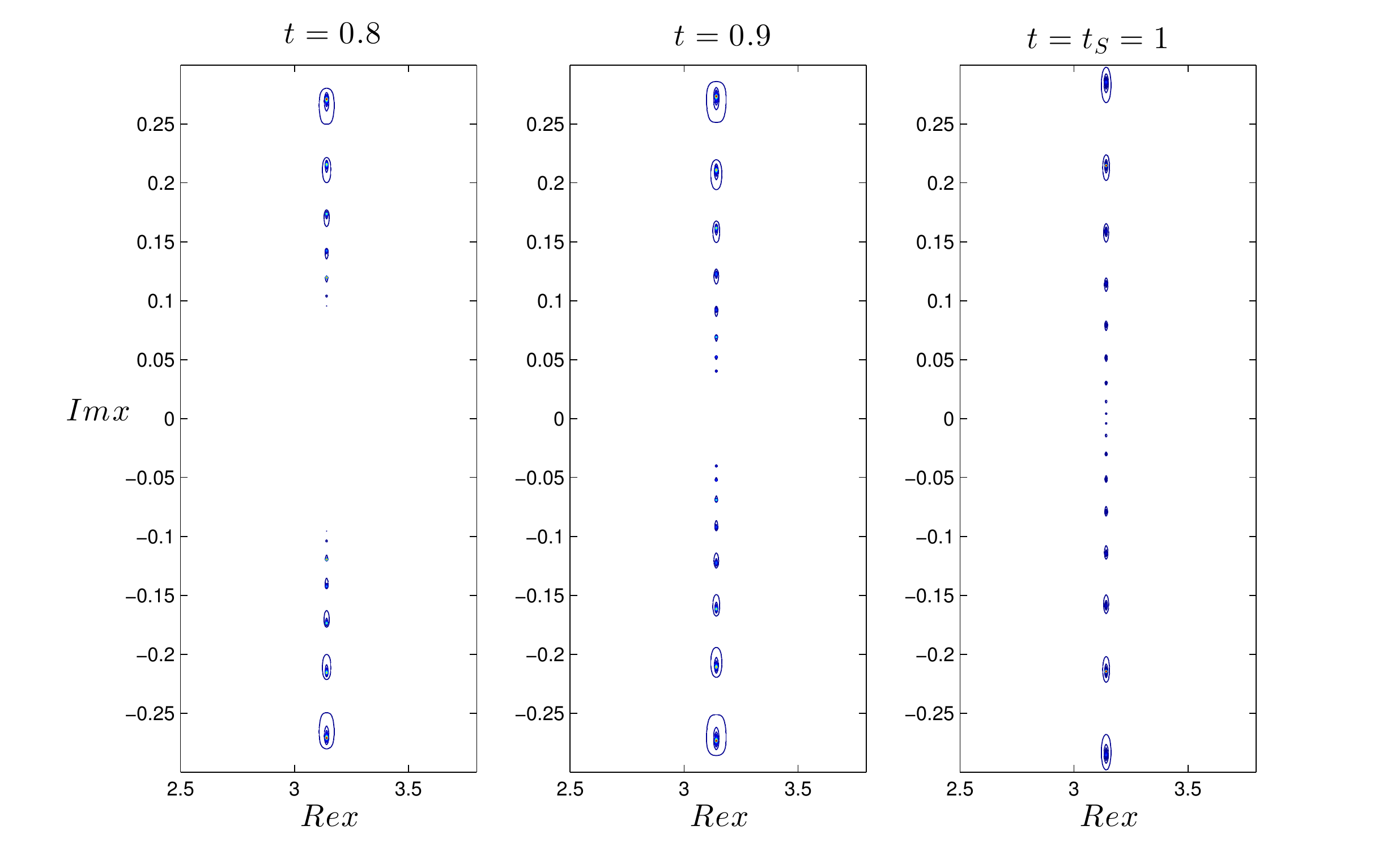}
\caption{Inviscid branch point and branch cuts at different times for the inviscid Burgers equation.}
\label{burgers_pade}
\end{figure}

 In the rest of this  paper we shall see several instances where the Pad\'e approximants are
 an effective tool to detect singularities which are outside the strip of analyticity of a Fourier series.
Here we present as an example, the Pad\'e approximants of the solution
of the inviscid Burgers equation \eqref{burgers} with initial datum
\eqref{IC-burgers}. We compute the coefficients of the Pad\'e
approximant $P_{L/M}$, with $M=L=50$,  solving the algebraic system
\eqref{padematrix} and considering the Fourier coefficients of the
numerical solution of Burgers equation.
In Fig.\ref{burgers_pade} it is shown the absolute value, at different
times, of the analytic continuation $P_{L/M}(z)$ in the complex plane of
the solution of Burgers equation.
As we said in Section 2, the solution has two square root singularities,
placed symmetrically
on the imaginary axis with respect to the origin. These singularities
move from infinity at the initial time toward the real axis, where they
meet at the singularity time $t_S=1$.
The singularities of $P_{L/M}(z)$ appears in Fig.\ref{burgers_pade} as a
string of poles located at the corresponding branch cut which is the
imaginary axis. The singularities approach the real axis according to
the results of the singularity tracking method in Fig.\ref{burgers_fitt}.

%%%%%%%%%%%%%%%%%%%%%%%%%%%%%%%%%%%%%%%%%%%%%%%%%%%%%%%%%%%%%%%
\subsection{Kida technique}\label{subsec:2.2}

The method introduced by Kida in \cite{Ki86} consists in filtering the function $u(x)$ with

\begin{equation}
v(x|X,d)=u(x)G(x|X,d),
\end{equation}
where $G(x|X,d)$ is a Gaussian-type filtering function which as a peak in $x=X$ with standard deviation $d$:
\begin{equation}\label{gaussian}
G(x|X,d)=\frac{1}{d\sqrt{\pi}}\exp \left[ -\frac{(x-X)^2}{d^2}\right].
\end{equation}
Taking the Fourier transform
\begin{equation}
\hat{v}(k|X,d)=2 \pi \int_{-\infty}^{\infty}\hat{u}(\xi)\hat{G}(k-\xi|X,d)d \xi,
\end{equation}
with
\begin{equation}
\hat{G}(k|X,d)=\exp \left[ -\frac{k^2 d^2}{4}+ i k X\right] \, ,
\end{equation}
and using the Laplace asymptotic formula \eqref{laplace} to the previous relation one has:
\begin{equation}\label{kida_filtered_spectrum}
\hat{v}(k|X,d) \sim \sum_{j} A_j |k|^{-\alpha_j-1} \exp \left[ ikz_j-\frac{(X-z_j)^2}{d^2} \right],
\end{equation}
while the exponential decay is given by
\begin{equation}
\Delta_j(k|X,d)=-\delta_j k-\frac{-(X-x^*_j)^2-\delta_j^2}{d^2}
\end{equation}
where $z_j=x^*_j+i \delta_j$ are the  complex singularities.

Note that $\Delta_j(k|X,d)$ is larger for $x_j^*$ closer to $X$ or for smaller $\delta_j$ as long as $k\leq \delta_j/d^2$.

If one denotes by $j=J$ the term which gives the maximum of $\Delta_j(k|X,d)$ over a certain range of $k$, then the previous formula can be approximate by

\begin{equation}\label{kida_filtered_spectrum_max}
\hat{v}(k|X,d) \sim A_J |k|^{-\alpha_J-1} \exp \left( -k \delta_J \right),
\end{equation}
which decrease exponentially in $k$.
One can therefore estimate the $\delta$ of the most relevant singularities which exist within
distance $d$ from $X$, by estimating the exponential decay rate of the spectrum.
%%%%%%%%%%%%%%%%%%%%%%%%%%%%%%%%%%%%%%%%%%%%%%%%%%%%%%%%%%%%%%%
\subsection{Borel-Polya-Van der Hoeven  method}\label{subsec:2.3}

In this section we review the Borel-Polya-Van der Hoeven  method (BPH method in the sequel) proposed in \cite{PF07} to retrieve more information about the  singularities outside the width
of the analyticity strip of the Burgers equation for different initial conditions. This method is useful when one deals with a finite
number of distinct complex singularities (poles or branches).

\begin{center}
 \begin{figure}
 \includegraphics[width=9.0cm]{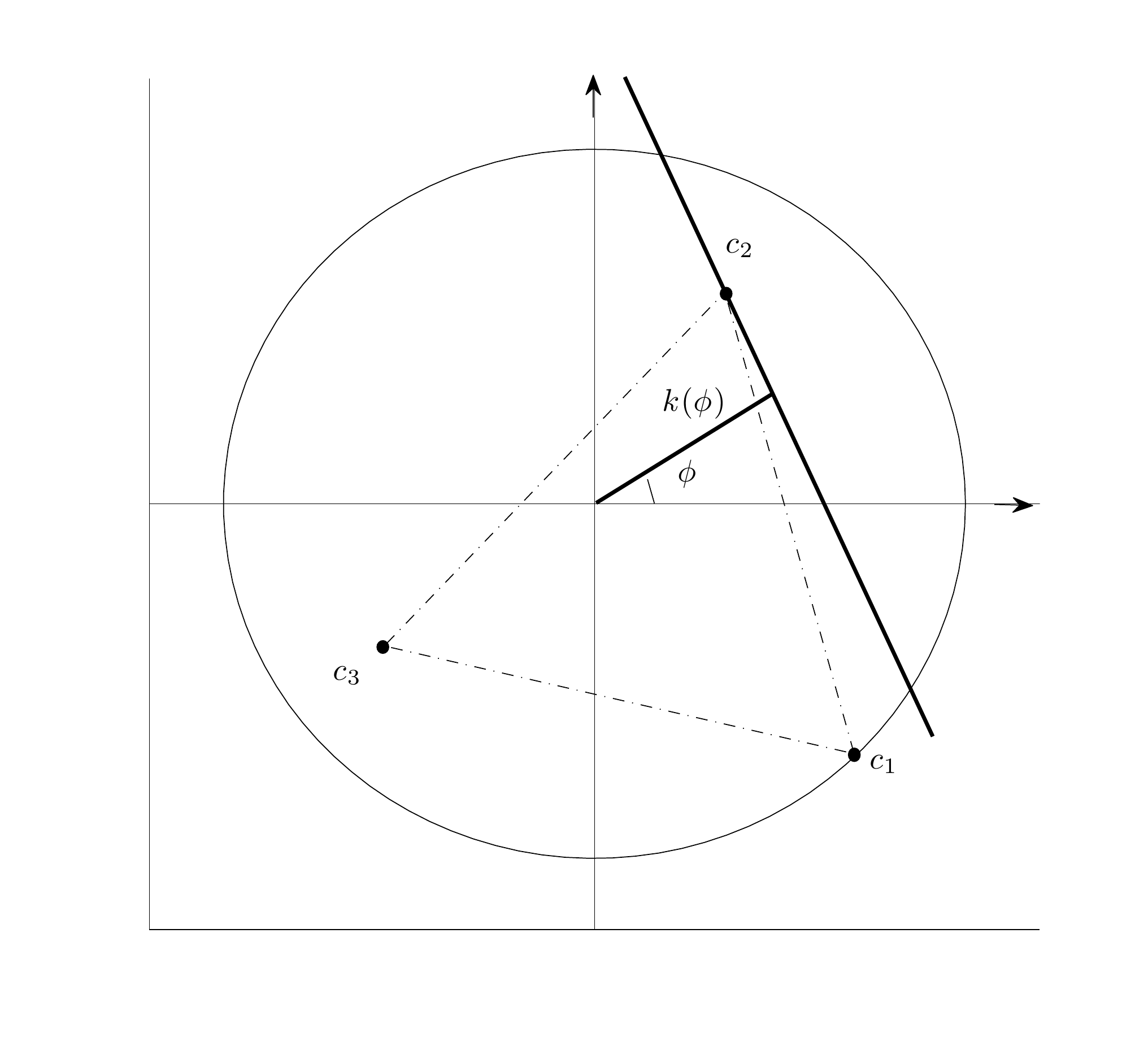}
 \caption{The convex hull of a discrete set of complex singularities is the
 smallest convex polygon containing all the singularities $c_i$.
 The \textit{supporting} function $k(\phi)$ is the distance from the origin to
 the supporting line, normal to the direction $\phi$, and
 touching a singularity.}
 \label{indicatrix_fig}
 \end{figure}
 \end{center}

In particular, given the inverse Taylor series $f(z) = \sum_{k=0}^N f_k/z^{k+1}$
that has $n$ complex singularities
$c_j=|c_j|$e$^{-i\gamma_j}$ for
$j=1,2,\ldots, n$, its Borel transform is given by $U_B(\zeta)=\sum_{k=0}^N
f_k\zeta^{k}/n!$.
Evaluating the modulus of the Borel series $G(r)=|U_B(r\textrm{e}^{i\phi})|$
along the rays $r\textrm{e}^{i\phi}$, one obtains, through a steepest descent
argument, the following asymptotic behavior
\begin{equation}
G(r)\approx C(\phi)r^{-(\alpha(\phi)+1)}\textrm{e}^{h(\phi)r} \quad
\textrm{for}\quad r \rightarrow \infty.
\label{borel_as}
\end{equation}
The function $h(\phi)$ is called the \textit{indicatrix} function of the Borel
transform.
To better understand the role
 of the indicatrix function, we consider the set $K=\{c_1,\ldots,c_n\}$ of
 all the singularities, and we define the \textit{supporting line} of $K$ as a line that has at least one point in common with $K$ and such that its points
 are in the same half space with respect to the supporting line of $K$.
 The intersection of all these half spaces is the \textit{convex hull} of $K$,
 which in the case of separate poles or branches reduces to the smallest
 convex polygon containing all the singularities as illustrated in Figure~\ref{indicatrix_fig}. The \textit{supporting} function $k(\phi)=h(-\phi)$ is the distance from the
 origin to the supporting line normal to $\phi$.
 In \cite{PF07}, it has been shown that, in the case of isolated singularities, the indicatrix function
 is the piecewise cosine function
\begin{equation}\label{indicatrix}
h(\phi)=|c_j|\cos(\phi-\gamma_j) \quad  \textrm{for} \quad
\phi_{j-1}<\phi<\phi_j,
\end{equation}
where the angular intervals $(\phi_j,\phi_{j+1})$ are depending on the complex positions of the singularities (we refer to \cite{PF07} for a deeper explanation on how
the set $\phi_j$, $j=1,2,\ldots,n$ is determined).
Therefore,
through numerical interpolation we can determine
the parameters $|c_j|$ and $\gamma_j$ that give the locations of the complex
singularities $c_j$. In practice, for each direction $\phi$ we need to determine
the exponential rate of \eqref{borel_as} that allows for construction of the indicatrix
function $h$. Moreover, an estimate of $\alpha(\gamma_j)$ in \eqref{borel_as}
returns the characterization of the singularity $c_j$.
The BPH method can be easily applied to the Fourier series $f(z)=\sum_{k=-K/2}^{k=K/2}f_k\textrm{e}^{ikz}$
by introducing the complex variables $Z_+=e^{iz},Z_-=e^{-iz}$ so that
 \begin{eqnarray}
 u(z)=\sum_{k=0}^{K/2}u_ke^{ikz}+\sum_{k=1}^{K/2}u_ke^{-ikz}=\sum_{k=0}^{K/2}
 u_k/Z_-^k+\sum_{K=1}^{K/2}u_k/Z_+^k.
 \label{padefourier}
 \end{eqnarray}

 The advantage of this methodology in comparison to the singularity-tracking
method lies in the fact that
it is possible to capture information on  the singularities located
outside the radius
of convergence of a Taylor series (or the strip of analyticity of a
Fourier series). However, there are some
drawbacks. In particular, singularities that are close to each other can be
difficult to distinguish, mainly because a cosine function  relative to a singularity $s_1$ can be \textit{hidden} by another cosine function relative to a singularity $s_2$,
if $s_2$ is closer to the real domain than $s_1$.  Using high numerical precision in conjunction with the asymptotic extrapolation method proposed in \cite{vDH09} can only in part
overcome this issue. Moreover, the computational cost
is heavier in comparison to the singularity-tracking method, as a numerical
interpolation must be performed in various directions containing all of the singularities.

%%%%%%%%%%%%%%%%%%%%%%%%%%%%%%%%%%%%%%%%%%%%%%%%%%%%%%%%%%%%%%%
\section{Applications}\label{sec:3}

%%%%%%%%%%%%%%%%%%%%%%%%%%%%%%%%%%%%%%%%%%%%%%%%%%%%%%%%%%%%%%%
\subsection{Dispersion and dissipation}\label{sec:3.1}

In this section we shall apply   the techniques explained in the previous sections to analyze the complex singularities of some nonlinear dissipative and nonlinear dispersive PDEs.
Many nonlinear dispersive systems, in the regime of small dispersion, exhibit rapid oscillations
in their spatio-temporal dependence.
Although a fascinating mathematical phenomenon, these oscillations are generally
quite difficult to describe and control and are an obstacle to the efficiency of numerical
and analytical methods.
A complete rigorous description of these oscillatory behavior would necessitate multiple scale analysis
(with the introduction of  fast variable to resolve the oscillatory structure)
and an asymptotic matching procedure. However the oscillatory structure has
been successfully analyzed only in cases like the KdV equation \cite{EGLS94,ELZ97,LL87I,LL83II,LL83III,GK12}
and the nonlinear Schrodinger equation \cite{JLM99,KMM03,TVZ04}.

The first examples we consider here is the viscous Burgers equation:
\begin{equation}\label{burgers_visc}
u_t+u u_x =\nu u_{xx},
\end{equation}
and the dispersive Burgers equation introduced in \cite{SCE96}:
\begin{equation}\label{burgers_disp}
u_t+u u_x =\epsilon \exp(i \theta) u_{xx}.
\end{equation}

\begin{figure}
\includegraphics[width=13.5cm]{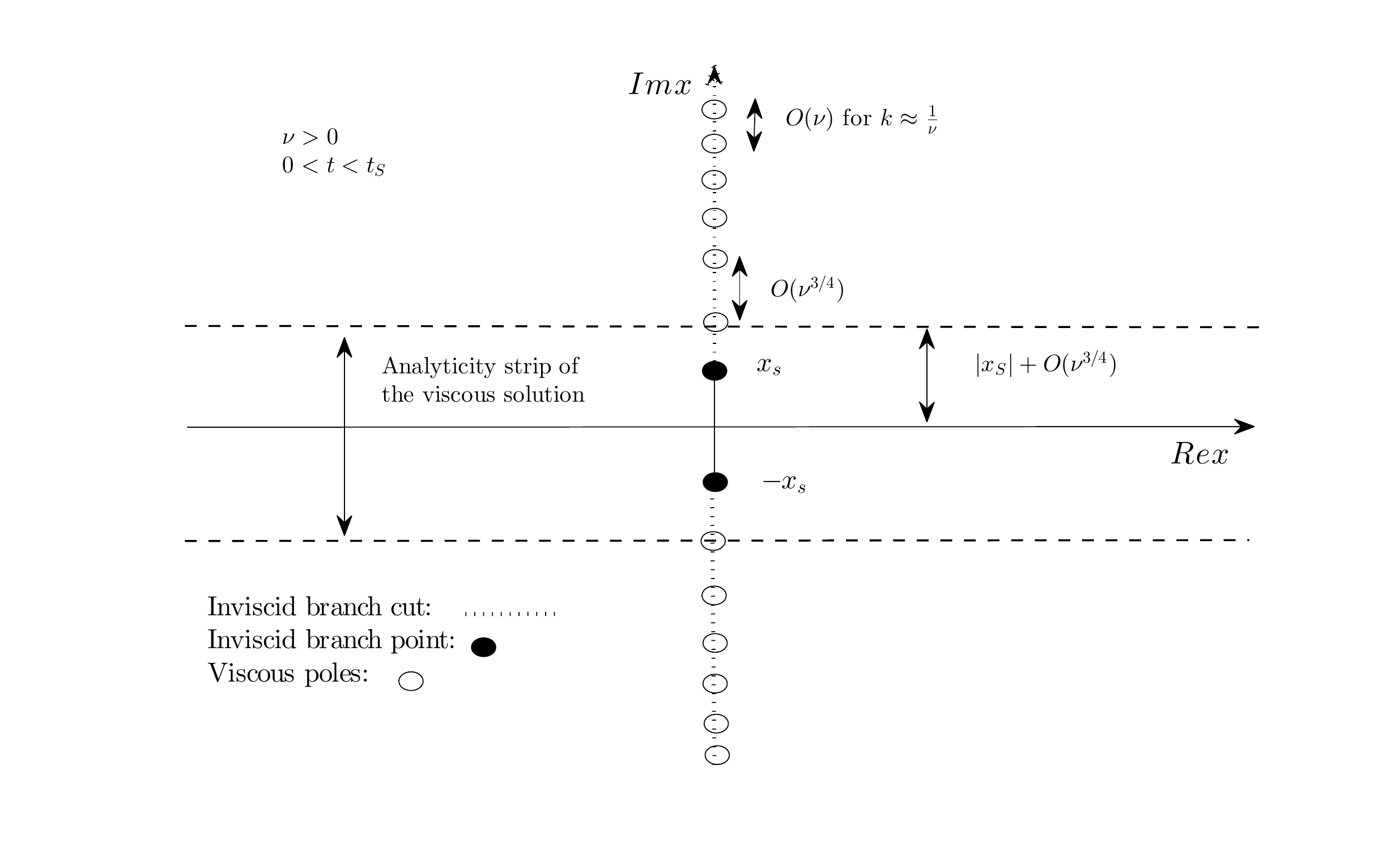}
\caption{Inviscid branch points, branch cuts and viscous poles for the viscous Burgers equation \eqref{burgers_visc} with $\nu>0$ and $0<t<t_S$.}
\label{pole_visc_1}
\end{figure}

\begin{figure}
\includegraphics[width=13.5cm]{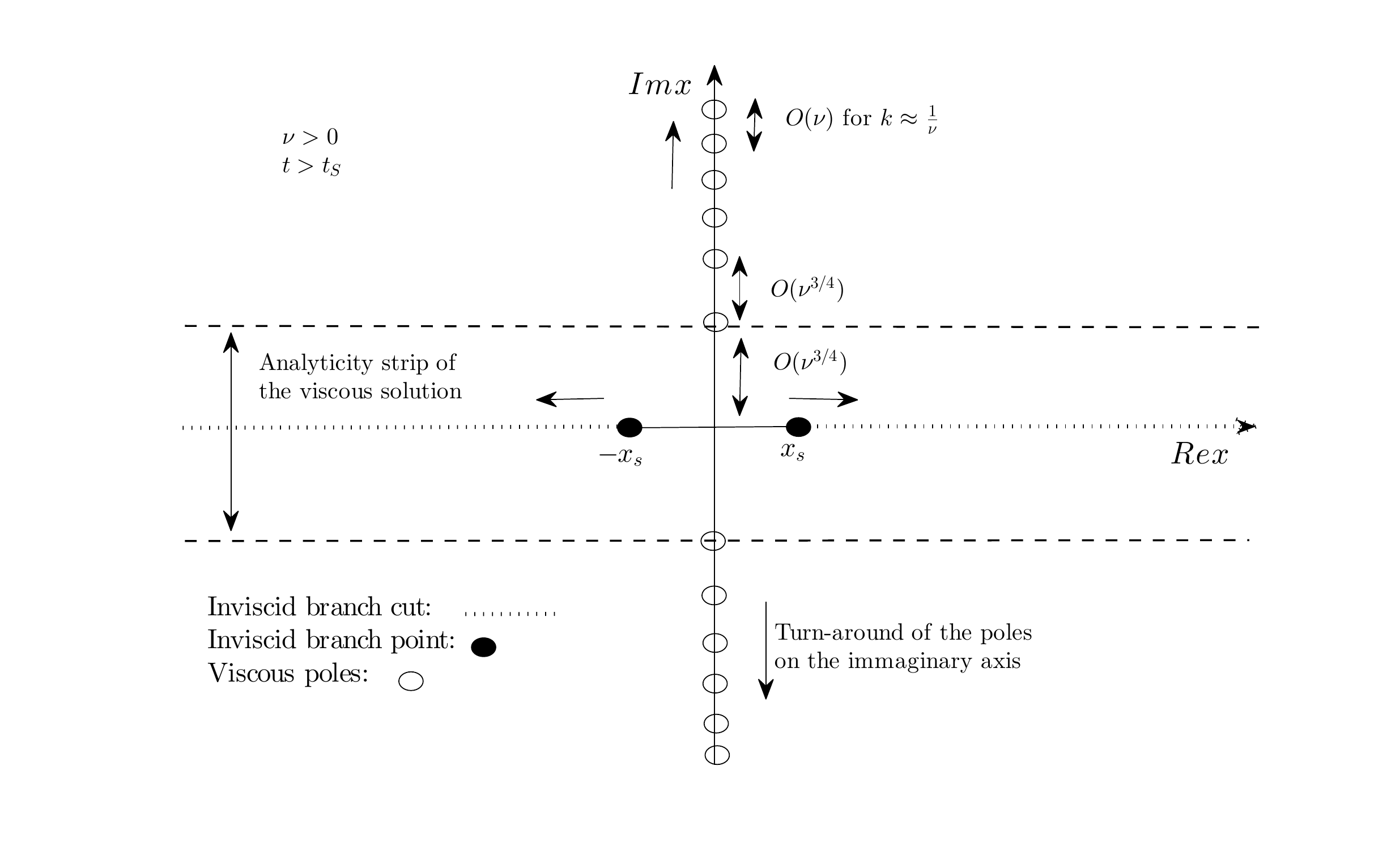}
\caption{Inviscid branch points, branch cuts and viscous poles for the viscous Burgers equation \eqref{burgers_visc} with $\nu>0$ and $t>t_S$.}
\label{pole_visc_2}
\end{figure}

\begin{figure}
\includegraphics[width=13.5cm]{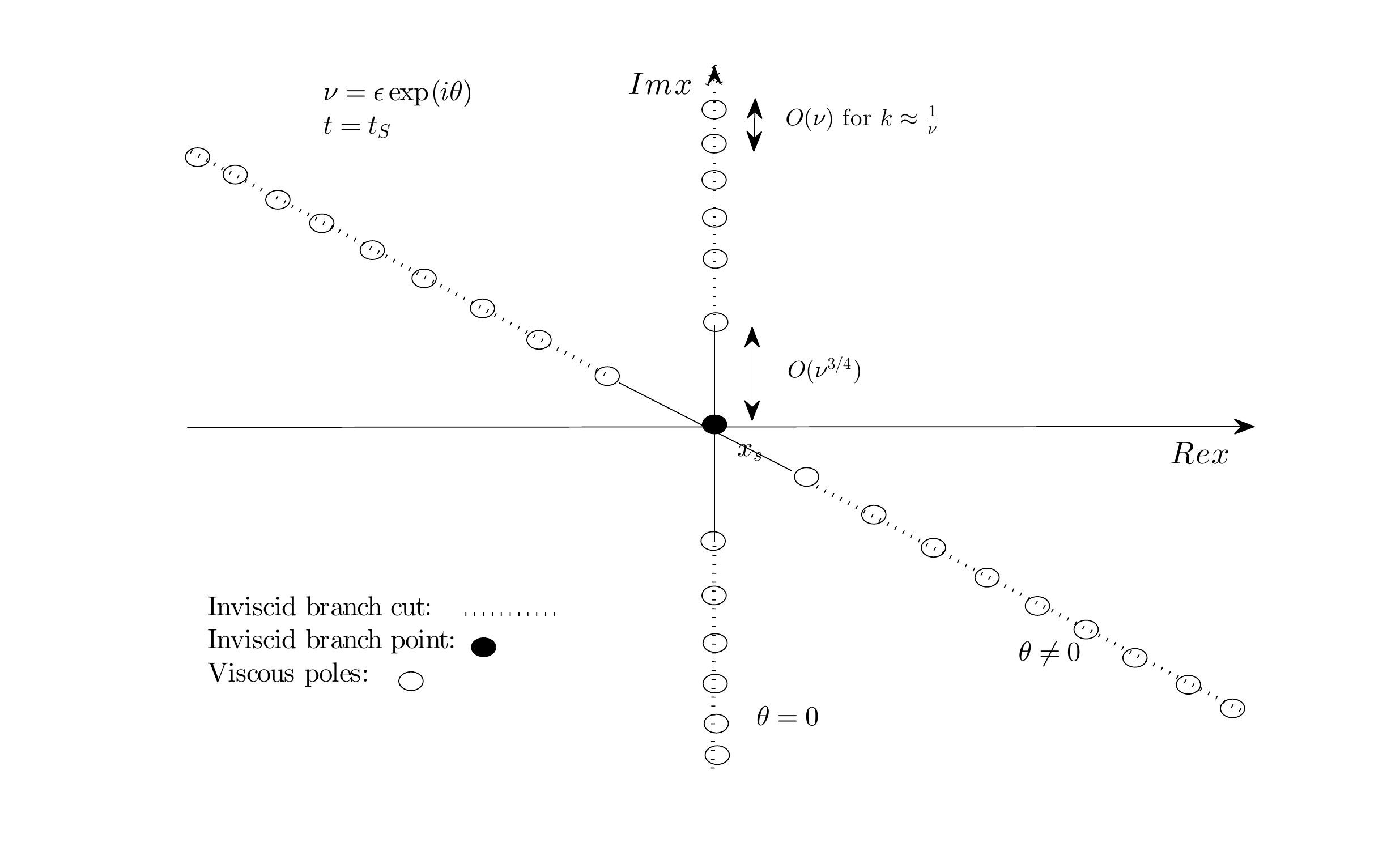}
\caption{Inviscid branch point, branch cuts and viscous poles at the singularity time $t=t_S$ for the dispersive Burgers equation \eqref{burgers_disp} with $\nu=\epsilon \exp(i\theta)$.}
\label{pole_disp}
\end{figure}

In \cite{Sen97I,Sen97II,SCE96}, the authors analyzed the pole dynamics of the above equations.
They showed the different behavior of the poles in presence of dissipation versus the presence of dispersion.
Their results are summarized in Figs \ref{pole_visc_1} and  \ref{pole_visc_2} and in Fig. \ref{pole_disp}.

In the zero-dispersion (or zero-viscosity limit), the complex poles coalesce onto a branch cut, and the zero-dispersion
solution is described by branch-cut dynamics.
As shown in the previous section, the cube root singularity is known to be a generic
singularity for the inviscid Burgers equation. It is due to the coalescence of two conjugate
branch points of order two in the complex plane \cite{SCE96}.
In the purely dispersive case, the solution of \eqref{burgers_disp} or \eqref{kdv}
develops rapid oscillations. These
oscillations are caused by the presence of complex poles in the solution
which have moved close to the
real axis.
This result is important in providing a tangible cause for
the formation of the oscillations.

The above results can be recovered through the application of the techniques presented in the
previous Section.
To \eqref{burgers_visc} and to \eqref{burgers_disp} we shall impose the initial datum $u(x,t=0)=\sin{x}$.

\begin{figure}
\includegraphics[width=13.5cm]{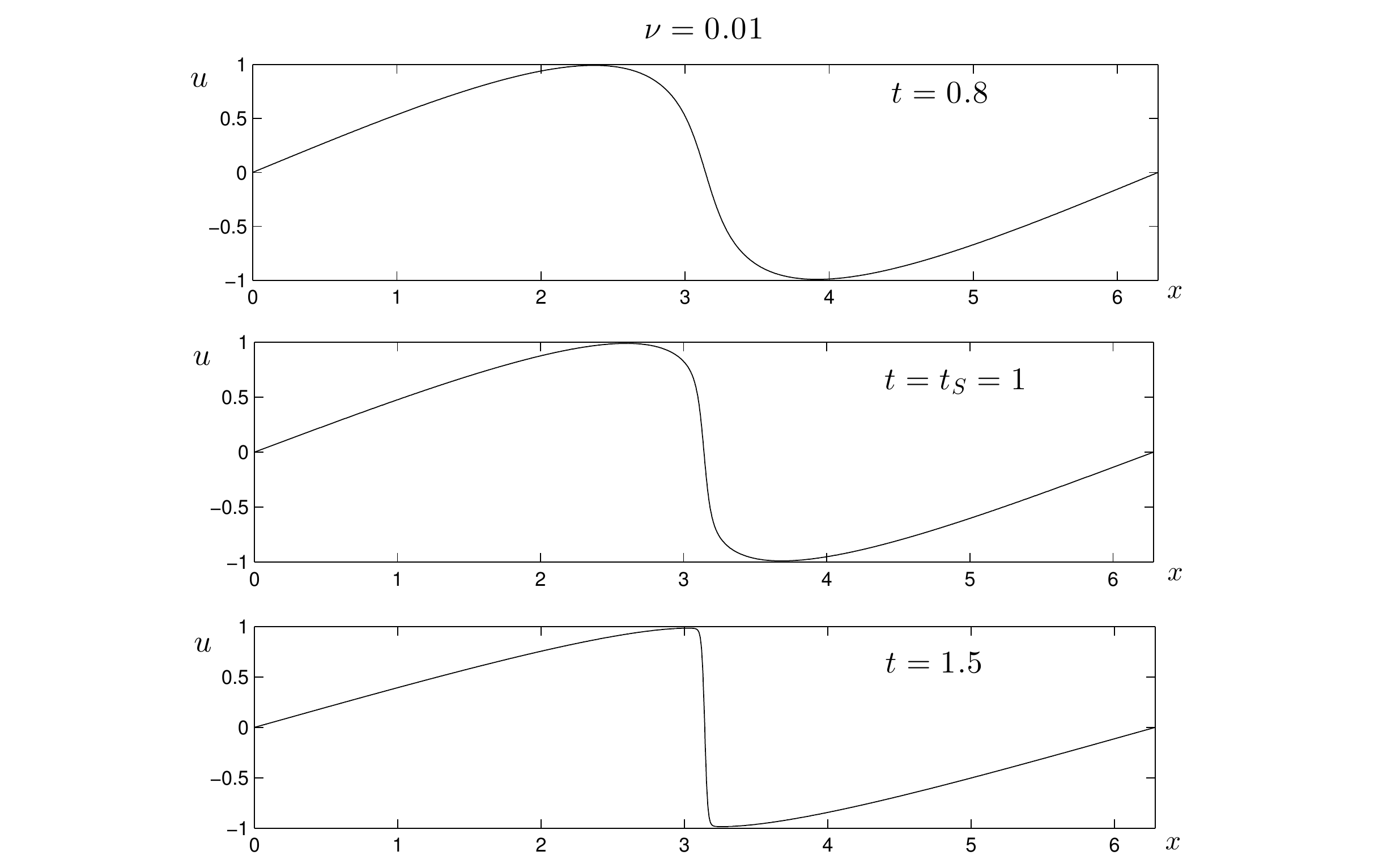}
\caption{The evolution in time for the viscous Burgers equation \eqref{burgers_visc} with $\nu=0.01$.}
\label{burgers_visc_fig}
\end{figure}

\begin{figure}
\includegraphics[width=13.5cm]{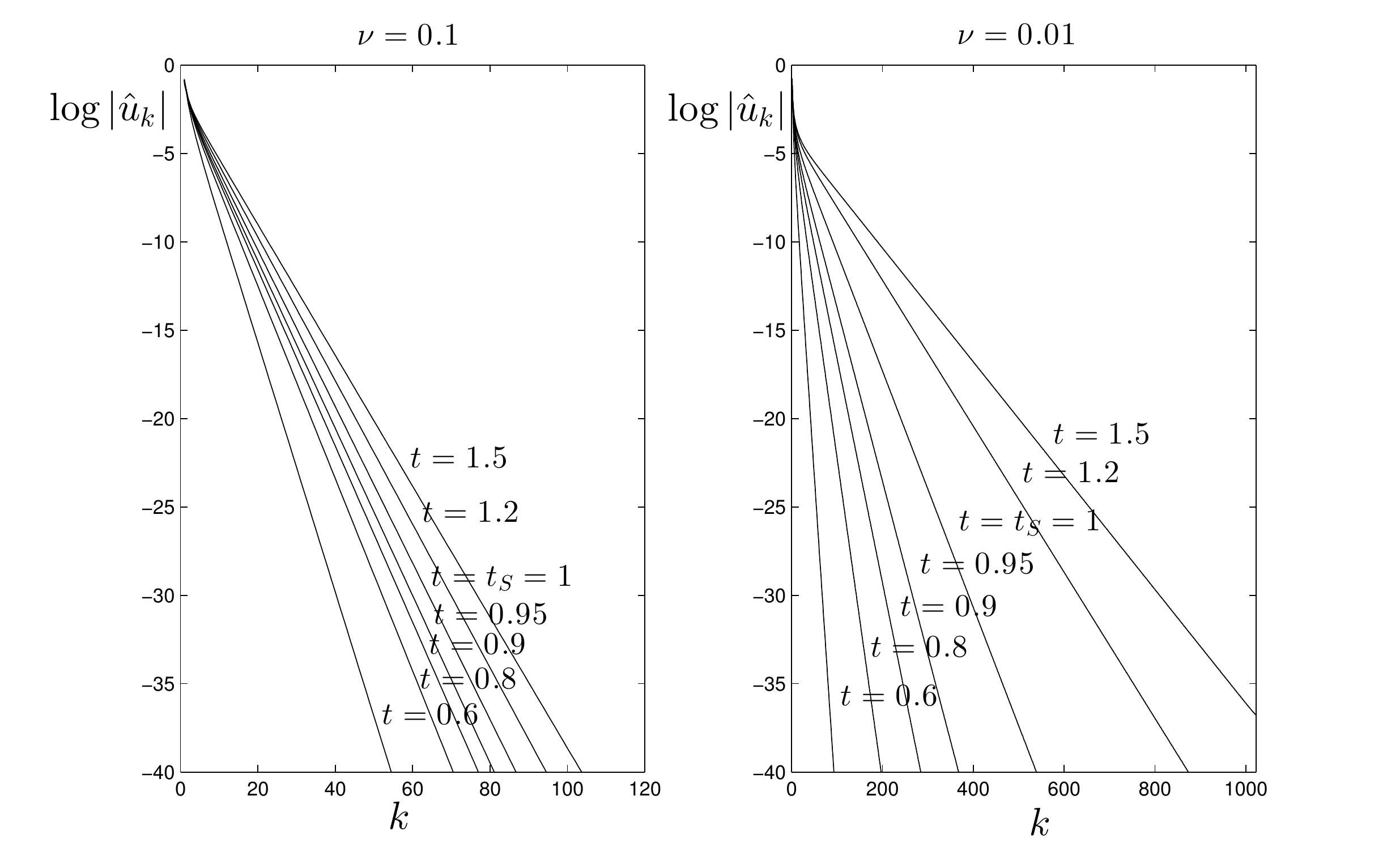}
\caption{The spectrum at different times for the viscous Burgers equation \eqref{burgers_visc} with $\nu=0.01$ and $\nu=0.1$.}
\label{burgers_visc_spectrum}
\end{figure}

In Fig.\ref{burgers_visc_fig} it is shown the evolution in time for the viscous Burgers equation \eqref{burgers_visc} with $\nu=0.01$ and in fig. \ref{burgers_visc_spectrum} the behavior of its spectrum at different times and different viscosity.

\begin{figure}
\includegraphics[width=13.5cm]{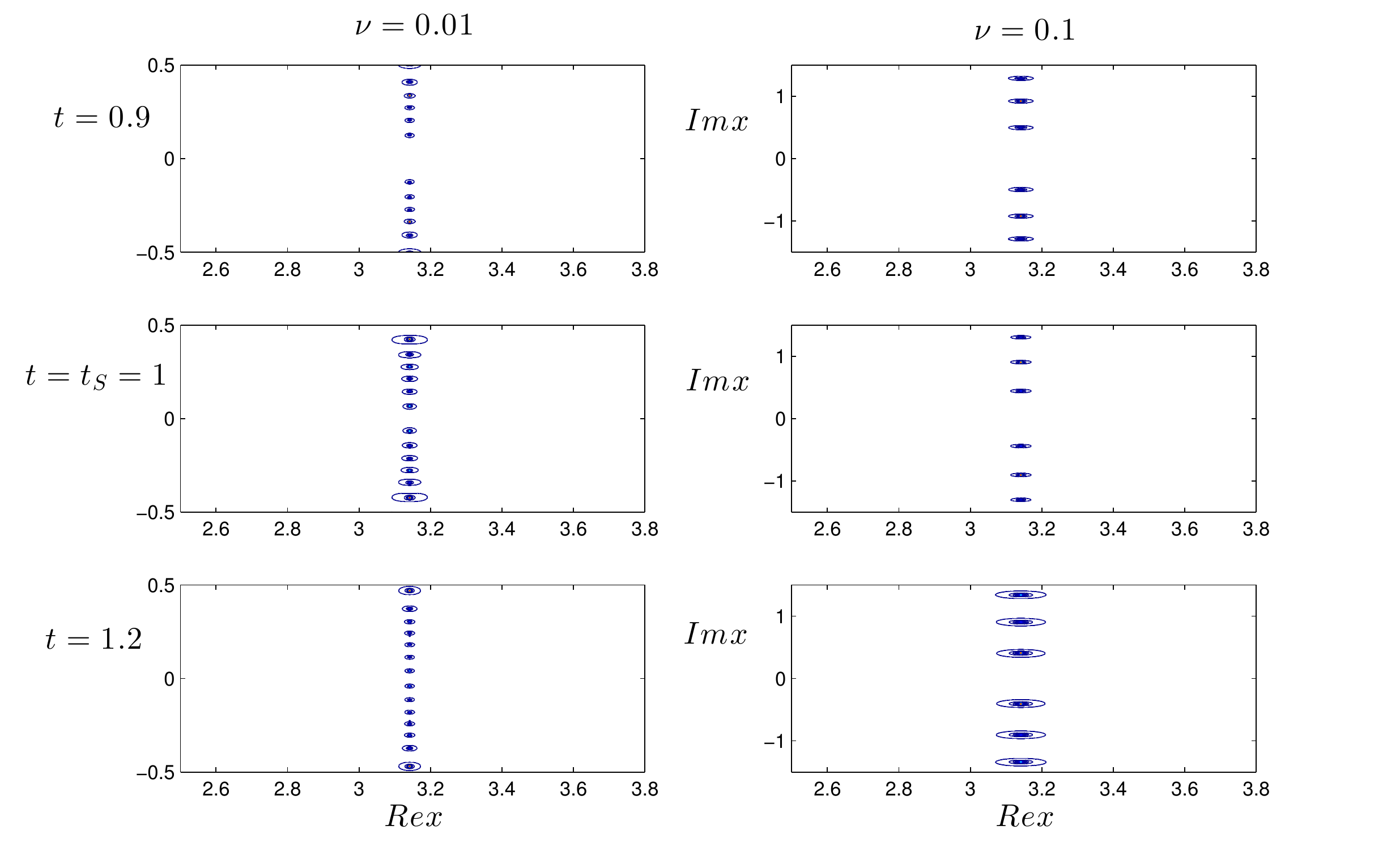}
\caption{Viscous poles at different times for the viscous Burgers equation \eqref{burgers_visc} with $\nu=0.1$ and $\nu=0.01$. The Pad\'e analysis results.}
\label{burgers_visc_pade}
\end{figure}

In Fig.\ref{burgers_visc_pade} it is shown the Pad\'e approximants for the viscous Burgers equation \eqref{burgers_visc} with two different viscosity $\nu=0.01$ and $\nu=0.1$.

\begin{figure}
\includegraphics[width=13.5cm]{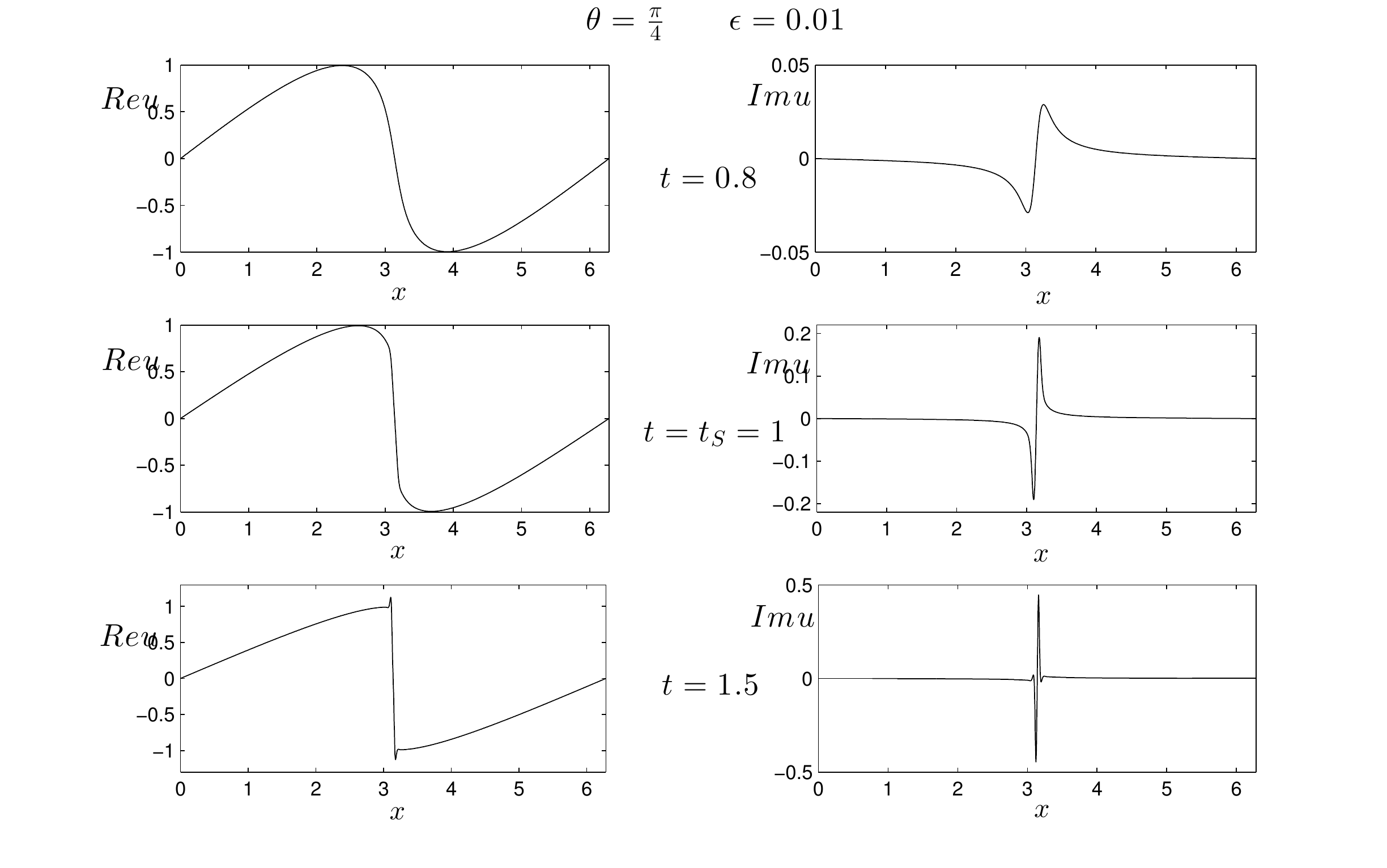}
\caption{The evolution in time for the dispersive Burgers equation \eqref{burgers_disp} with $\epsilon=0.01$ and $\theta=\pi/4$.}
\label{burgers_disp_fig}
\end{figure}

\begin{figure}
\includegraphics[width=13.5cm]{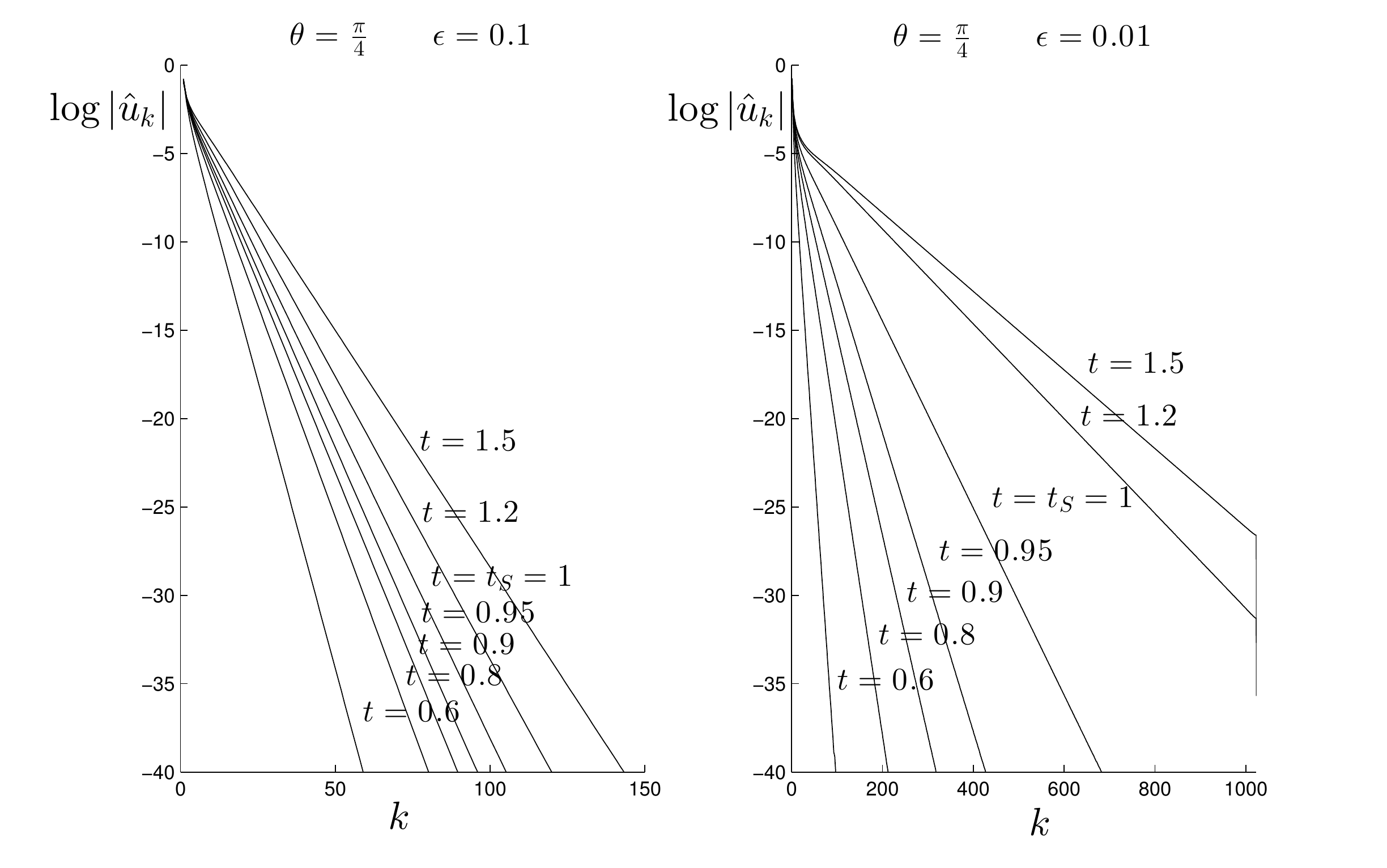}
\caption{The spectrum at different times for the dispersive Burgers equation \eqref{burgers_disp} with $\epsilon=0.01$ and $\epsilon=0.1$ and $\theta=\pi/4$.}
\label{burgers_disp_spectrum}
\end{figure}

In Fig.\ref{burgers_disp_fig} it is shown the evolution in time for the dispersive Burgers equation \eqref{burgers_disp} with $\epsilon=0.01$ and $\theta=\pi/4$ and in Fig.\ref{burgers_disp_spectrum} the behavior of its spectrum at different times and different dispersion value $\epsilon=0.1$ and $\epsilon=0.01$.

\begin{figure}
\includegraphics[width=13.5cm]{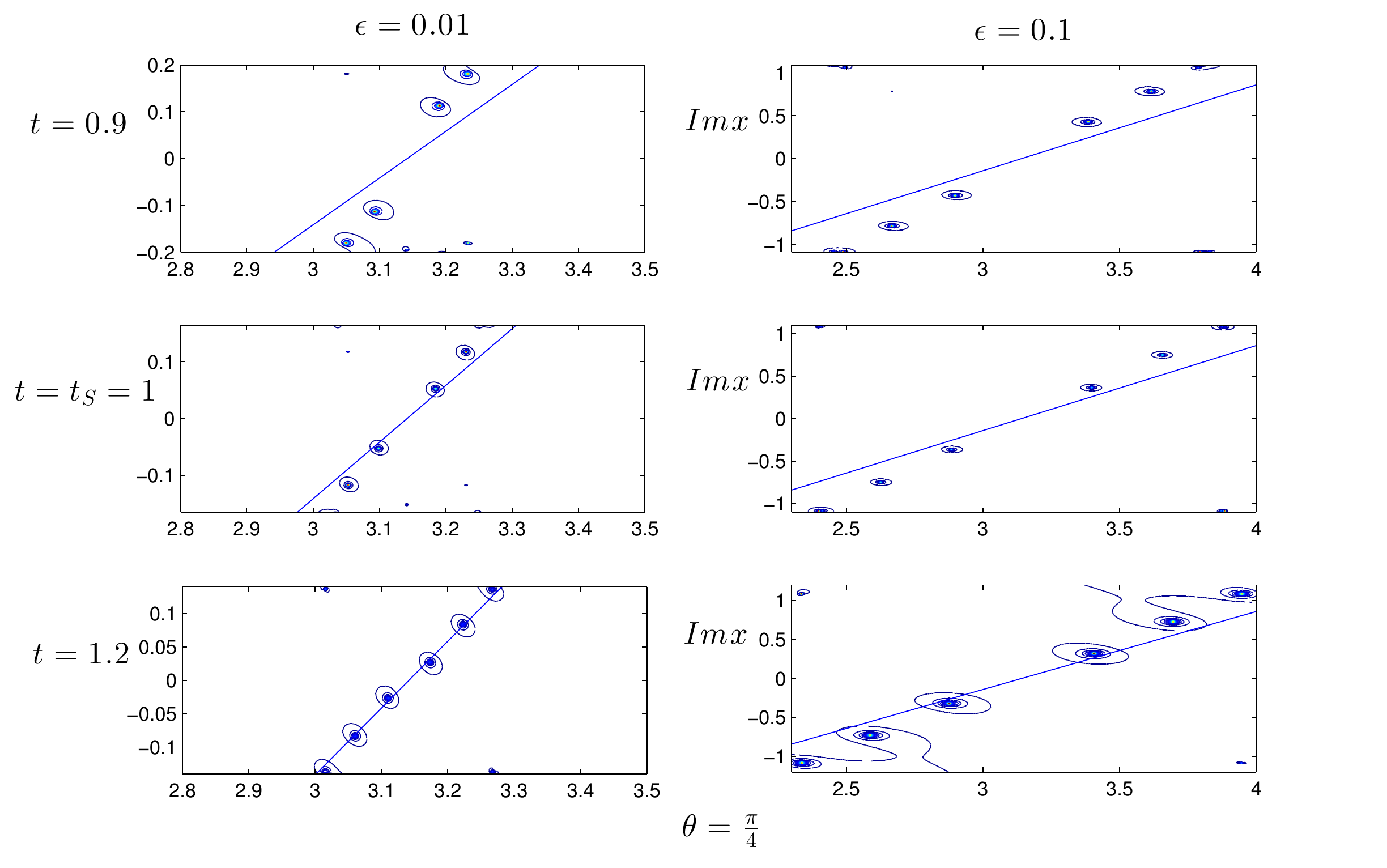}
\caption{Complex poles at different times for the dispersive Burgers equation \eqref{burgers_disp} with $\nu=\epsilon \exp(i\theta)$. The Pad\'e analysis results.}
\label{burgers_disp_pade}
\end{figure}

In Fig.\ref{burgers_disp_pade} it is shown the Pad\'e approximants for the dispersive Burgers equation \eqref{burgers_disp} with $\epsilon=0.01$ and $\epsilon=0.1$, and $\theta=\pi/4$.

One can notice an agreement between  the results of \cite{SCE96} sketched in Figs \ref{pole_visc_1},
\ref{pole_visc_2} and \ref{pole_disp}
and our analysis based on the Pad\'e approximants, reported in Fig.\ref{burgers_visc_pade} for the viscous case and in  Fig.\ref{burgers_disp_pade} for the dispersive cases.

We now pass to the analysis of the KdV equation:
\begin{equation}\label{kdv}
u_t+uu_x+\epsilon^2 u_{xxx}=0 \, .
\end{equation}
which is considered to be the canonical example of dispersive equation.
To the above equation we shall impose the initial datum $u(x,t=0)=\sin{x}$.
We shall see that the solution presents a series of complex singularities.
Moreover, in the zero dispersion limit, the singularities approache the real axis and tend to coalesce.
The dynamics of the KdV complex singularities seems therefore to be  analogous to what we have seen for
the dispersive Burgers equation.

\begin{figure}
\includegraphics[width=13.5cm]{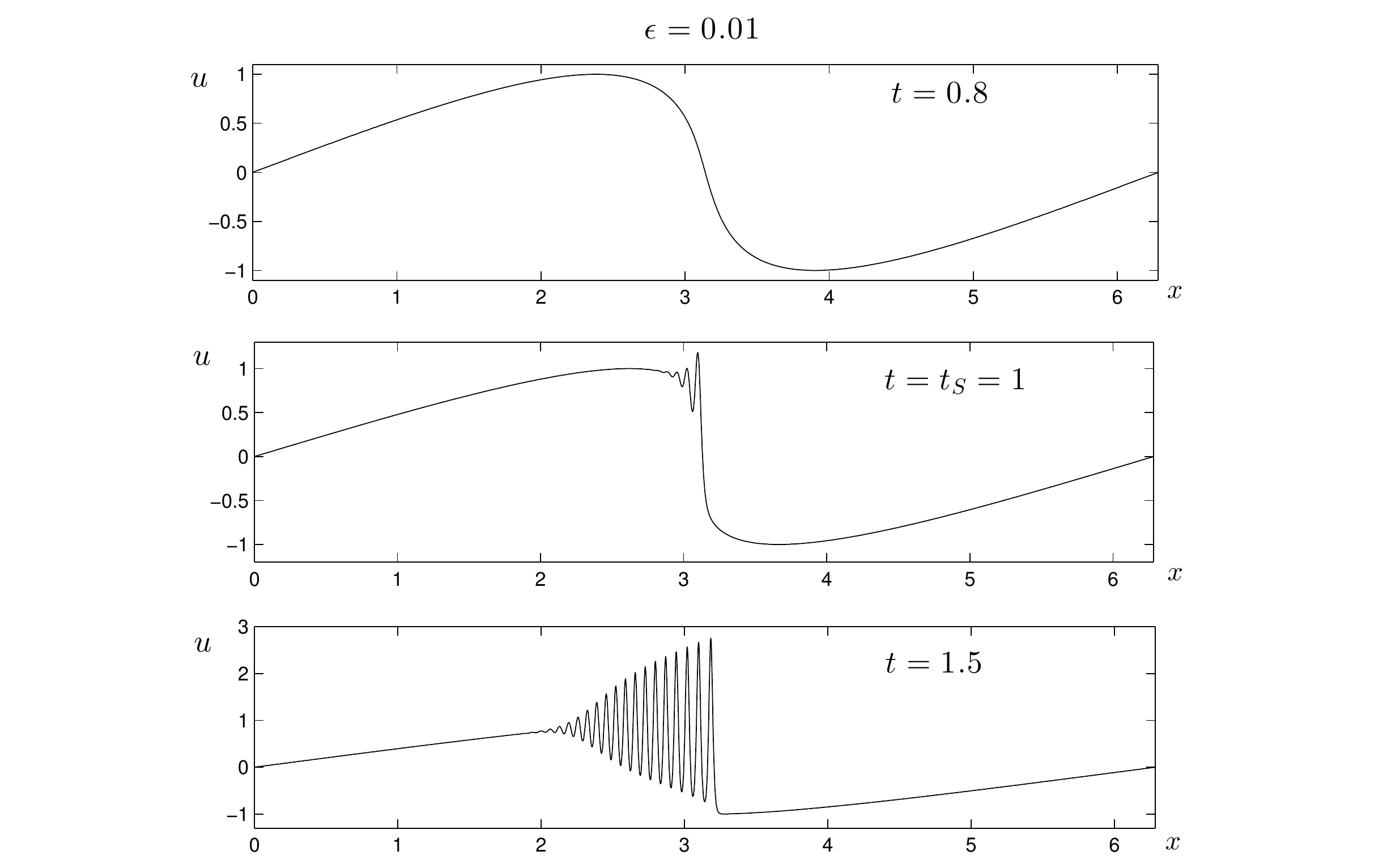}
\caption{The evolution in time for the dispersive KdV equation \eqref{kdv} with $\epsilon=0.01$.}
\label{kdv_fig}
\end{figure}

\begin{figure}
\includegraphics[width=13.5cm]{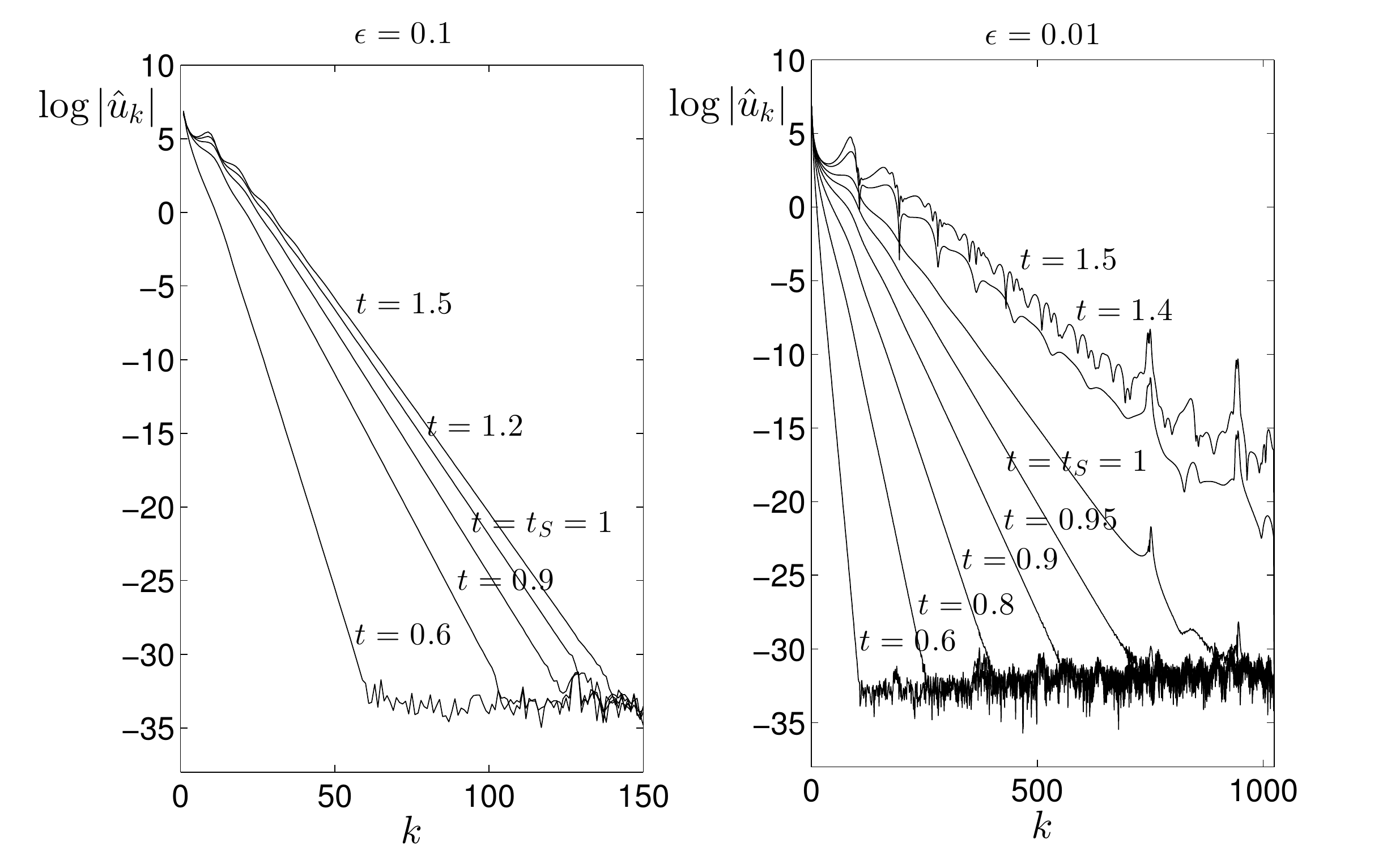}
\caption{The spectrum at different times for the dispersive KdV equation \eqref{kdv} with $\epsilon=0.01$ and $\epsilon=0.1$.}
\label{kdv_spectrum}
\end{figure}

In Fig.\ref{kdv_fig} it is shown the evolution in time for the KdV equation \eqref{kdv} with $\epsilon=0.01$ and in Fig.\ref{kdv_spectrum} the behavior of its spectrum at different times and different dispersion value $\epsilon=0.1$ and $\epsilon=0.01$.

\begin{figure}
\includegraphics[width=13.5cm]{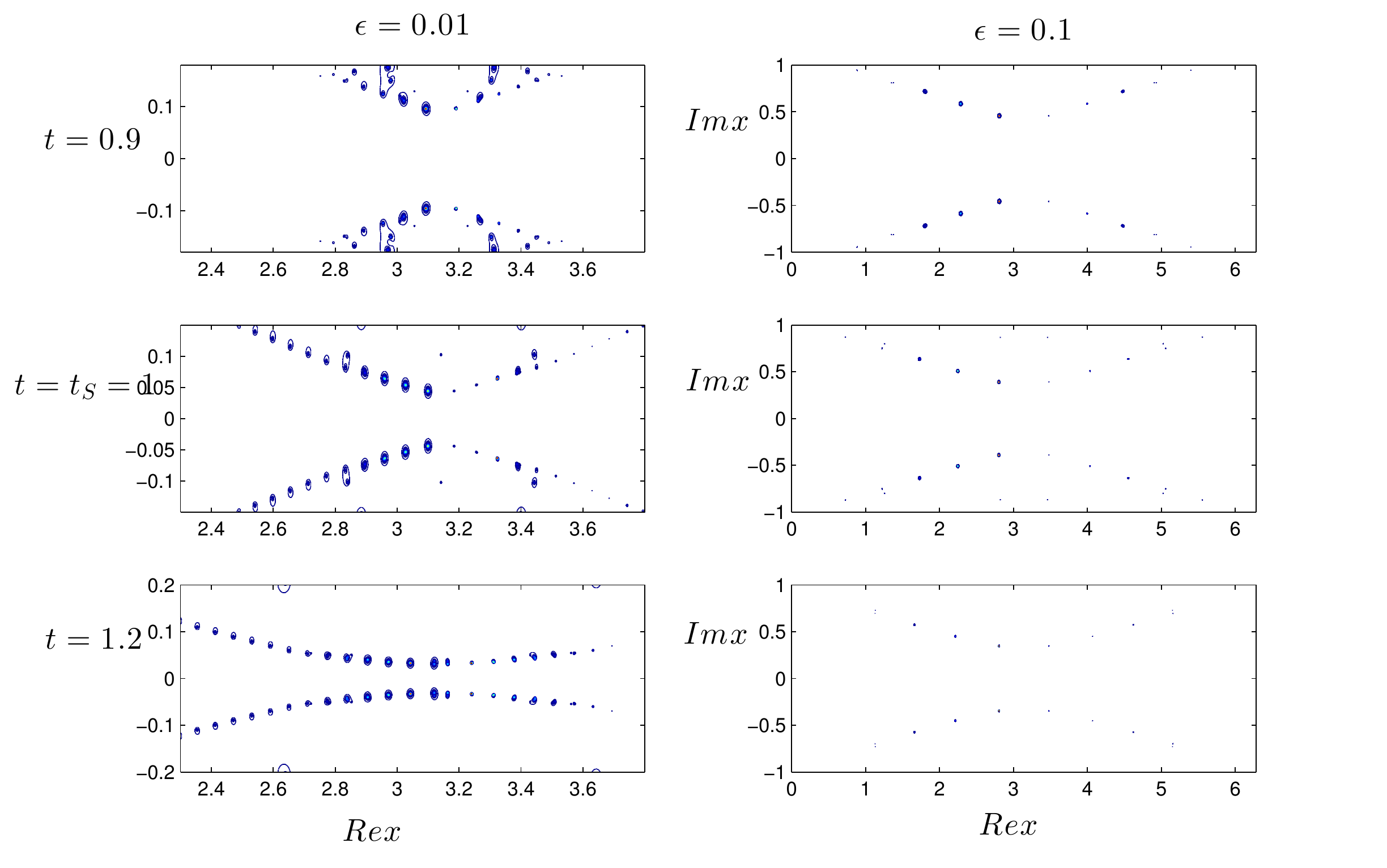}
\caption{Poles at different times for the dispersive KdV equation \eqref{kdv} with $\epsilon=0.01$ and $\epsilon=0.1$. The Pad\'e analysis results.}
\label{kdv_pade}
\end{figure}

In Fig. \ref{kdv_pade} it is shown the Pad\'e approximants for the KdV equation \eqref{kdv} with $\epsilon=0.01$ and $\epsilon=0.1$.

\begin{figure}
	\includegraphics[width=13.5cm]{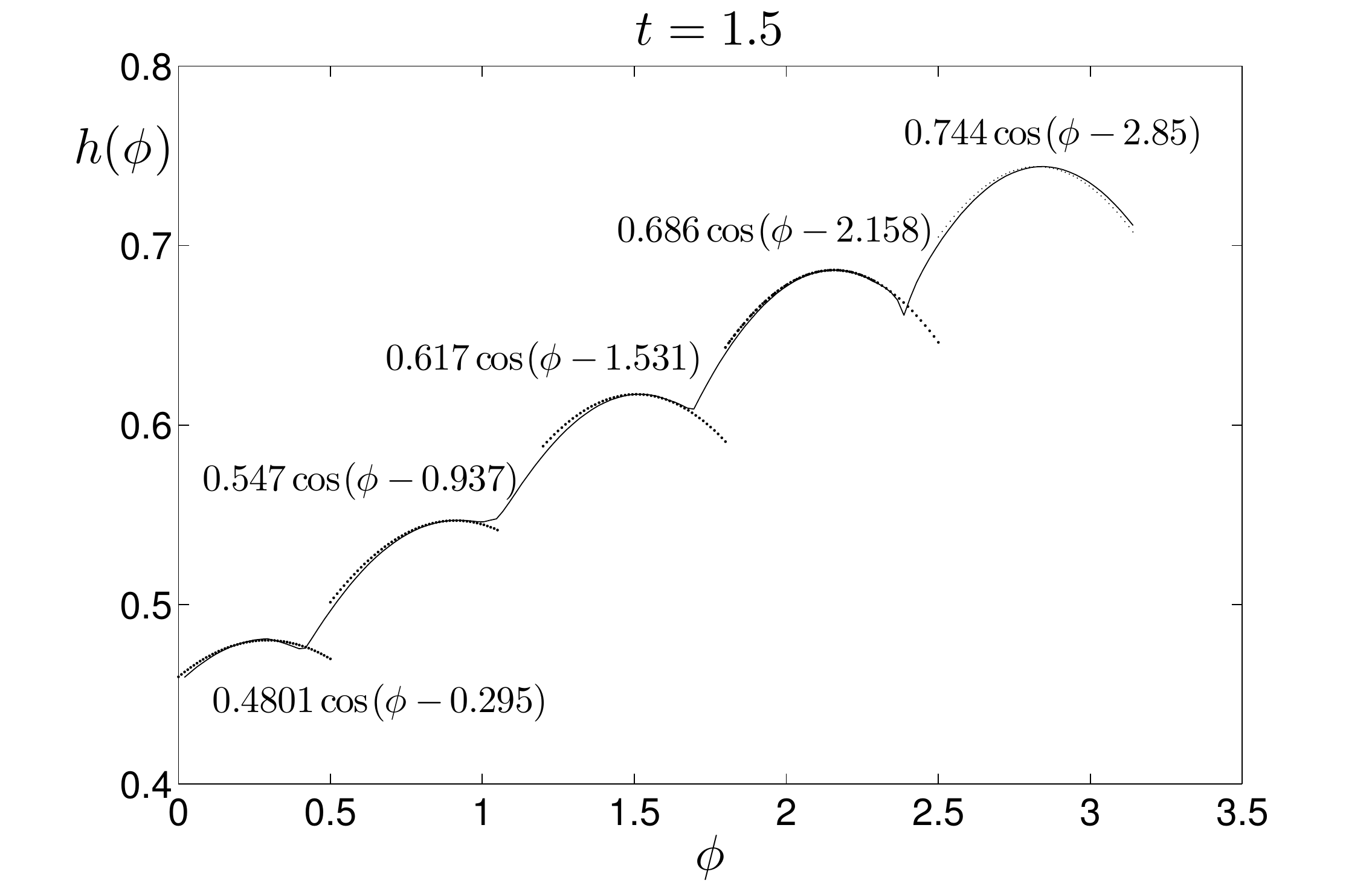}
	\caption{The indicatrix \eqref{indicatrix} for the dispersive KdV equation \eqref{kdv} with $\epsilon=0.1$ at time $1.5$. Dotted lines represent the local fitted cosine functions}
	\label{kdv_indicatrix_1p5}
\end{figure}

We perform an analysis of the hidden singularities for the KdV equation applying the BPH and Kida methods. In Fig. \ref{kdv_indicatrix_1p5} it is shown the indicatrix function given by \eqref{indicatrix}
for the dispersive KdV equation \eqref{kdv} with $\epsilon=0.1$ at time $t=1.5$.
It is also shown  the results of the fitting of the piecewise cosine function which gives the location of the singularities. The results are in agreement with the Pad\'e approximant analysis of Fig.\ref{kdv_pade}.

\begin{figure}
	\includegraphics[width=13.5cm]{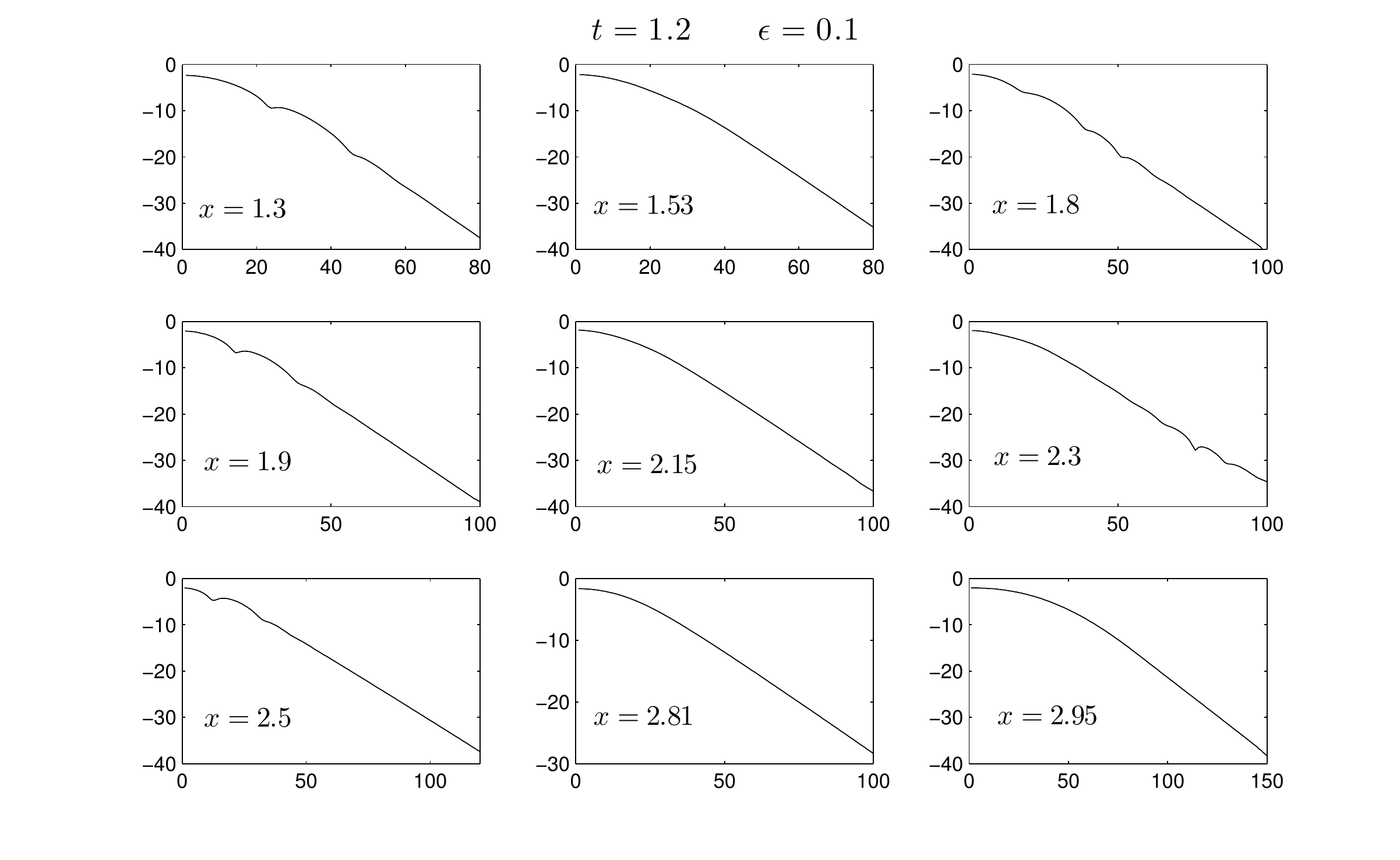}
	\caption{The Kida filtered spectrum (\eqref{kida_filtered_spectrum} and \eqref{kida_filtered_spectrum_max}) for the dispersive KdV equation \eqref{kdv} with $\epsilon=0.1$ at time $1.2$.}
	\label{kdv_kida}
\end{figure}

We perform also an analysis using the Kida technique, with $d=0.2$ in the Gaussian \eqref{gaussian}. In Fig.\ref{kdv_kida} it is shown the filtered spectrum \eqref{kida_filtered_spectrum} at different locations of $x$. One can notice that at the singularity locations $x=1.53$, $x=2.15$ and $x=2.81$ the respective filtered spectrum has a linear behavior as predicted by formulas \eqref{kida_filtered_spectrum_max}.

The fitting results at time $1.2$ of the Kida filtered spectrum  \eqref{kida_filtered_spectrum_max}
for the KdV equation with $\epsilon=0.1$  at the singularity location $x=1.53$, $x=2.15$ and $x=2.81$
are shown in Table \ref{kdv_table}.
The singularities are complex poles and the respective distances from the real axis are in agreement with
the Pad\'e approximant results shown in Fig.\ref{kdv_pade}.
The fitting to determine the algebraic characters $\alpha$ of the singularities are performed in the range $1<k<5$, while the fitting to calculate the distances $\delta$ form the real axis of the singularities are performed in the range $1<k<80$.

\begin{table}[ht]
	\begin{center}
		\begin{tabular}{|l|l|l|l|}
			\hline
			& $x=1.53$ &  $x=2.15$ & $x=2.81$ \\
			\hline
			$\alpha+1$ & $\;\;0.081  $ &  $\;\;0.098 $ &  $\;\;0.039 $     \\
			\hline
			$\delta$ & $\;\;0.56 $ & $\;\;0.45$ &   $\;\;0.34 $\\
\hline
		\end{tabular}
		\vskip0.5cm
		\caption{The fitting results at time $1.2$ of the Kida filtered spectrum (\eqref{kida_filtered_spectrum} and \eqref{kida_filtered_spectrum_max}) for the KdV equation with $\epsilon=0.1$  at the singularity location $x=1.53$, $x=2.15$ and $x=2.81$. \label{kdv_table}}
	\end{center}
\end{table}

%%%%%%%%%%%%%%%%%%%%%%%%%%%%%%%%%%%%%%%%%%%%%%%%%%%%%%%%%%%%%%%
\subsection{Singularity formation for Prandtl equation }\label{sec:3.2}

In this section we apply the complex singularity tracking method to investigate the singularity formation for 2D Prandtl equations, and its link with the  separation phenomena occurring when an incompressible viscous flow interacts with a rigid boundary.

Prandtl equations are used to describe the boundary layer flow  in the zero viscosity limit. These equations are obtained by introducing the
following scaling into the Navier-Stokes equations and taking the limit as $Re \to \infty$ (see \cite{SC}):
\begin{equation}
	y= Re^{-1/2} \, Y, \qquad \qquad v= Re^{-1/2} \, V,
\end{equation}
where $y$ is the normal coordinate, $v$ is the normal component of the velocity, and
$Y$ and $V$ are the rescaled  coordinate  and normal velocity.
The equations obtained at first order of the asymptotic expansion are:
\begin{eqnarray}
\frac{\partial u}{\partial t}+u\frac{\partial u}{\partial x}+V
\frac{\partial u}{\partial Y}-U_{\infty}\frac{d
	U_{\infty}}{d x} = \frac{\partial^2 u}{\partial Y^2},
\label{pramomentum}\\
\frac{\partial u}{\partial x}+\frac{\partial V}{\partial Y} = 0,\label{consmass}
\end{eqnarray}
with initial and boundary conditions given by
\begin{eqnarray}
u(x,Y,0)=U_{\infty},\label{initialprandtl}\\
u(x,0,t)=V(x,0,t)=0, \quad
u(x,Y\rightarrow\infty,t)=U_{\infty},\label{boundaryprandtl}
\end{eqnarray}
where $U_\infty(x)$ is the inviscid Euler solution at the boundary.

We consider here the classical case of an impulsively started circular cylinder immersed in an uniform background flow. In this case the inviscid Euler solution at the boundary is $U_\infty(x) = 2 \sin x$ and the streamwise coordinate $x$ is measured along the cylinder surface from the front stagnation point, and the normal coordinate $y$ is measured from the cylinder surface (see \cite{vDS80}).

The occurrence of a singularity in Prandtl's solution  was first proved numerically by van Dommelen \& Shen
in \cite{vDS80} by using a numerical lagrangian method. For that reason, in the sequel, we call the singularity in Prandtl solution also as the VDS (van Dommelen \& Shen) singularity.
The work of van Dommelen \& Shen was improved by Cowley in \cite{COW83} where it was investigated
the singularity formation for the  displacement thickness $\beta_{vDS}$ and the normal velocity at infinity $V_{\infty}$ using power time series expansion and approximating them
with a special case of Pad\'e approximants.
Singularity formation was also analyzed in \cite{GSS09} through the singularity-tracking method applied on the streamwise velocity component $u$ of Prandtl
equation, and it was found that for the initial condition $U_{\infty}(x)/2=\sin x$,
a cubic-root singularity forms at $2t_s\approx3$
with the blow up of $\partial_x u$ at $(x_s,Y_s)\approx(1.94,7)$.

In this section we investigate the singularity formation for Prandtl wall shear $\tau_w^P =\partial_Y u_{|Y=0}$, in order also to compare the results with the singularity analysis performed on the Navier-Stokes wall shear at different  $Re$ numbers (Section \ref{SingWallNS}). These results are originally presented in \cite{GSSC14}.

\begin{figure}
	\includegraphics[width=13cm]{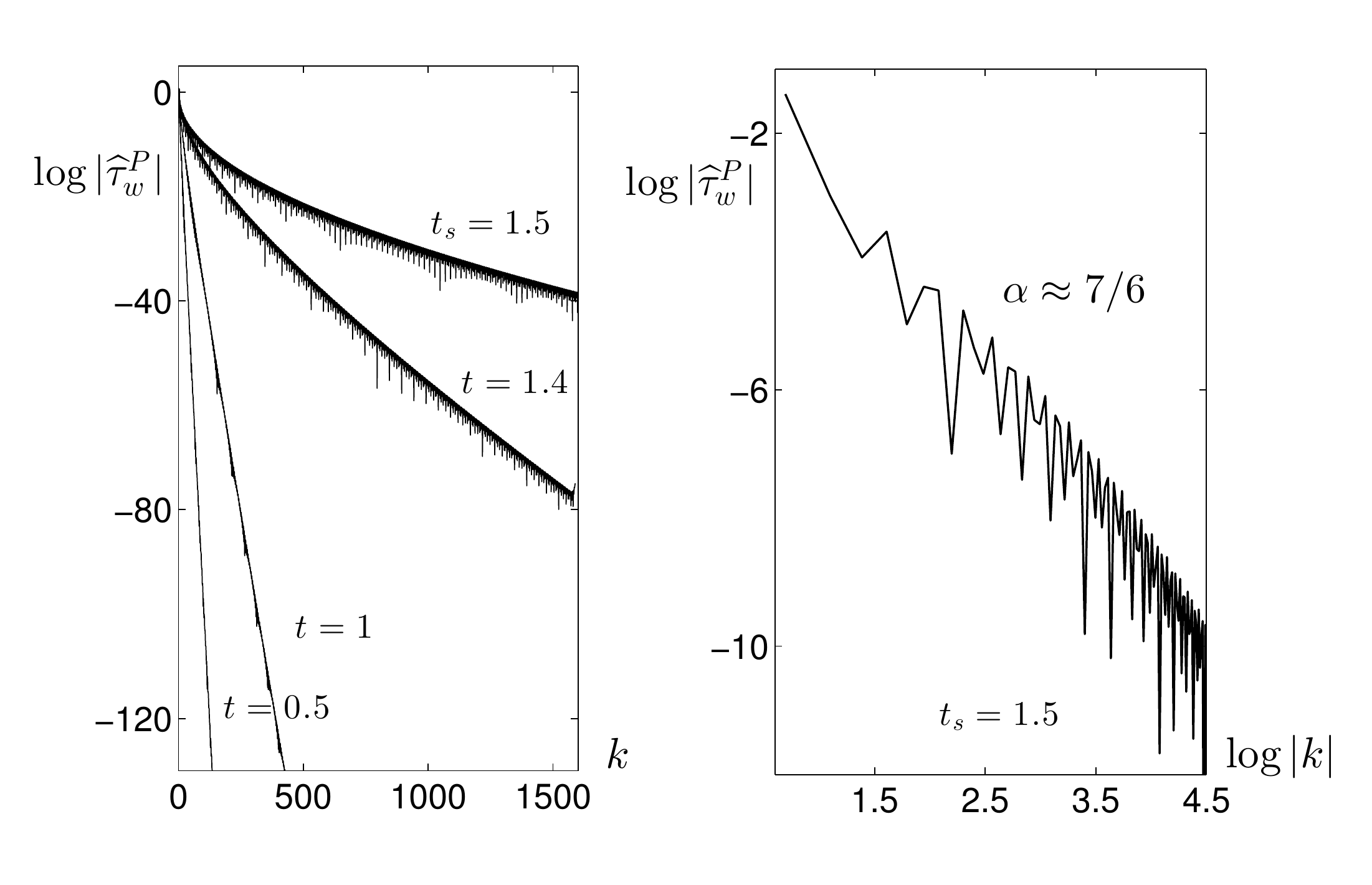}
	\caption{On the left the Fourier spectrum $\widehat{\tau}_w^P$ of Prandtl's wall shear at various times. On the right, the Fourier spectrum of $\tau_w^P$ for Prandtl's wall shear at $t_s=1.5$
		in log-log coordinates: the rate of algebraic decay  behaves like $\alpha \approx7/6$. }
	\label{fig_spettro_pra_new}
\end{figure}

In Fig.\ref{fig_spettro_pra_new} on the left it is shown the evolution of the spectrum $\widehat{\tau}_w^P$ of the Prandtl wall shear. At the singularity time $t_s$ the spectrum loses the exponential decay and the rate of its algebraic decays at $t_s$  gives the algebraic character of the singularity  $\alpha\approx7/6$,  revealing a blow-up in the second derivative of the wall shear as already shown in Fig.\ref{fig_pra_ws1p5_new}.

In Fig.\ref{fig_spettro_pra_new} on the right
it is shown the Fourier spectrum of $\tau_w^P$  at time $t_s$
in log-log coordinates where a slope of $7/6+1$ is visible.

\begin{figure}
	\includegraphics[width=13cm]{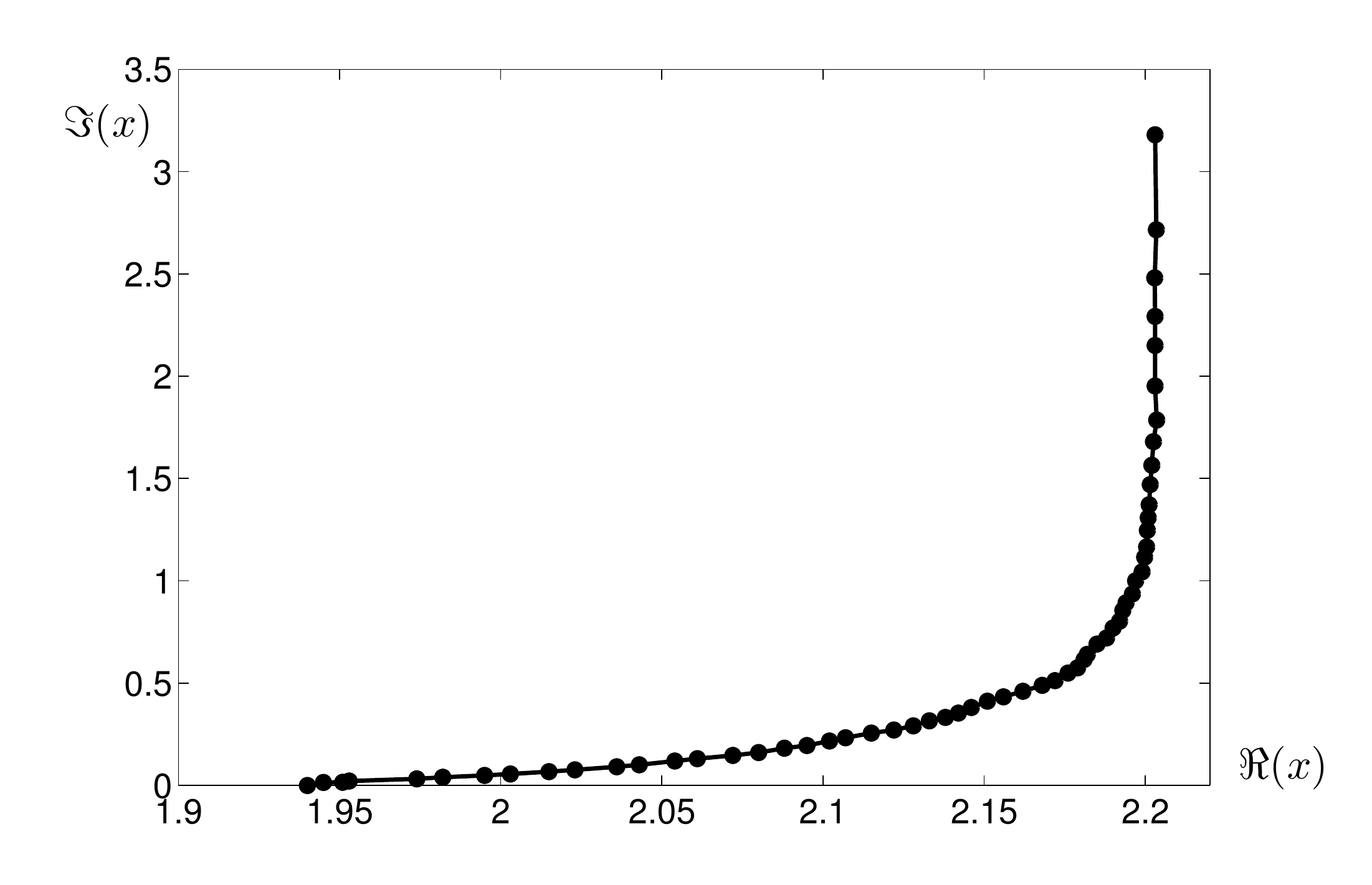}
	\caption{The time evolution in the complex plane
		$(\Re(x),\Im(x))$ of the complex singularity of $\tau_w^P$ from time 0.1 up to time 1.5. At $t_s=1.5$, the singularity of $\tau_w^P$ hits the real
		axis at $x_s\approx1.94$.}
	\label{fig_pra_sing}
\end{figure}

\begin{figure}
	\includegraphics[width=12cm]{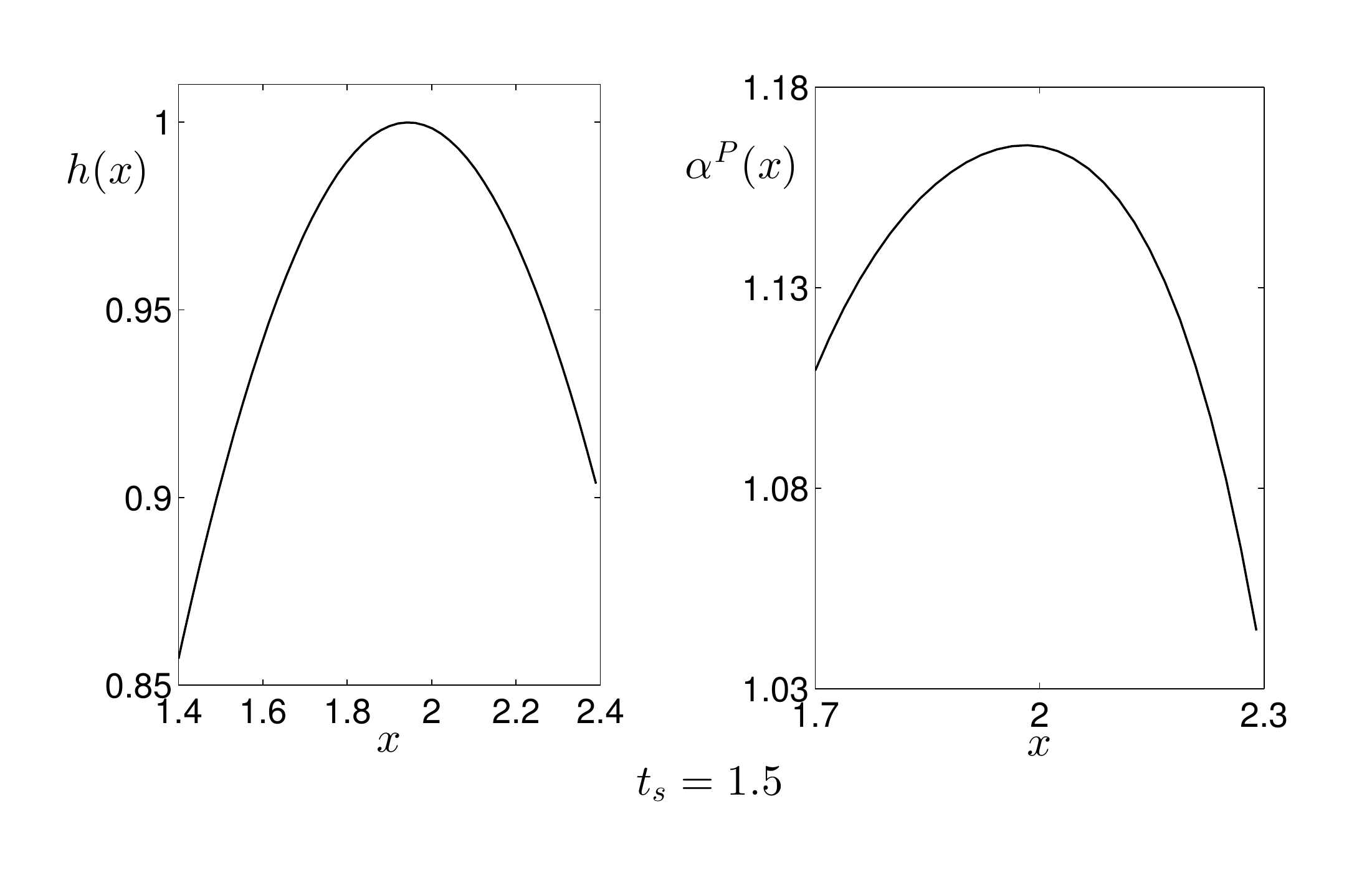}
	\caption{The indicatrix function $h(x)$ in \eqref{borel_as} at $t_s=1.5$ on the left: $h$ behaves like a cosine centered at $x_s\approx1.94$. On the right, the rate of algebraic decay evaluated from equation \eqref{borel_as} at $t_s=1.5$. In $x_s=1.94$, where  the singularity forms, this decay behaves like $\alpha \approx7/6$.}
	\label{fig_prandtl_borel_new}
\end{figure}

We apply the BPH method to Prandtl's wall shear to track the complex singularity in the complex plane, and in Fig.\ref{fig_pra_sing} it is shown the time evolution of the singularity of
$\tau_w^P$, from $t=0.1$ to $t_s=1.5$ with a time step of $0.05$.
After a transient time in which the
singularity is characterized  by a movement parallel to the imaginary axis, the singularity moves
toward the position $x_s=1.94$, hitting the real axis at time $t_s \approx 1.5$. At this time the Fourier spectrum loses exponential decay and the indicatrix function $h$ in \eqref{borel_as} behaves like a cosine function of amplitude 1 centered in $x_s\approx1.94$, as it is visible in  Fig.\ref{fig_prandtl_borel_new} on the left. In Fig.\ref{fig_prandtl_borel_new} on the right,
the rate of algebraic decay $\alpha^P(x)$ from \eqref{borel_as} is shown at $t=t_s$ showing that $\alpha(x=1.94)\approx7/6$.

\begin{figure}
\begin{center}
		\includegraphics[width=12cm]{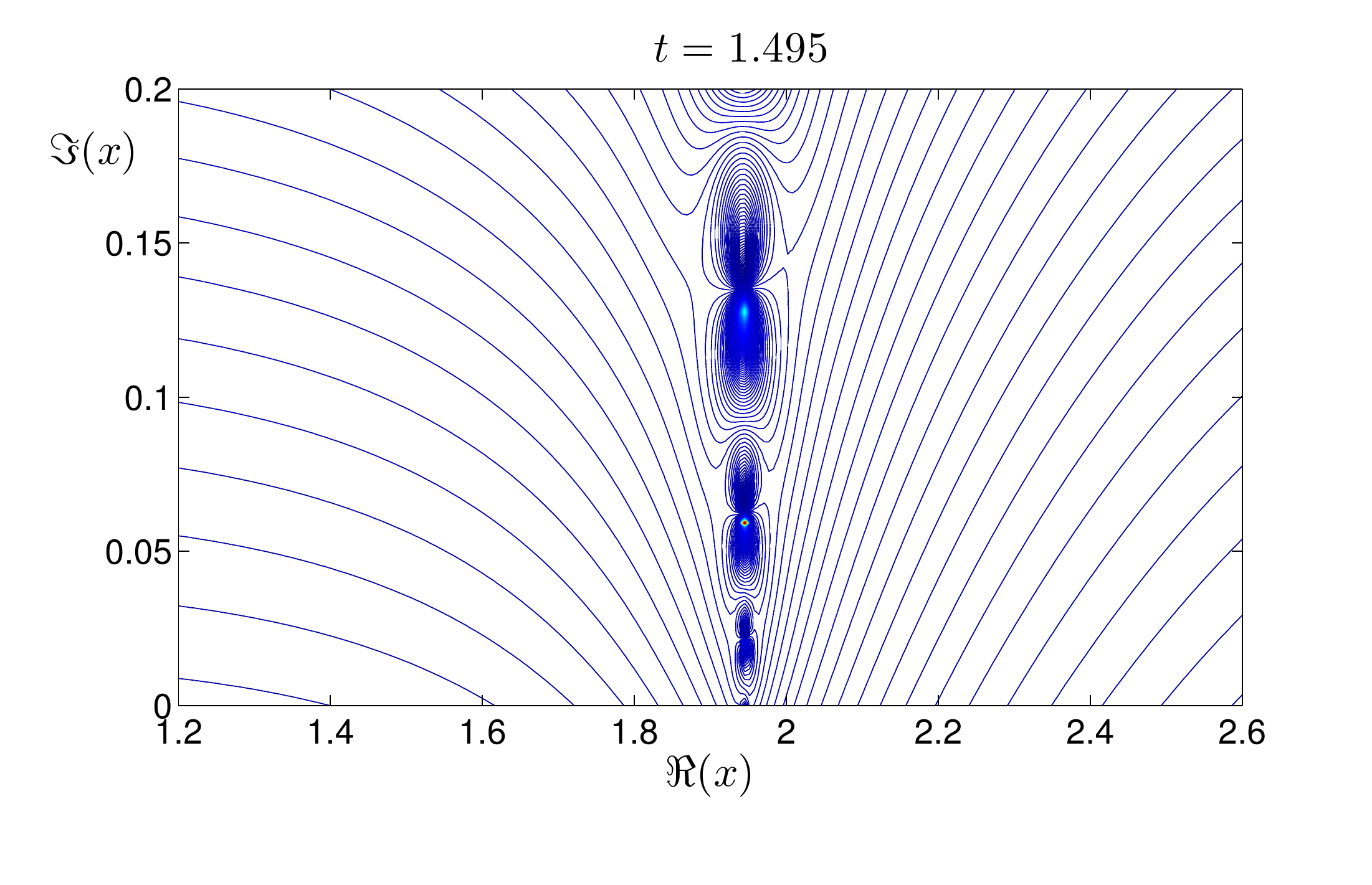}
		\caption{The contour levels of the modulus of the Pad\'{e} approximant $P_{250/250}$ of
			$\tau_w^P$ at $t=1.495$.}
		\label{pra_pade_ws1p5_new}
	\end{center}
\end{figure}

We conclude this analysis evaluating the modulus of the Pad\'{e} approximant $P_{250/250}$ of $\tau_w^P$ at time $t$ close to $t_s=1.5$. In Fig.\ref{pra_pade_ws1p5_new} one can see the algebraic branch cut, which is visible as a series of poles and zeros  along a cut parallel to the imaginary axis and located at $x_s\approx 1.94$.

\subsection{Singularity analysis for  Navier Stokes solutions }
\label{SingWallNS}
In \cite{OC02,OC05,GSS11,GSSC14} it was shown that the  wall shear $\tau_w^{NS}$ of Navier-Stokes solution is a relevant indicator
of the onset of the various viscous-inviscid interactions characterizing the separation process in Navier-Stokes solutions.
In fact, after the formation of the back-flow, the first relevant interaction visible in Navier-Stokes solutions, i.e. the so
called large-scale interaction, leads to the disagreement between the Navier-Stokes and Prandtl wall shear.
The subsequent small-scale interaction, observable only for moderate-high $Re$ numbers, characterizes the typical turbulent chaotic regime of the high $Re$ number flow with high gradients formation in $\tau_w^{NS}$. In \cite{GSSC14} it was found a relationship between these interactions
and the presence of complex singularities in $\tau_w^{NS}$. We present in this section the results due to the singularity analysis of the wall shear $\tau_w^{NS}$ for the impulsively started disk case, allowing also a direct comparison with the Prandtl case.

The Navier-Stokes equations  in the
vorticity-streamfunction formulation for the impulsively started disk read as:
\begin{eqnarray}
\frac{\partial \omega}{\partial t}+\frac{u}{r}\frac{\partial \omega}{\partial
	\theta}+v\frac{\partial \omega}{\partial
	r}=\frac{1}{Re} \left( \frac{1}{r^2}\frac{\partial^2 \omega}{\partial
	\theta^2}+\frac{1}{r}\frac{\partial \omega}{\partial r}+\frac{\partial^2
	\omega}{\partial r^2} \right), \label{NSequationd}\\
\frac{1}{r^2}\frac{\partial^2 \psi}{\partial \theta^2}+\frac{1}{r}\frac{\partial
	\psi}{\partial r}+\frac{\partial^2 \psi}{\partial r^2}=-\omega,
\label{poissond}\\
u=\frac{\partial \psi}{\partial r},\quad v=-\frac{1}{r}\frac{\partial
	\psi}{\partial \theta},\label{velocityewd}\\
u=v=0 , \qquad r=1, \label{noslippd}\\
\omega\rightarrow0, \qquad r\rightarrow\infty,\label{inftycond}\\
\omega(\theta,r,t=0)=0,\label{NSinitdd_vorticitybcd}\\
\psi(\theta,r,t=0)= \left( r - \frac{1}{r} \right) \sin \theta. \label{NSinitdd_PSI}
\end{eqnarray}

Equation \eqref{NSequationd} is the vorticity-transport equation, equation
\eqref{poissond} is the Poisson equation for the streamfunction, and
equations \eqref{velocityewd} relate the velocity components to the streamfunction.
\eqref{noslippd} and \eqref{inftycond} are the no-slip and impermeability conditions on the circular cylinder and the irrotational
condition at infinity, respectively. The initial
condition \eqref{NSinitdd_vorticitybcd} expresses the irrotationality
condition of the flow at the initial time, \eqref{NSinitdd_PSI} is the initial condition for the streamfunction. The walls shear is as usually defined as $\tau_w^{NS}=-\omega_{r=1}/\sqrt{Re}$.
The problem is solved in the domain $[0,\pi]\times[1,\infty)$, and only the upper
part of the circular cylinder is considered owing to symmetry. Details on the numerical scheme used can be found in \cite{GSSC14}

\begin{figure}
	\begin{center}
		\includegraphics[width=10.5cm]{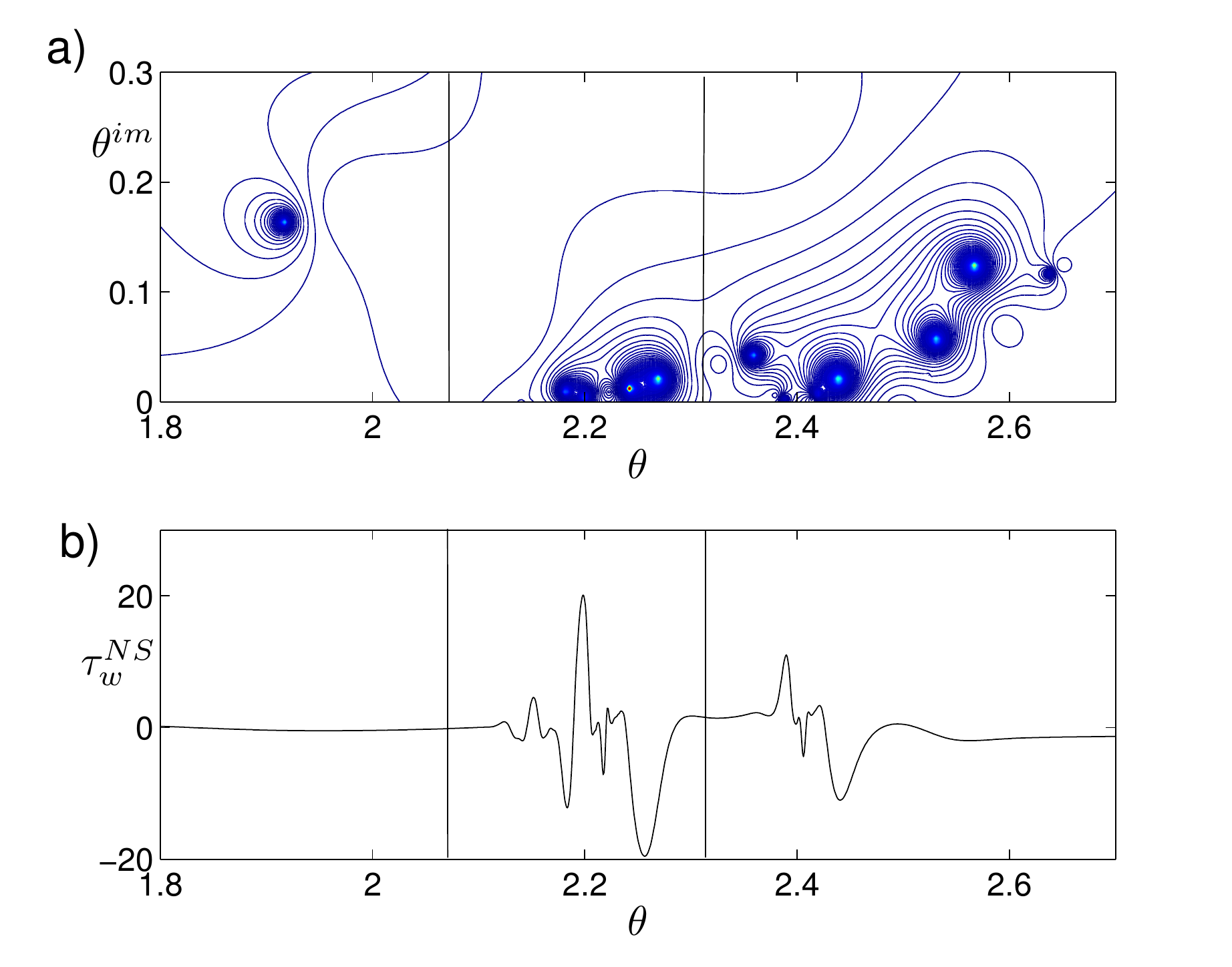}
		\caption{\textbf{a)}  The contour levels of the modulus of the Pad\'{e} approximant $P_{300/300}$ of
			$\tau_w^{NS}$ for $Re=10^5$ at $t=1.58$.  Three distinct groups of
			complex singularities are present. Each group correspond to a different viscous-inviscid interaction. 	\textbf{b)} The wall shear $\tau_w^{NS}$: a strong correspondence between the high gradients in $\tau_w^{NS}$ and the positions of the complex singularities in \textbf{a)} is  visible. }
		\label{singgroup}
	\end{center}
\end{figure}

\begin{figure}
	\includegraphics[width=13cm]{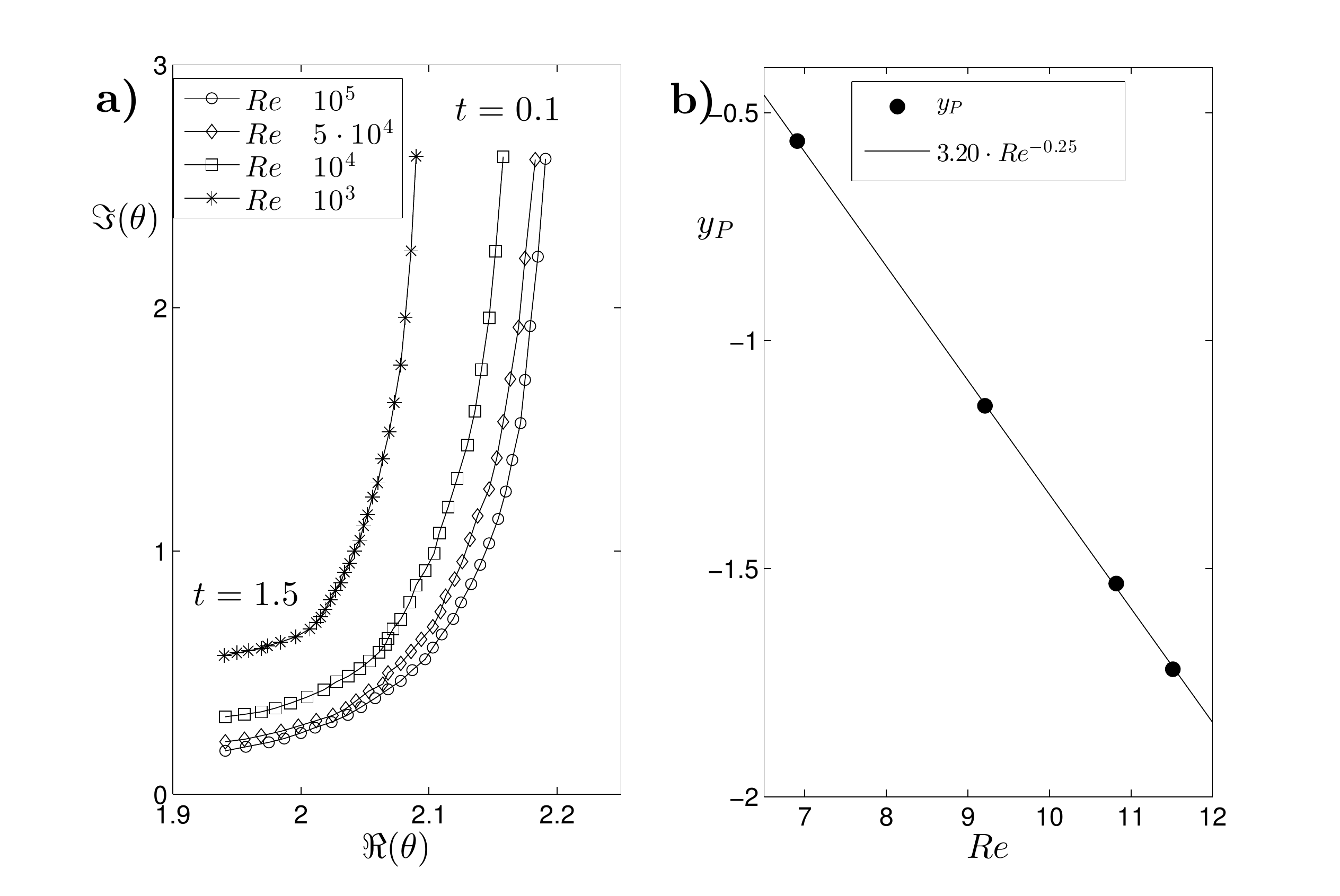}
	\caption{\textbf{a)} The time evolutions in the complex plane
		$(\Re(\theta),\Im(\theta))$ of the  complex
		singularities $s_P$ of $\tau_w^{NS}$ for the various Reynolds numbers from time 0.1 up to time 1.5 with
		time step of 0.05. At $t_s=1.5$ the singularity remains at a
		distance $y_P$ from the real axis which goes like $3.2\cdot Re^{-0.25}$.\textbf{ b)}
		The distance $y_P$ is shown versus the Reynolds numbers in log-log coordinates. }
	\label{sing_PRA_ALL_2}
\end{figure}

In \cite{GSSC14} it was shown that  $\tau_w^{NS}$ has several singularities that can be divided
into three distinct groups.  These three groups of singularities are visible in Fig.\ref{singgroup} in  which it is shown the modulus of the Pad\'{e} approximant $P_{300/3000}$ of $\tau_w^{NS}$ for
$Re=10^5$ at $t=1.58$: at this time the large and small scale interactions have already formed for $Re=10^5$ .
The left group comprehends only a singularity.  The middle group comprehends
several complex singularities  corresponding to the large-scale interaction. The right group is visible only for moderate-high $Re$ number, and consists of complex singularities that correspond to the small-scale interaction. 		

\begin{figure}
	\begin{center}
		\includegraphics[width=10.5cm]{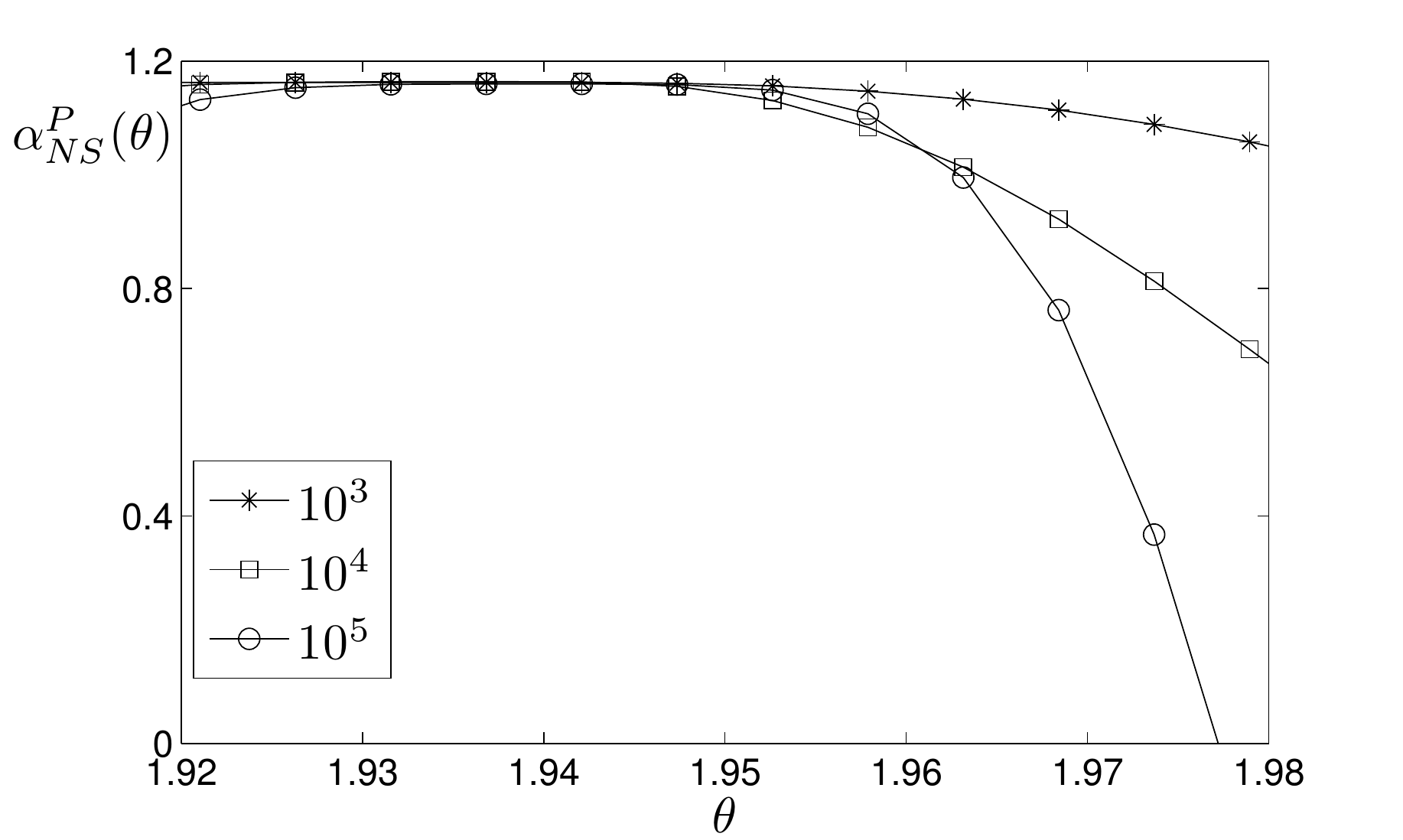}
		\caption{The characterization $\alpha_{NS}^P$ of the complex singularity $s_P$
			of $\tau_w^{NS}$ evaluated through the BPH method.  At $t_s=1.5$ in $\theta \approx1.94$ we retrieve
			$\alpha_{NS}^P\approx7/6$ for all the $Re$ considered. }
		\label{borel_NS_ALPHA_P}
	\end{center}
\end{figure}

The first singularity of $\tau_w^{NS}$ in the left group as shown in Fig.\ref{singgroup}a is comparable with the singularity of Prandtl wall shear $\tau_w^P$ (hereafter we shall denote this singularity as $s_P$). The main similarity between $s_{P}$ and the singularity of
$\tau_{w}^P$ lies in
their characterization. In fact, through the BPH method we have obtained that the algebraic characterization
of $s_{P}$ at the time $t_s=1.5$ is
$\alpha_{NS}^P\approx7/6$ for each Reynolds number (see Fig.~\ref{borel_NS_ALPHA_P}) which is the same characterization observed for $\tau_{w}^P$ as shown in the previous section. We stress that as
compared
to the Prandtl case, the characterization of $\alpha_{NS}^P$ has been more difficult
to evaluate. This was mainly due to the presence of the various complex singularities of the other groups that affect the indicatrix function leading to some difficulties in handling numerically the evaluation of the algebraic decay rate. The second similarity  between $s_{P}$ and the singularity of
$\tau_{w}^P$ is given by the similar time evolution of their  positions in the complex plane as shown in Figure~\ref{sing_PRA_ALL_2}a ( singularities are tracked from $t = 0.1$ to  $t_s = 1.5$ with time step of $0.05$, see also Fig.\ref{fig_pra_sing} for the time evolution of the Prandtl wall shear in the complex plane).  All
singularities rapidly move toward the real axis slightly shifting upstream on the circular cylinder.
At time  $t_s=1.5$, when singularity forms in Prandtl solution, all the singularities have the real part of their
position close to $x_s\approx1.94$, and the imaginary part $y_P$ which follows the rule
$y_{P}=C_{P}Re^{\lambda_P}$, where $\lambda_P\approx-0.25$ and $C_P\approx3.2$
(see in Figure~\ref{sing_PRA_ALL_2}b, where $y_{P}$ is shown versus the
Reynolds number in log-log
scale). While it is clearly expected that $s_{P}$ get closer to the real domain ad $Re$ increases, it was less predictable, mainly due to the influences of the other singularities, that $s_P$  moves toward the position $x_s=1.94$ for all the $Re$.

\begin{figure}
	\begin{center}
		\includegraphics[width=11.5cm]{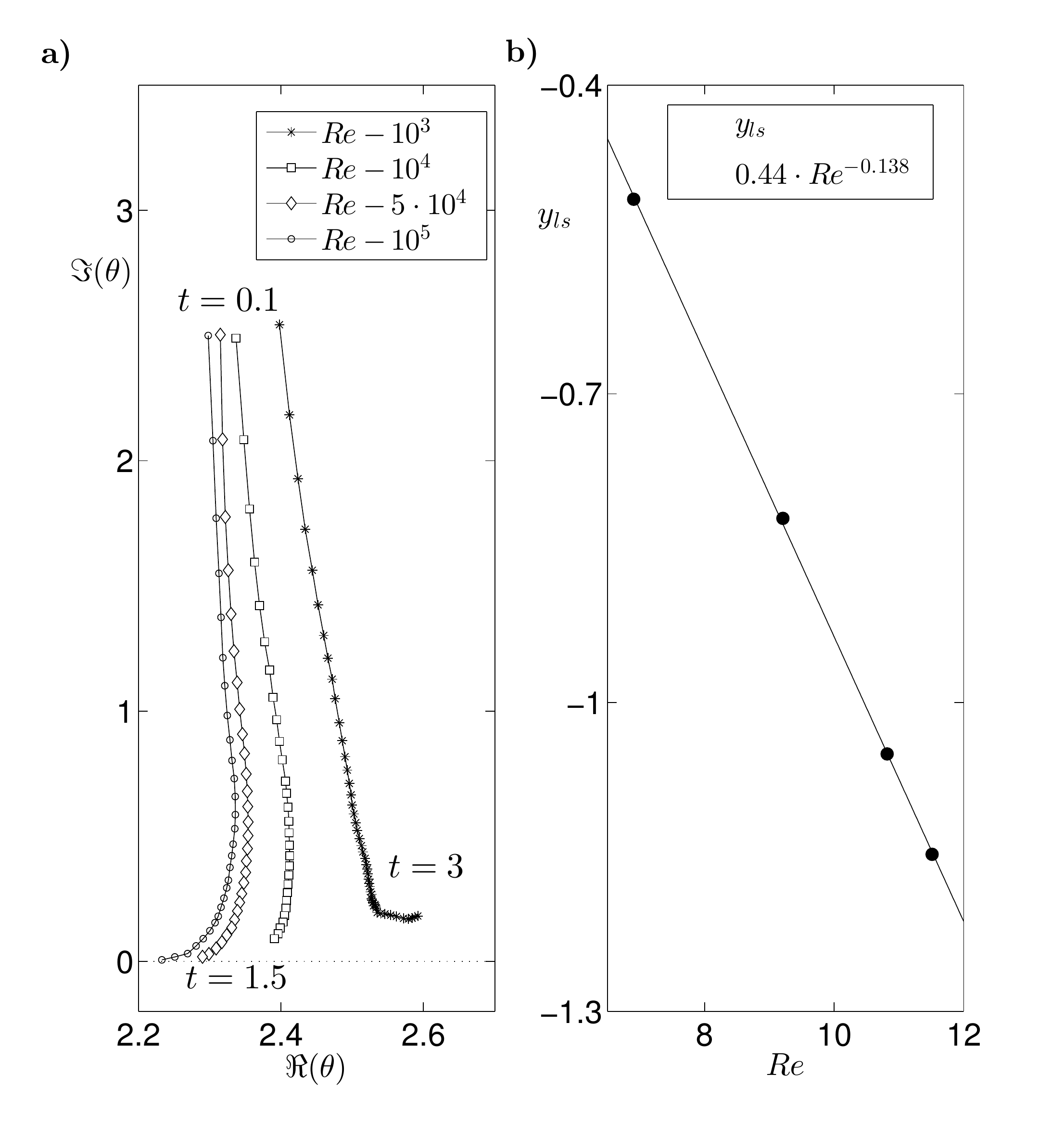}
		\caption{\textbf{a) }The time evolution in the complex plane
			$(\Re(\theta),\Im(\theta))$ of the complex singularity $s_{ls}$
			of $\tau_w^{NS}$ for $Re=10^3$ from time 0.1 up to time 3 with time step of
			0.05, and for $Re=10^4,5\cdot 10^4,10^5$
			from time 0.1 up to time 1.5 with time step of 0.05. \textbf{b)} The distance $y_{ls}$ of $s_{ls}$ from the real domain versus the Reynolds number in
						log-log coordinates at the
						time  at which large-scale interaction begins. The singularity is at a distance $y_{ls}$ from the real
			axis that goes like $0.44\cdot Re^{-0.138}$. }
		\label{sls}
	\end{center}
\end{figure}

\begin{figure}
	\begin{center}
		\includegraphics[width=10.5cm]{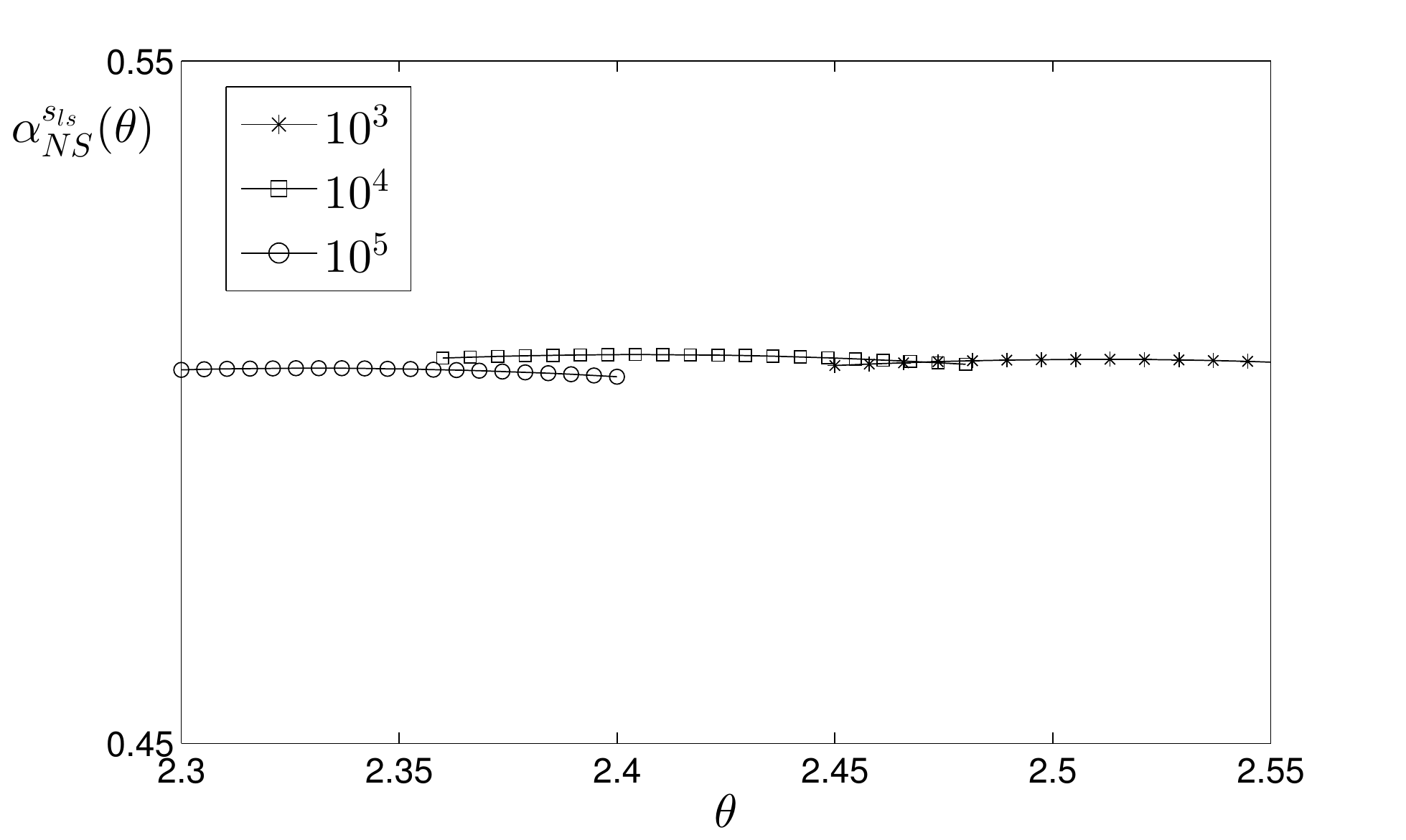}
		\caption{The characterization $\alpha_{NS}^{s_{ls}}$ of the complex singularity
			$s_{ls}$ of $\tau_w^{NS}$ evaluated from the BPH method.  At $t=1.05$, $\alpha_{NS}^{s_{ls}}\approx0.5$ for $Re=10^3,10^4,10^5$, and $s_{ls}$ is located at $(2.501,0.54),(2.40,0.37),(2.32,0.28)$, respectively.}
		\label{borel_NS_ALPHA_ls}
	\end{center}
\end{figure}

The second group of complex singularities in $\tau_w^{NS}$ is related to the large-scale interaction developing in the separation process, and it exists for all the
Reynolds number considered. The presence of several singularities close to each other leads to numerical difficulties in resolving all the singularity positions and characterizations.
Only the singularity closest to the real axis for all time (hereafter we shall call this singularity $s_{ls}$) can be well resolved. Also for $s_{ls}$ we have tracked in time the position in the complex plane for all the $Re$ (see Fig.\ref{sls}). While the evolution of $s_{ls}$ is quite similar for all $Re>10^3$, a different behavior  is observed for $Re=10^3$ . In this case, in fact, $s_{ls}$  continues to shift downstream along the circular cylinder even after total
detachment of the boundary layer. For the other $Re$ cases, instead, $s_{ls}$ changes its motion by shifting upstream along the circular cylinder during the small-scale interaction phase. At the time in which large scale interaction begins (from the analysis performed in \cite{GSSC14} this interaction forms at $t\approx 0.908, 0.916,0.94,0.952$ for $Re=10^3,10^4,5\cdot10^4,10^5$) the distance $y_{ls}$ from the real axis of the singularity $s_{ls}$ follows
the rule $y_{ls}=C_{ls}Re^{\lambda_{ls}}$, where $\lambda_{ls}\approx-0.138$ and
$C_{ls}\approx0.447$, as one can see in Figure~\ref{sls}b, where $y_{ls}$ is shown versus the $Re$ number in
log-log coordinates. By applying the BPH method, we have well resolved the characterization of the
singularity $s_{ls}$ in the range of time between  $t=0.95$ and $t=1.1$
when the other complex singularities are still far enough away from the singularity $s_{ls}$. The charachterization  is $\alpha_{NS}^{s_{ls}}\approx0.5$ for all the Reynolds numbers considered as shown in
Fig.~\ref{borel_NS_ALPHA_ls}, where the rate of algebraic decay
$\alpha_{NS}^{s_{ls}}$,
obtained from equation \eqref{borel_as}, is shown at $t=1.05$ for the various Reynolds numbers.

\begin{figure}
	\begin{center}
		\includegraphics[width=11.5cm]{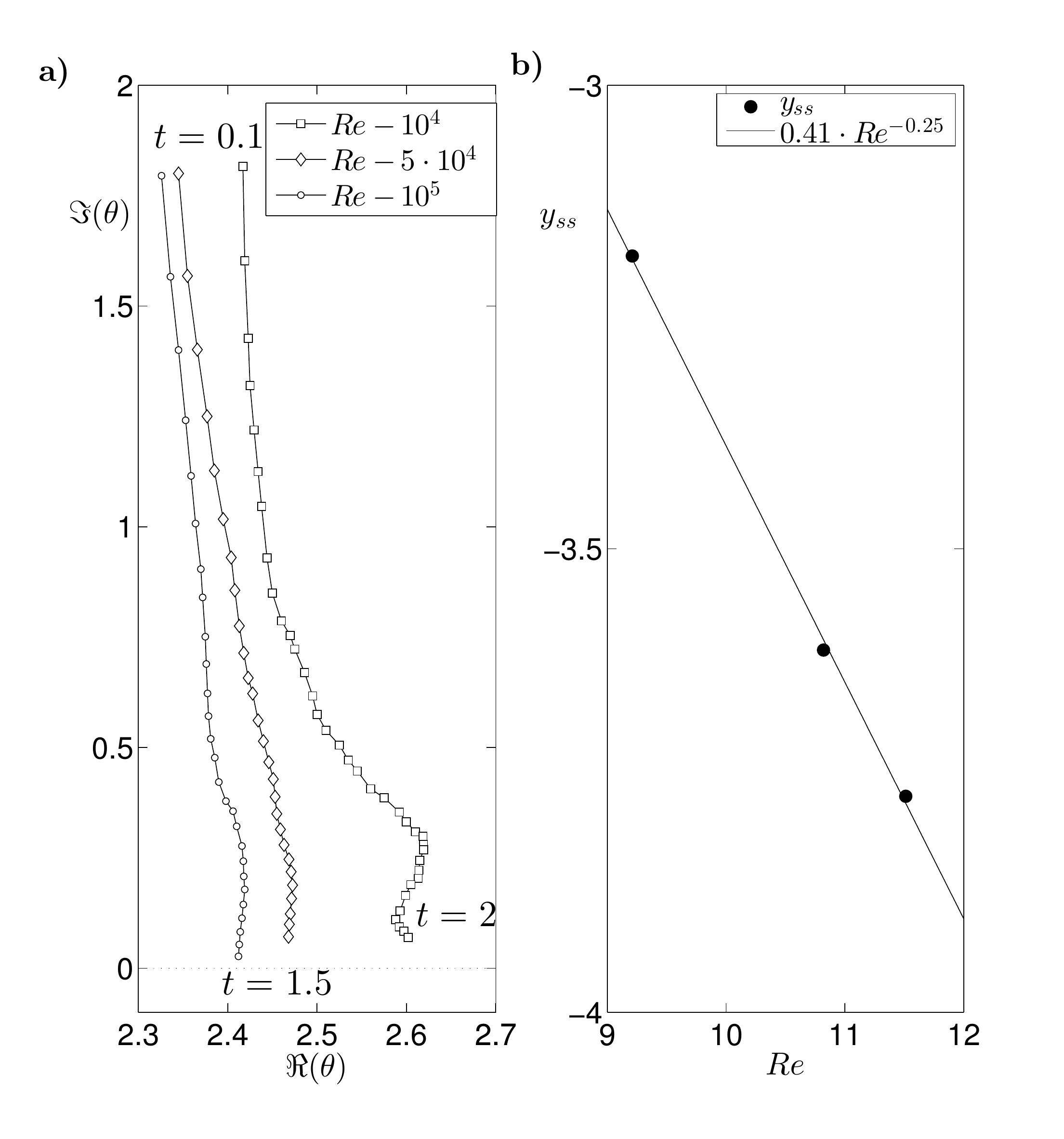}
		\caption{\textbf{a)} The time evolution in the complex plane
			$(\Re(\theta),\Im(\theta))$ of the complex singularity $s_{ss}$
			of $\tau_w^{NS}$ for $Re=10^4$ from time 0.1 up to time 2 with time step of
			0.05 and for $Re=5 \cdot 10^4,10^5$ from time 0.1 up to time 1.5 with
			time step of 0.05.\textbf{ b)} The distance $y_{ss}$ of $s_{ss}$ from the real domain is shown versus the Reynolds number in
			log-log coordinates at the time in which the small-scale interaction begins. The singularity is at a distance from the real
			axis that goes like $0.41\cdot Re^{-0.25}$.}
		\label{sss}
	\end{center}
\end{figure}

The third group of singularities  are related to  the small-scale interaction, and as for the second group of singularities we were able to well resolve only the primary singularity of this group that is always closest to
the real axis (hereafter this singularity is called $s_{ss}$).
In Fig.\ref{sss}a the time evolution of the position of $s_{ss}$ in the
complex plane for $Re = 10^4$ from time $t = 0.1$ up to time $t = 2$, and for $Re=5 \cdot 10^4$
and $Re=10^5$ for $t=0.1$ up to $t=1.5$ with time step of $0.05$.
This evolution is not as smooth as compared to that of
$s_P$ and $s_{ls}$, and in \cite{GSSC14} the physical events affecting the time evolution of $s_{ss}$ were explained.
At the time at which small-scale interaction begins (from the analysis performed in \cite{GSSC14} this interaction forms at $t\approx 1.505,1.29,1.26$ for $Re=10^4,5\cdot10^4,10^5$), it has been observed that the
distance $y_{ss}$ from the real axis
of the singularity $s_{ss}$ follows the rule $y_{ss}=C_{ss}Re^{\lambda_{ss}}$, where
$\lambda_{ss}\approx-0.25$ and $C_{ss}\approx0.41$.  This can be seen in Fig.\ref{sss}b,
where $y_{ss}$ is shown versus the Reynolds number in log-log
coordinates. The characterization of $s_{ss}$ was quite well resolved at the time in which small scale interaction begins, and we have obtained the value $\alpha^{ss}_{NS}\approx0.5$. This characterization is compatible with the kind of
gradient that forms in $\tau_w^{NS}$ as it clearly shows a growth in the first derivative.

%%%%%%%%%%%%%%%%%%%%%%%%%%%%%%%%%%%%%%%%%%%%%%%%%%%%%%%%%%%%%%%
\section{Complex singularity tracking method for multivariable function}\label{sec:4}

In this section the singularity-tracking method is extended to a bi-variate function
(see \cite{MBF05,PMFB06} for details)

Given a periodic function that can be expressed as a Fourier series
\begin{equation}
u(x_1,x_2)=\sum_{k_1,k_2} u_{k_1 k_2} e^{i k_1 x_1} e^{i k_2 x_2},
\label{bi_fourier}
\nonumber
\end{equation}
if one considers those modes $(k_1, k_2)$ such that
$k_1= k \cos \theta$ and $k_2= k \sin \theta $, where $k=|(k_1,k_2)|$,
then the asymptotic behavior of the Fourier coefficients in the Fourier
$\mathbf{k}$-space with $k\rightarrow\infty$ have the following asymptotic behavior:
\begin{equation}
u_{k_1 k_2}\approx  k^{-\left(\alpha(\theta)+1\right)} e^{-\delta(\theta) k}
e^{i k x^*(\theta)} \quad \mbox{where} \quad
(k_1, k_2)= k( \cos \theta,  \sin \theta ).
\label{Fourier_asymp_theta}
\end{equation}
The width of the analyticity strip $\delta^*$ is the minimum over all
directions $\theta$, i.e.
$\delta^*=\min_{\theta}\delta(\theta)$.

A second way to extend the singularity tracking method to bi-variate functions is to define
the shell-summed Fourier amplitudes,
are defined as
\begin{equation}
A_K \equiv \sum_{K\leq |(k_1,k_2)| < K+1} \left|u_{k_1 k_2} \right|,
\label{sing_shellSUMampl}
\end{equation}
which are a kind of discrete angle average of the Fourier coefficients.
The asymptotic behavior of these amplitudes is
\begin{equation}
A_K\approx CK^{-\left(\alpha_{Sh}+1/2\right)} \exp{\left(-\delta_{Sh} K\right)}
\quad \mbox{when}\quad
K\rightarrow\infty,
\label{ShellaAsym}
\end{equation}
where $\delta_{Sh}$ gives the width of the analyticity strip, while the algebraic
prefactor $\alpha_{Sh}$ gives information on the nature
of the singularity. As pointed out in \cite{PMFB06}, using a steepest descent argument, one can see that
the two techniques are equivalent.
In fact, if one denotes with $\theta^*$ the angle where  $\delta(\theta)$
takes its minimum (i.e. $\delta^*=\delta(\theta^*)$),
one has that $\delta_{Sh}=\delta(\theta^*)$ and that $\alpha_{Sh}=\alpha(\theta^*)-1/2$.

An interesting situation is when the most singular direction
coincides with one of the coordinate axes, e.g. $\theta^*=0$ (similar is the case when $\theta^{*}=\pi/2$), which means that
(see \eqref{Fourier_asymp_theta}):
\begin{equation}
u_{k_10}\approx k_1^{-\left(\alpha(\theta^*)+1\right)} e^{-\delta(\theta^*) k_1} \qquad
\mbox{with} \quad \theta^*=0 \; . \label{thetastareqzero}
\end{equation}
In this case it is easy to see that, to evaluate the width of the strip
of analyticity, one can consider the variable  $x_2$ as a
parameter (when $\theta^{*}=\pi/2$ one can be consider instead $x_1$ as a parameter) and adopt the following procedure.
First take the Fourier expansion relative to the variable $x_1$:
$$
u(x_1,x_2)=\sum_{k_1} u_{k_1}(x_2) e^{-ik_1x_1} \, ;
$$
second, given that for fixed $x_2$ the function $u(x_1,x_2)$ is
analytic in $x_1$, use that the spectrum has the asymptotic
behavior:
\begin{equation}\label{formula_cut}
u_{k_1}(x_2) \approx k_1^{-(\alpha(x_2)+1)} e^{-\tilde{\delta}(x_2)k_1} \; ;
\end{equation}
third use the definition of $u_{k_1 0}$ to write:
\begin{equation}
u_{k_1 0}=\int u_{k_1}(x_2) e^{ik_2 x_2} \, dx_2 \approx
\int k_1^{-(\alpha(x_2)+1)} e^{-\tilde{\delta}{(x_2)}k_1}  e^{ik_2 x_2} \, dx_2
\approx e^{-\min_{x_2}{\tilde{\delta}{(x_2)}}k_1} \; , \label{computation}
\end{equation}
where, to get the last estimate, we have used a steepest descent argument.
Comparing \eqref{thetastareqzero} with \eqref{computation} one finally derives that, when
$\theta^*=0$:
$$
\delta(\theta^*)=\min_{x_2}{\tilde{\delta}{(x_2)}}  \; .
$$

The procedures needed to capture the asymptotic behavior of the spectrum require high numerical
precision and in fact, in the calculations we shall present,
we have used a 32--digits precision (using the
ARPREC package).
For more details on
the method and on the various techniques introduced in the
literature to fit the spectrum, see
\cite{Caf93,FMB03,GPS98,MBF05,PF07,Sh92,SSF83}.

%%%%%%%%%%%%%%%%%%%%%%%%%%%%%%%%%%%%%%%%%%%%%%%%%%%%%%%%%%%%%%%

%%%%%%%%%%%%%%%%%%%%%%%%%%%%%%%%%%%%%%%%%%%%%%%%%%%%%%%%%%%%%%%
\subsection{Prandtl equation}\label{sec:4.1}

We apply the techniques of singularity
tracking method for multivariable function, explained in the previous section, to analyze the singularity of the Prandtl solution for the VDS initial datum $U_{\infty}=2\sin(x)$ (see Section \ref{sec:3.2}).

\begin{figure}
	\begin{center}
		\includegraphics[height=8.cm,width=11.5cm]{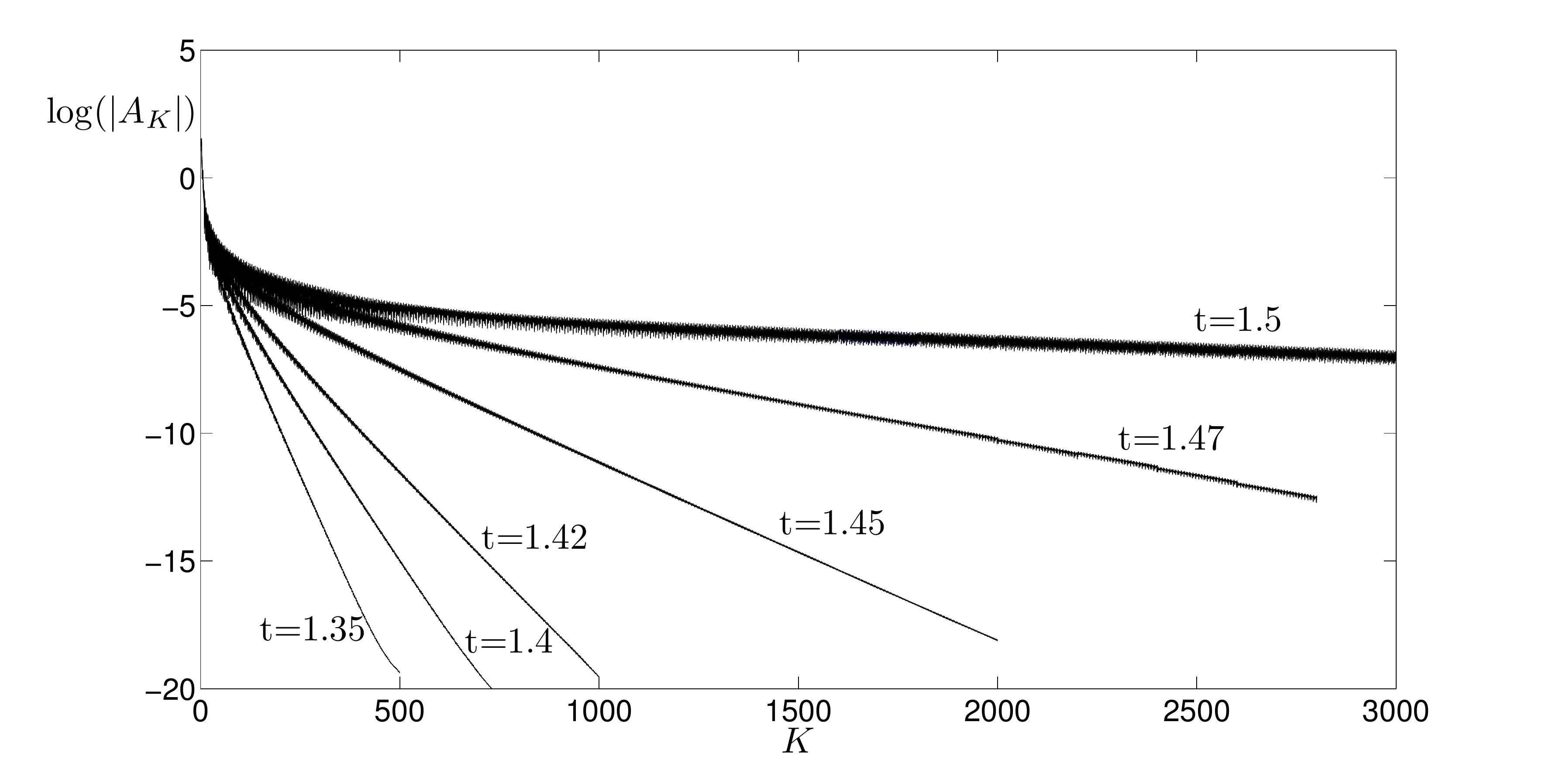}
		\caption{{The behavior in time of the shell summed Fourier amplitude up
				to the singularity time.}}\label{VDSshell}
	\end{center}
\end{figure}

\begin{figure}
	\begin{center}
		\includegraphics[height=7.cm,width=10.5cm]{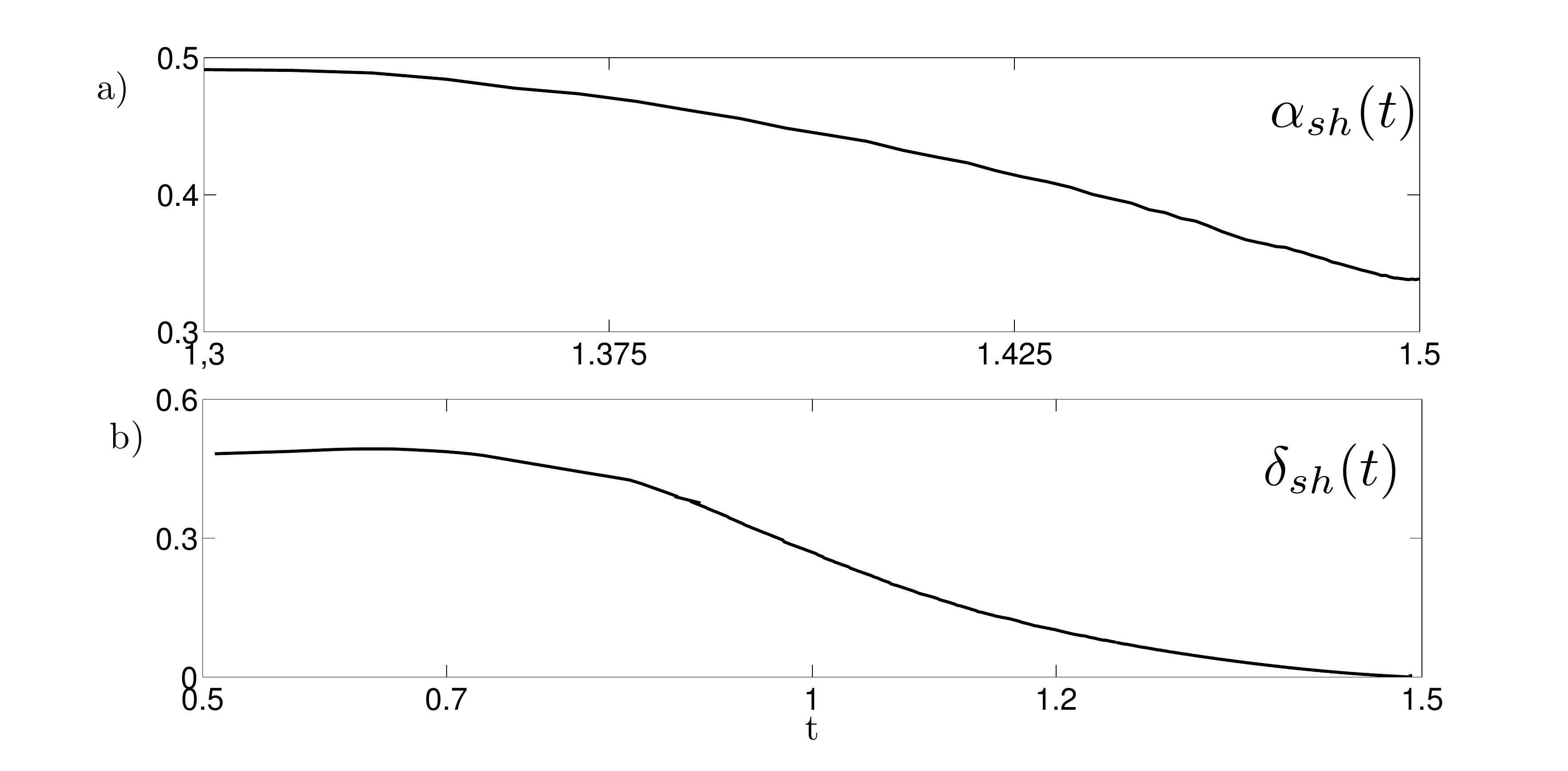}
		\vspace*{-0.5cm}\caption{$\bold{a)}$ The behavior of the algebraic character of the singularity from the shell summed Fourier amplitude of Prandtl's solution. The singularity is of cubic-root type.$\bold{b)}$ The behavior in time of the width of the analyticity strip.
			The singularity time is $t_s\approx 1.5$. }\label{VDSshellFITT}
	\end{center}
\end{figure}

\begin{figure}
	\begin{center}
		\includegraphics[height=10.cm,width=12cm]{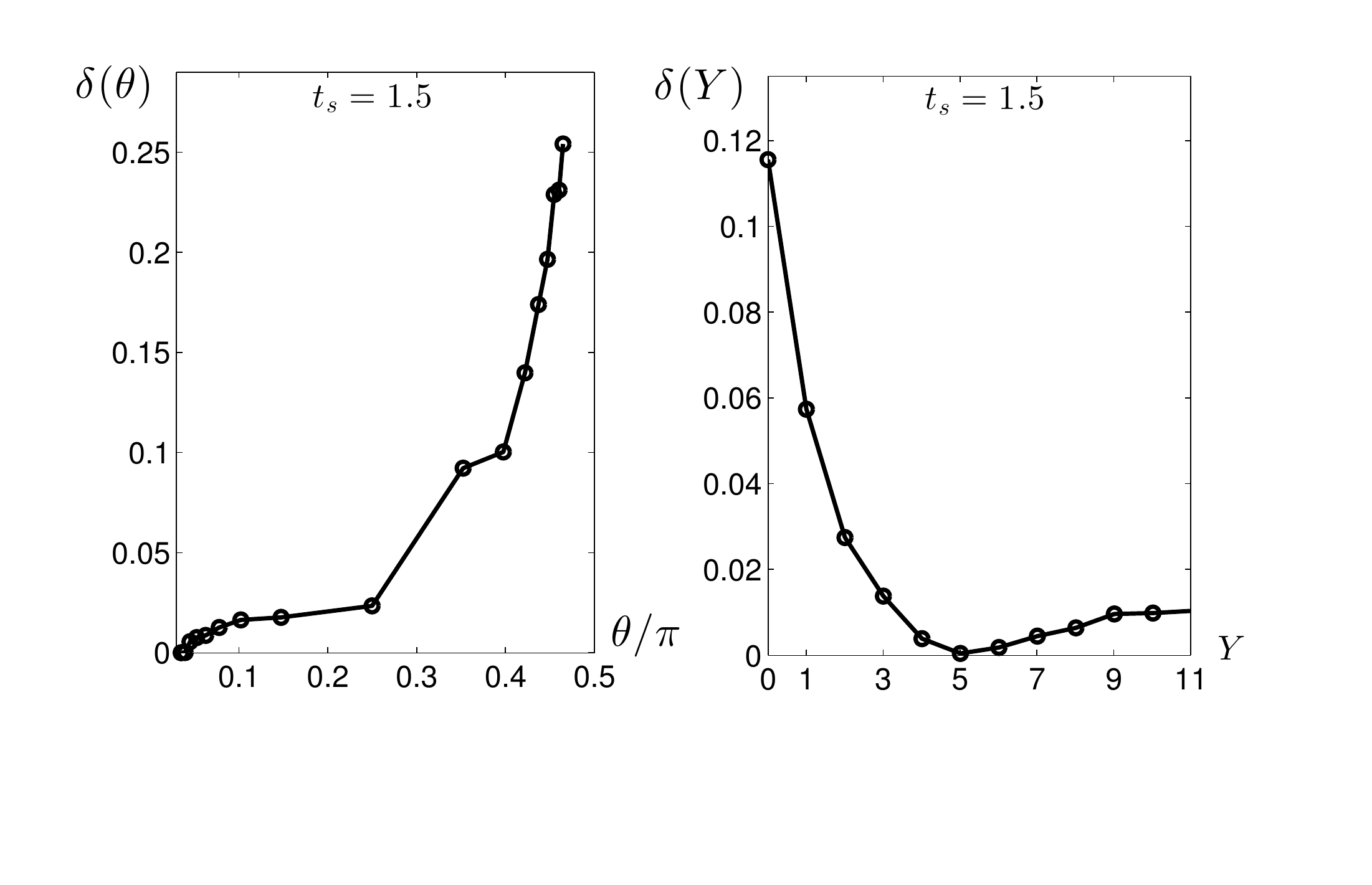}
		\caption{On the left the fitting of the exponential decay $\delta$ at different directions $\theta$ of the bi-dimensional Fourier spectrum of Prandtl's solution at the singularity time $t_s$. The most singular direction is at $\theta=0$.
		On the right, the estimates of the exponential decay $\delta$ of the Fourier spectrum of Prandtl's solution at time $t_s$, in terms of the normal variable $Y$. The location of the minimum,  $Y\approx 5$, is the location of the singularity.}\label{VDSdeltathetadeltaY}
	\end{center}
\end{figure}

\begin{figure}
	\begin{center}
		\includegraphics[height=8.cm,width=11.5cm]{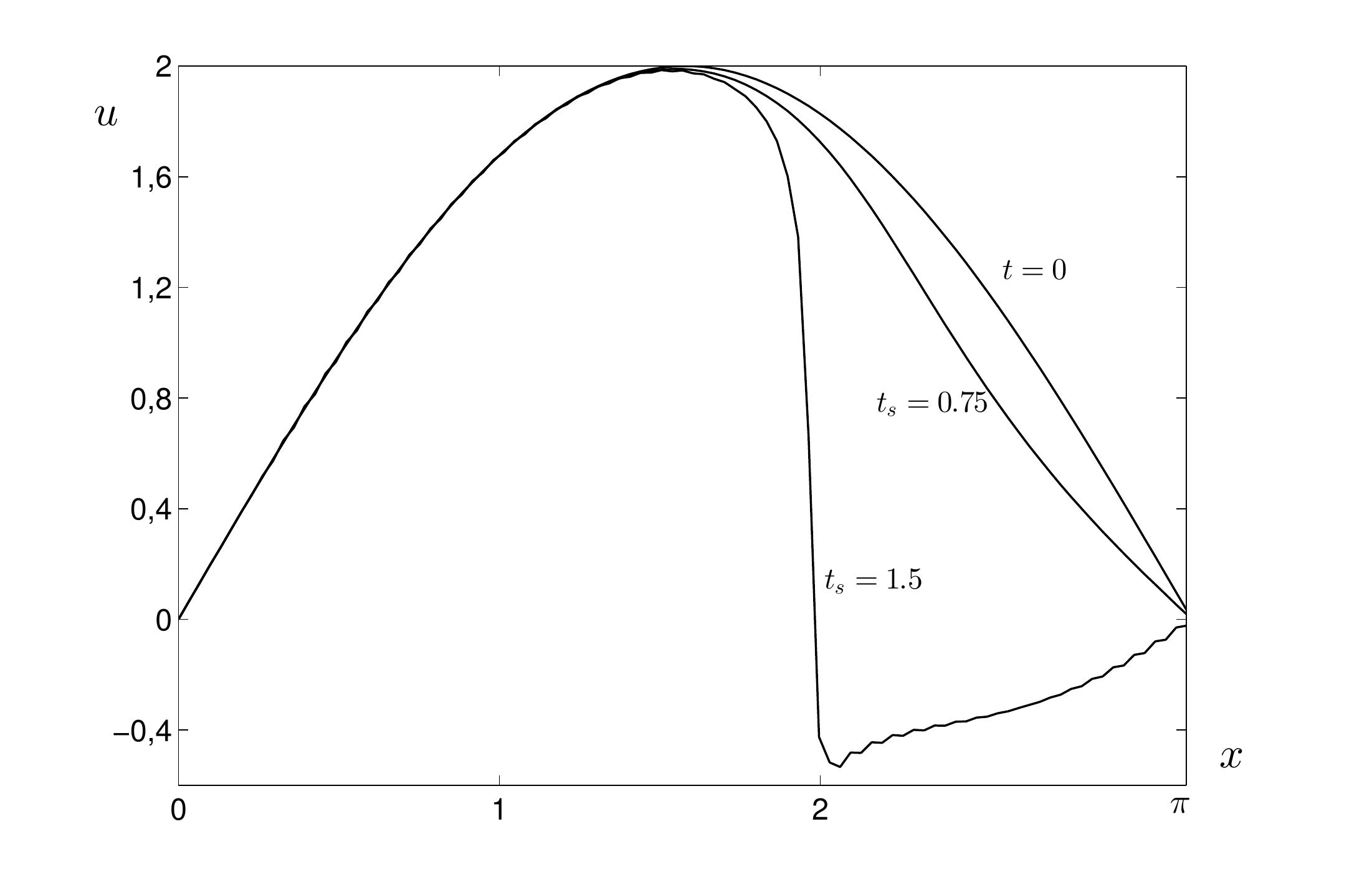}
		\caption{The
			solution of Prandtl's equation at the location $Y=5$ for van Dommelen and Shen
			initial datum $U_{\infty}=2\sin(x)$ at different times. It is visible the formation of a shock at $x\approx 1.94$ at the singularity time $t_s=1.5$.}\label{VDSfig}
	\end{center}
\end{figure}

In Fig. \ref{VDSshell} we show the shell-summed Fourier amplitudes, where it is evident
the loss in time of the exponential decay.
Fitting these data using formula \eqref{ShellaAsym}, we get the evolution in time of the width of the analyticity strip, shown in Fig.\ref{VDSshellFITT}(b), and the algebraic characterization of the singularity in Fig.\ref{VDSshellFITT}(a).
At the critical time $t_s=1.5$, the solution loses analyticity
as a cubic-root singularity.
Analyzing the Fourier spectrum of Prandtl solution $u$ at the singularity time $t_s$ using formula \eqref{Fourier_asymp_theta},
in Fig.\ref{VDSdeltathetadeltaY} on the right we show the angular dependence of $\delta$ where it is visible that the most singular direction is at $\theta=0$.
As explained in the previous section, this result allows to treat the normal variable $Y$ as a parameter, and on the right of the same Figure we show the dependence of $\delta$ on $Y$ at the singularity time $t_s$ using formula \eqref{formula_cut}. Because $\delta(Y)$ attains its  minimum at $Y\approx 5$,
this implies that the singularity is located at  $Y\approx 5$ and we apply the singularity tracking method to the one dimensional function $u(x,5)$, whose evolution in time is shown in Fig.\ref{VDSfig} where the shock at $x^\star\approx 1.94$ is visible at time $t_s$.

\begin{figure}
	\begin{center}
		\includegraphics[height=8.cm,width=11.5cm]{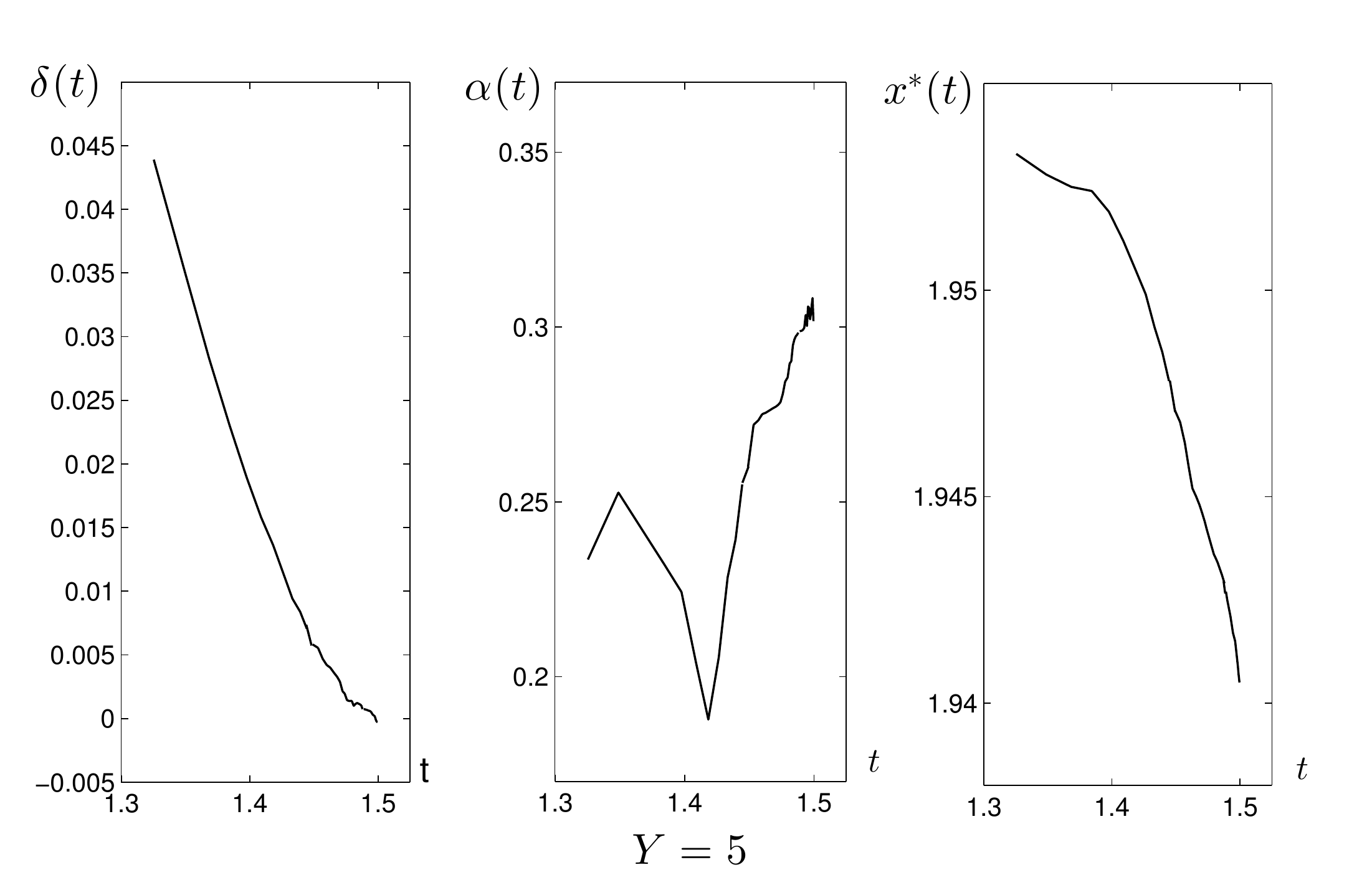}
		\caption{The results of the
			singularity tracking for the spectrum of the
			solution of Prandtl's equation at the location $Y=5$.}\label{VDSfit}
	\end{center}
\end{figure}

\begin{figure}
	\begin{center}
		\includegraphics[height=10.cm,width=13cm]{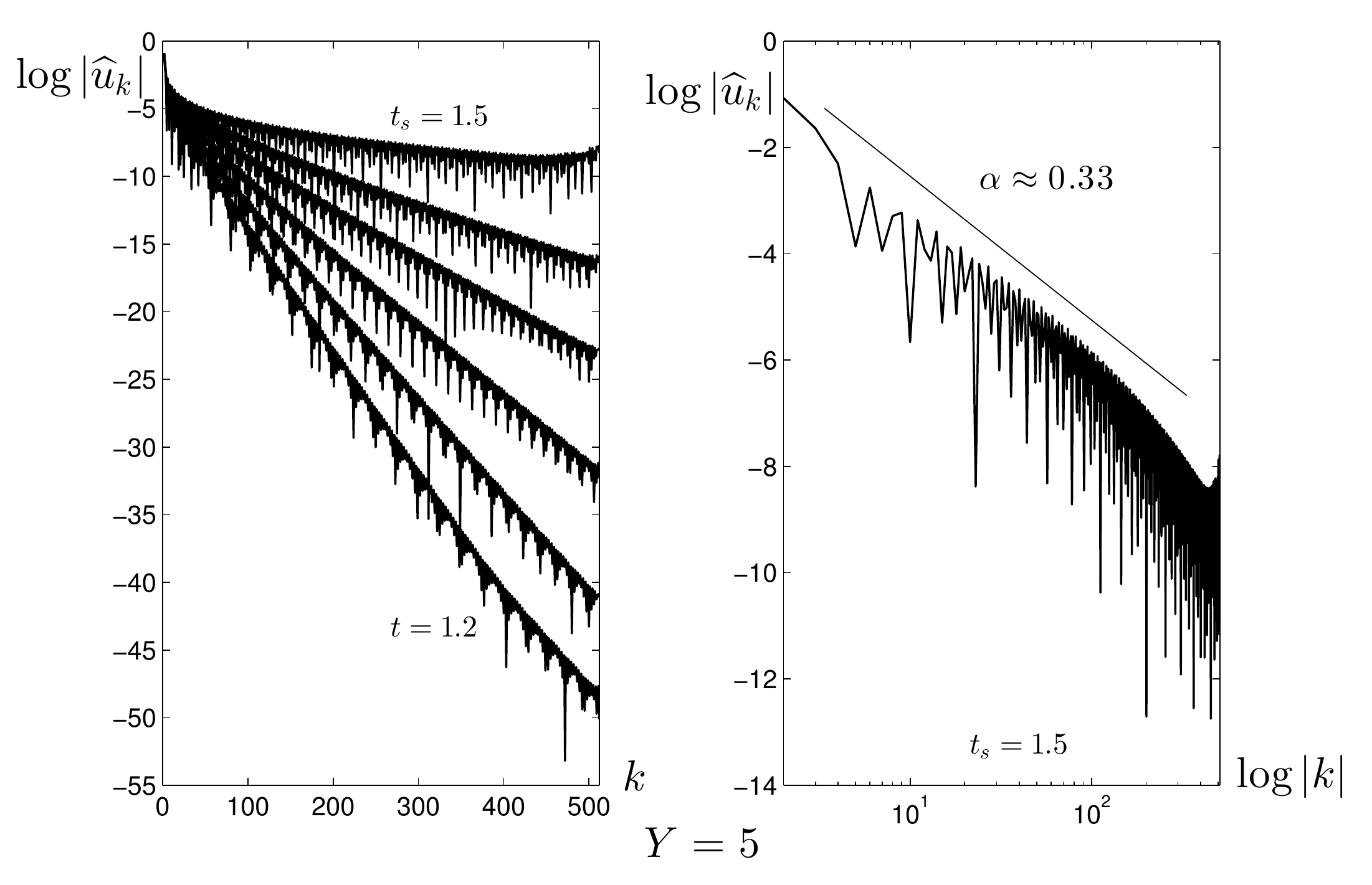}
	\vspace*{-0.5cm}\caption{On the left the spectrum of the
			solution of Prandtl's equation at the location $Y=5$ from time $t=1.2$ up to the singularity time $t_s=1.5$ with increments of $0.05$. On the right the Fourier spectrum of Prandtl's solution at the singularity time $t_s$ and $Y=5$ in log-log coordinates. It is visible the lose of the exponential decay while the rate of the algebraic decay is $\alpha\approx0.33$. }\label{VDSspec}
	\end{center}
\end{figure}

In Fig.\ref{VDSspec} it is shown the behavior in time of the Fourier spectrum at the location $Y=5$ of $u$ and in Fig.\ref{VDSfit} one can see the results of the singularity
tracking method  at the location
$Y=5$, showing again the formation of a cubic root singularity at time $t_s=1.5$. What is important now is the determination of the real tangential location
of the singularity $x^*$, which is founded with a study of the oscillatory behavior of the spectrum depurated by the exponential and algebraic decay, using formula \eqref{laplace}. This complete the analysis of the singularity of Prandtl's solution in the case of an impulsively started disk. The details of these results are in \cite{GSS09,DLSS06,GLSS09}.

%%%%%%%%%%%%%%%%%%%%%%%%%%%%%%%%%%%%%%%%%%%%%%%%%%%%%%%%%%%%%%%
\subsection{Navier-Stokes equation}\label{sec:4.2}

In this section we present the results obtained by applying the singularity analysis for the 2D spectrum of the velocity component
$u(r,\theta)$ of the Navier-Stokes solutions obtained from \eqref{velocityewd} for different Reynolds number(see \cite{GSSC14} for more details).
To perform this analysis we have mapped the physical domain of the various solutions to $\Lambda=[1,2]$ so that the points in $\Lambda$ are the Gauss-Lobatto points
$\kappa_i=\cos(i\pi/N)_{i=0,\ldots,M}$ in $\Lambda$. By introducing the points $\zeta_i=\arccos \kappa_i$ we can write Navier-Stokes solution as
\begin{equation}
u(\theta,\zeta,t)\approx
\sum\limits_{k=-K/2}^{k=K/2}\sum\limits_{j=0}^{j=M}u_{kj}(t)e^{ik\theta}
\cos{(j\zeta)},
\label{ChebCos}
\end{equation}
and the singularity-tracking method is applied on the Fourier coefficients
$u_{kj}$.

\begin{figure}
	\begin{center}
		\includegraphics[width=11.5cm]{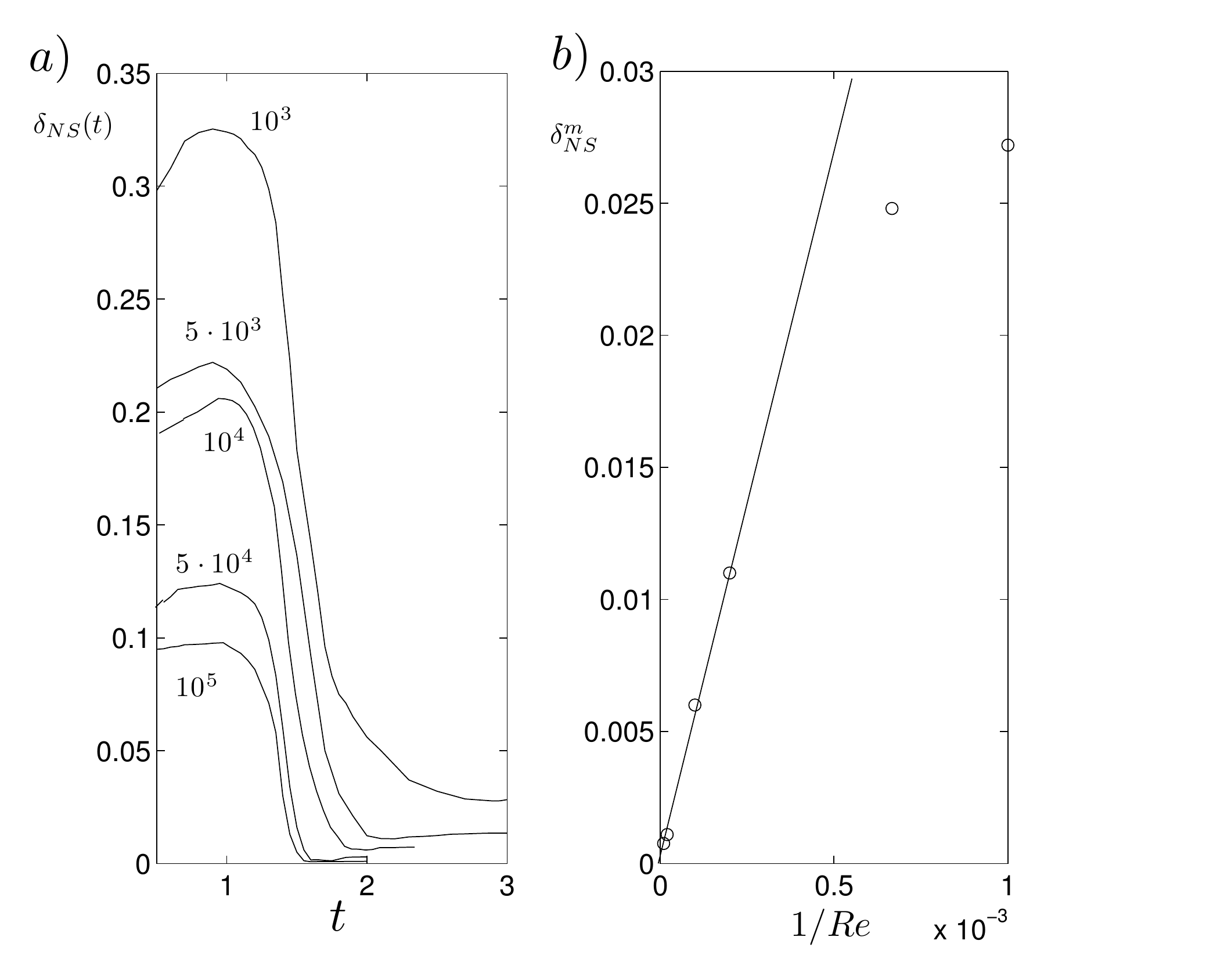}
		\caption{\textbf{a)} Time evolution of $\delta_{NS}$ for various Reynolds
			numbers.  $\delta_{NS}^m$ has a minimum in time which is of order  $O(1/Re)$ as shown in \textbf{b)} (at least
			for the Reynolds numbers for which small-scale interaction forms).}
		\label{delta_NS_inset}
	\end{center}
\end{figure}

\begin{figure}
	\begin{center}
		\includegraphics[width=9.25cm]{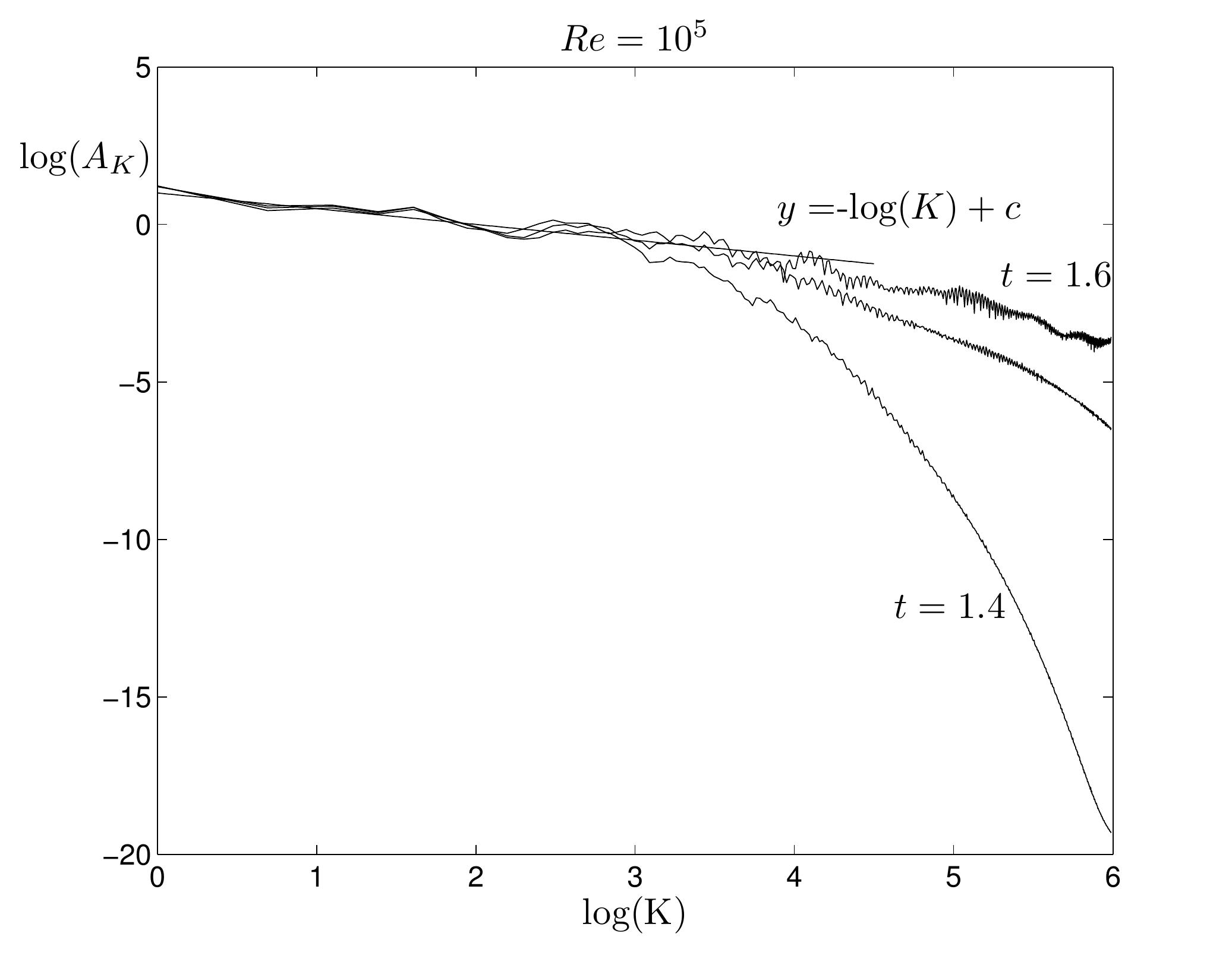}
		\caption{The behavior for different times (in increments of 0.1) of the shell-summed
			Fourier amplitudes $A_K$ for $Re=10^5$ in log-log coordinates.
			The first linear range of the amplitudes
			seems to have a slope equal to $-1$.}
		\label{alphafitting100000}
	\end{center}
\end{figure}

The time evolution of of the rate of exponential decay $\delta_{NS}$ in \eqref{ShellaAsym}  in shown in Fig.\ref{delta_NS_inset}a for various Reynolds numbers.
The various time evolutions are similar in all
cases, and after the formation of a maximum value $\delta_{NS}^M$ in time, $\delta_{NS}$ decreases as the gradients in the $\theta$ direction become intense. For all the Reynolds numbers considered, $\delta_{NS}$ has a local minimum $\delta_{NS}^m$ in time
after $t_s=1.5$, and after this event it then begins to
increase again. It is worth noting that the minimum in time
$\delta_{NS}^m$ seems to scale linearly with respect to $1/Re$ for the Reynolds
numbers for which small-scale interaction occurs (see also Fig.~\ref{delta_NS_inset}b).
Regarding the evaluation of $\alpha_{NS}$ we first  observe that all of the spectra analyzed have several structures leading to several difficulties in determining the correct value of this charachterization. At the onset of large-scale
interaction, when the spectrum is more easily handled, a fitting of the Fourier amplitudes always gives results in the range
$0.45<\alpha_{NS}<0.55$ for all of the Reynolds numbers considered, suggesting that the
value $\alpha_{NS}=1/2$ is the more probable characterization. For instance, in Fig.~\ref{alphafitting100000} the behavior of the Fourier amplitudes $A_K$ for $Re=10^5$,
is shown in log-log coordinates: the linear behavior of the first range of Fourier amplitudes, whose slope
returns the rate of algebraic decay,
compares with the straight line of slope $-1$, and this
supports the prediction that $\alpha_{NS}=1/2$.

\begin{figure}
	\begin{center}
		\includegraphics[width=11.5cm]{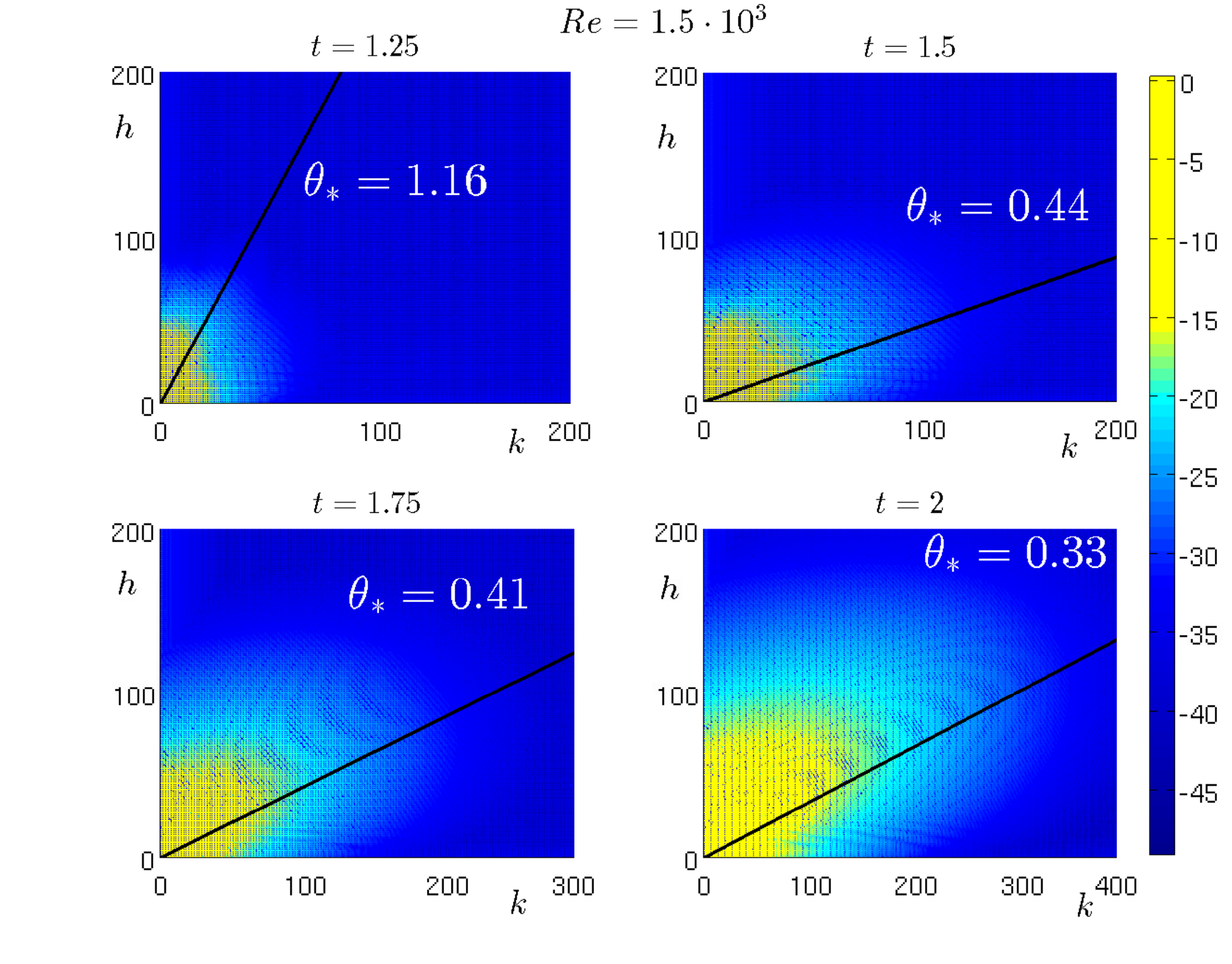}
		\caption{The spectrum (in log-scale) of $u$ for $Re=1.5\cdot10^3$ at various
			time.}
		\label{spetrum2d1500}
	\end{center}
\end{figure}

\begin{figure}
	\begin{center}
		\includegraphics[width=11.5cm]{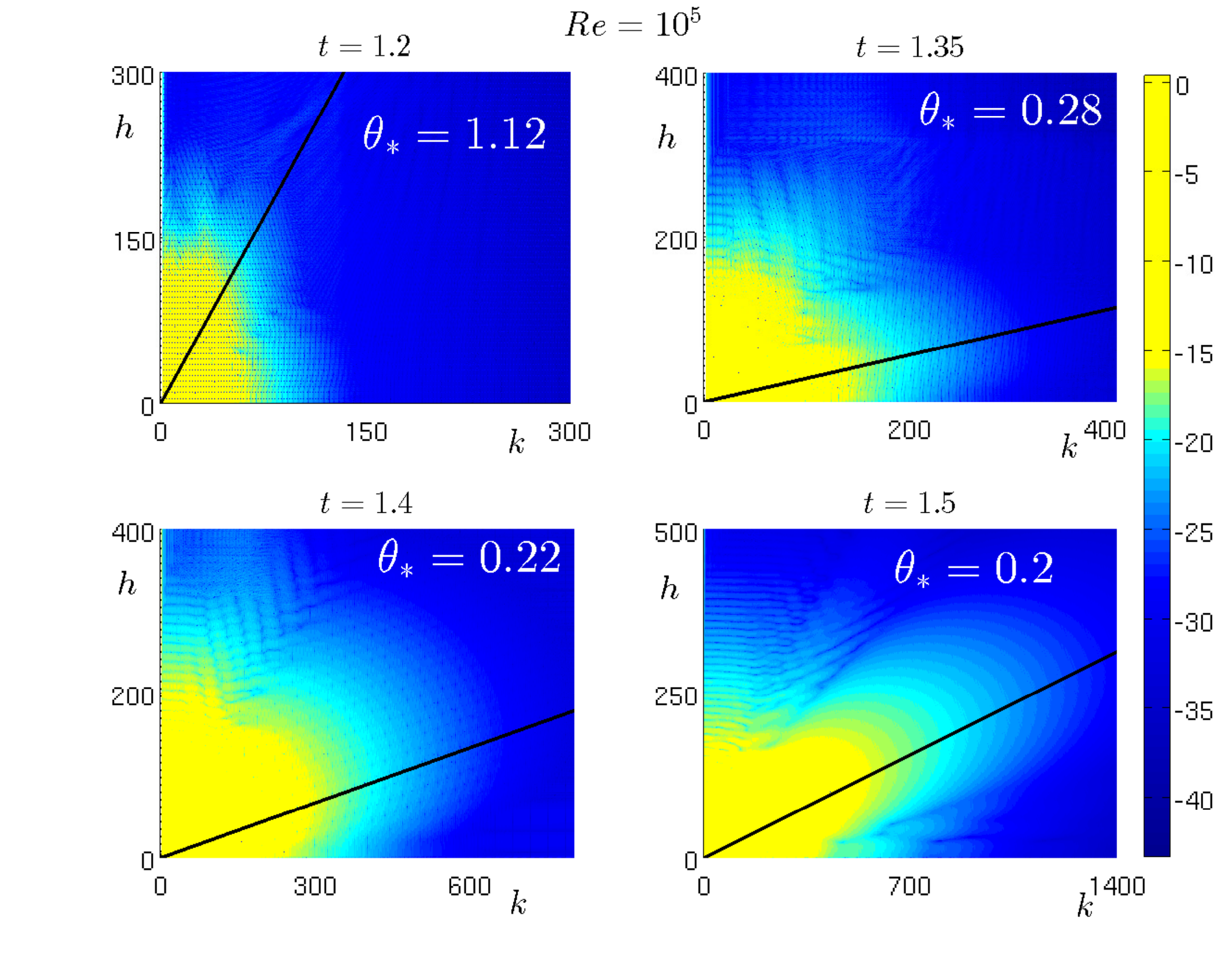}
		\caption{  The  spectrum (in log-scale) of $u$ for $Re=10^5$ at various time. At
			$t\approx 1.35$, a bulge forms in the spectrum that becomes more pronounced as time
			passes due to the effect of the small-scale interaction characterized  by appearance of strong gradients of $u$ in the variable $\theta$.}
		\label{spetrum2d100000}
	\end{center}
\end{figure}

A relevant feature of the small-scale interaction is the formation of a bulge in the 2D spectra of the solutions for the Reynolds numbers for which this interaction forms. In Figs~\ref{spetrum2d1500}-\ref{spetrum2d100000} the spectra are shown at different times for $Re=1.5\cdot10^3$ and $10^5$ with the most singular direction indicated by a straight line. For $Re=10^5$ (also for
$Re=5\cdot10^3,10^4,5\cdot10^4$ not shown here) we can observe that the bulge begins to appear at the time in which small-scale interaction begins ($t\approx 1.35$), while for the case $Re=1.5\cdot10^3$ the bulge never forms and the spectrum grows throughout a wider range
around the most singular direction. We can relate the presence of this bulge  to the effect of    the small-scale
interaction which reveals itself through the formation of large gradients in the angular direction
$\theta$ in the solution, leading to the excitement of the high wavenumber Fourier modes along the most singular direction.
Moreover as time passes, the most singular direction approaches $\theta^*=0$, which confirms that the
relevant gradients present in $u$ are those
relative to the coordinate $\theta$. This result is also compatible with
the result obtained for Prandtl's solution for which , at the singularity time, the most singular direction is $\theta^*=0$.

%%%%%%%%%%%%%%%%%%%%%%%%%%%%%%%%%%%%%%%%%%%%%%%%%%%%%%%%%%%%%%%%%%%%%%%%%%%%%%%%
\subsection{KP equation}

The KP equation can be put in the form:
\begin{equation}\label{KP_eq}
\partial_t u+6u \partial_x u + \epsilon^2 \partial_{xxx} u = -\lambda \partial_{x}^{-1} \partial_{yy} u\; ,
\end{equation}
with $\lambda=\pm 1$ (here we consider the case when $\lambda=1$ with defocusing effects)
with periodic boundary condition.
The variable $x$ and $y$ are in $[-L_x \pi,\, L_x \pi]$ and $[-L_y \pi,\, L_y \pi]$, and we choose $L_x=L_y=5$.

We consider the initial datum:
\begin{equation}\label{KP_ini}
u(x,y,0)=-\partial_x {\rm sech}^2\left(\sqrt{x^2+y^2}\right),
\end{equation}
in such a way that the Fourier mode at $k_x=0$ is null at the initial time and we can treat the $\partial_x^{-1}$ as
$-i/k_x$ with $k_x \neq 0$.

The analysis of the singularity formation when $\epsilon=0$ is very similar to the Prandtl case.

\begin{figure}
\begin{center}
\includegraphics[height=8.cm,width=11.5cm]{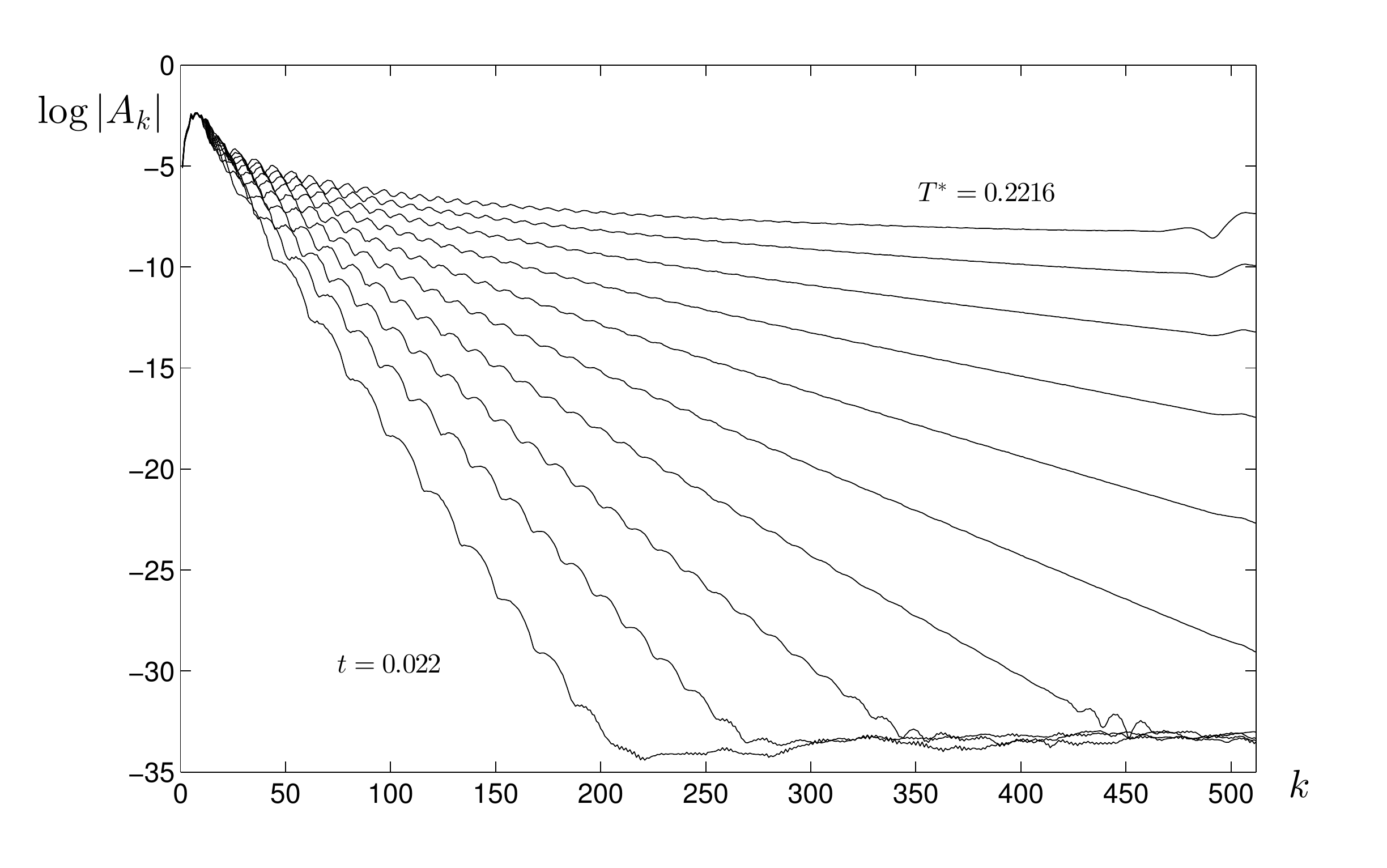}
\caption{{The behavior in time of the shell summed amplitude up
to the singularity time.}}\label{KPshell}
\end{center}
\end{figure}

\begin{figure}
\begin{center}
\includegraphics[height=8.cm,width=11.5cm]{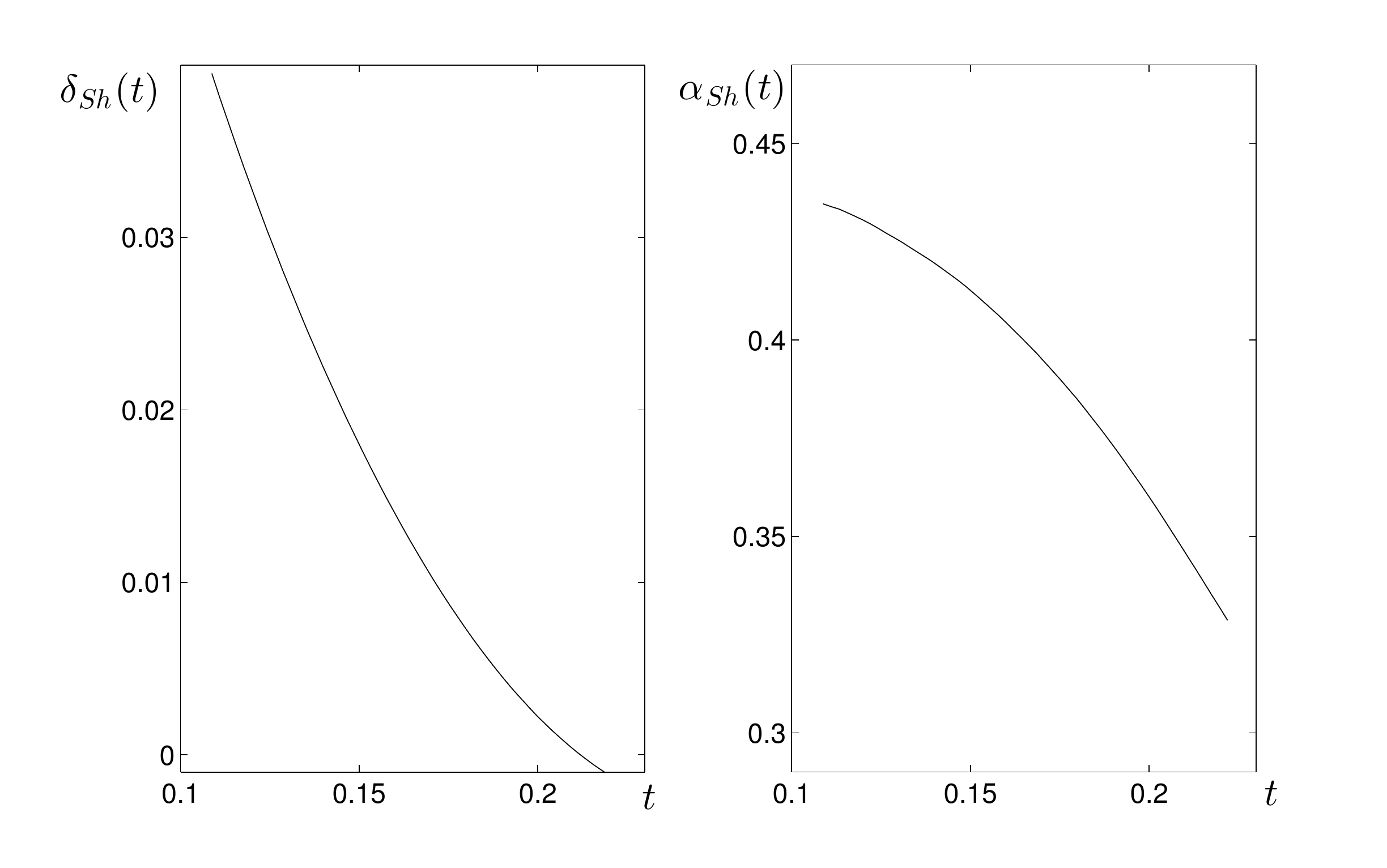}
\caption{The behavior in time of the width of the analyticity strip.
The singularity time is $T^*\approx 0.2216$. The singularity is of cubic-root
type.}\label{KPshellFITT}
\end{center}
\end{figure}

\begin{figure}
\begin{center}
\includegraphics[height=8.cm,width=11.5cm]{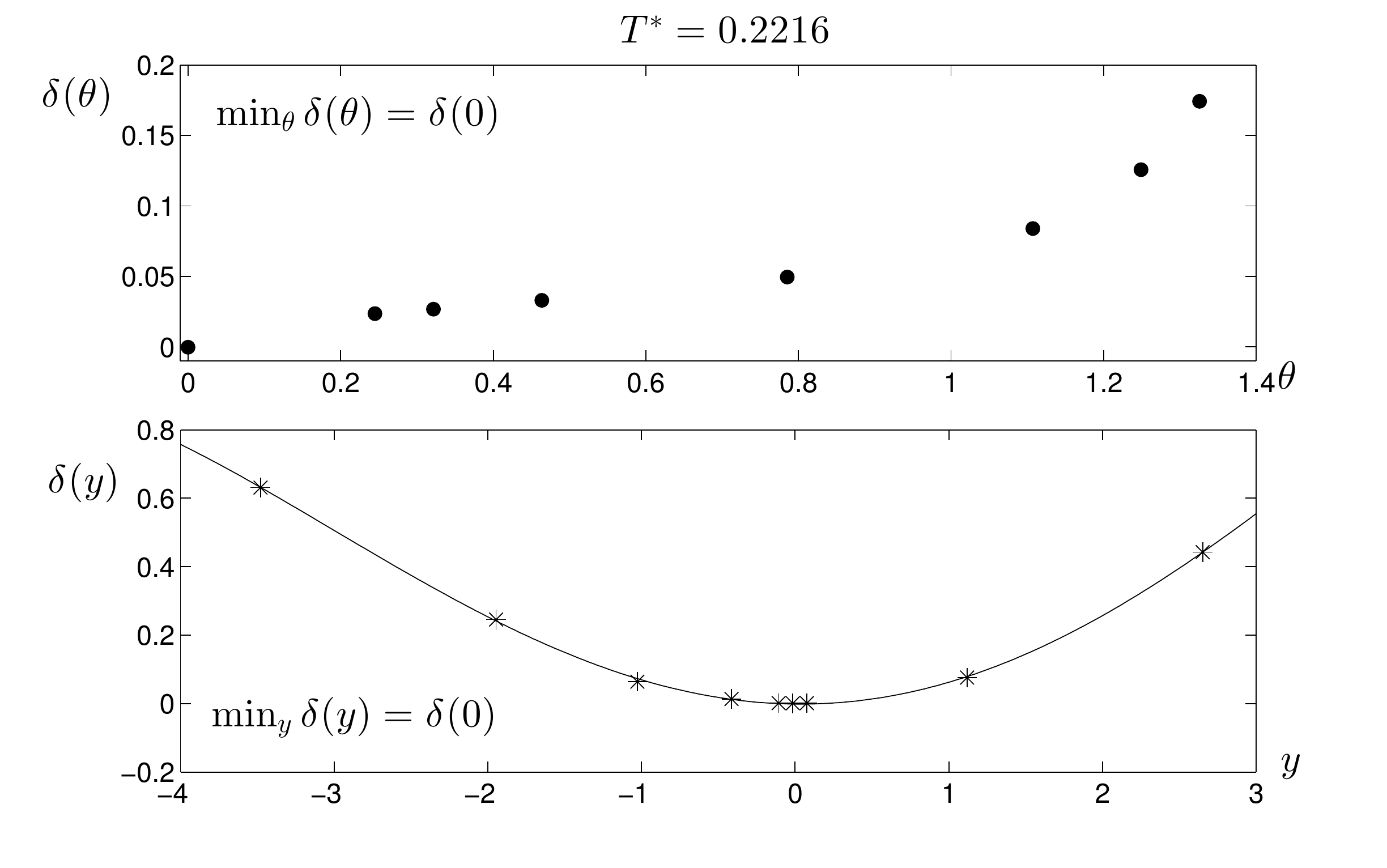}
\caption{The most singular direction is $\theta=0$.
If one estimates the $\delta$ in its dependence on the $y$ variable,
one finds that the location of the singularity is at $y\approx 0$.}\label{KPdeltaY}
\end{center}
\end{figure}

In Fig. \ref{KPshell} we show the shell-summed amplitudes, where it is evident
the loss of exponential decay.
Fitting these amplitudes we get the results shown in Fig. \ref{KPshellFITT},
where one can see that, at the critical time $T^*=0.2216$, the solution loses analyticity
as a cubic-root singularity hits the real axis.
In Fig. \ref{KPdeltaY} we show the angular dependence of $\delta$ and
one can see that the most singular direction is $\theta=0$.
This result allows to treat the normal variable as a parameter.
At the bottom of the same Figure we show the dependence of $\delta$ on  $y$,
and one can see that  $\delta(y)$ attains its  minimum at $y=0$.

\begin{figure}
\begin{center}
\includegraphics[height=8.cm,width=11.5cm]{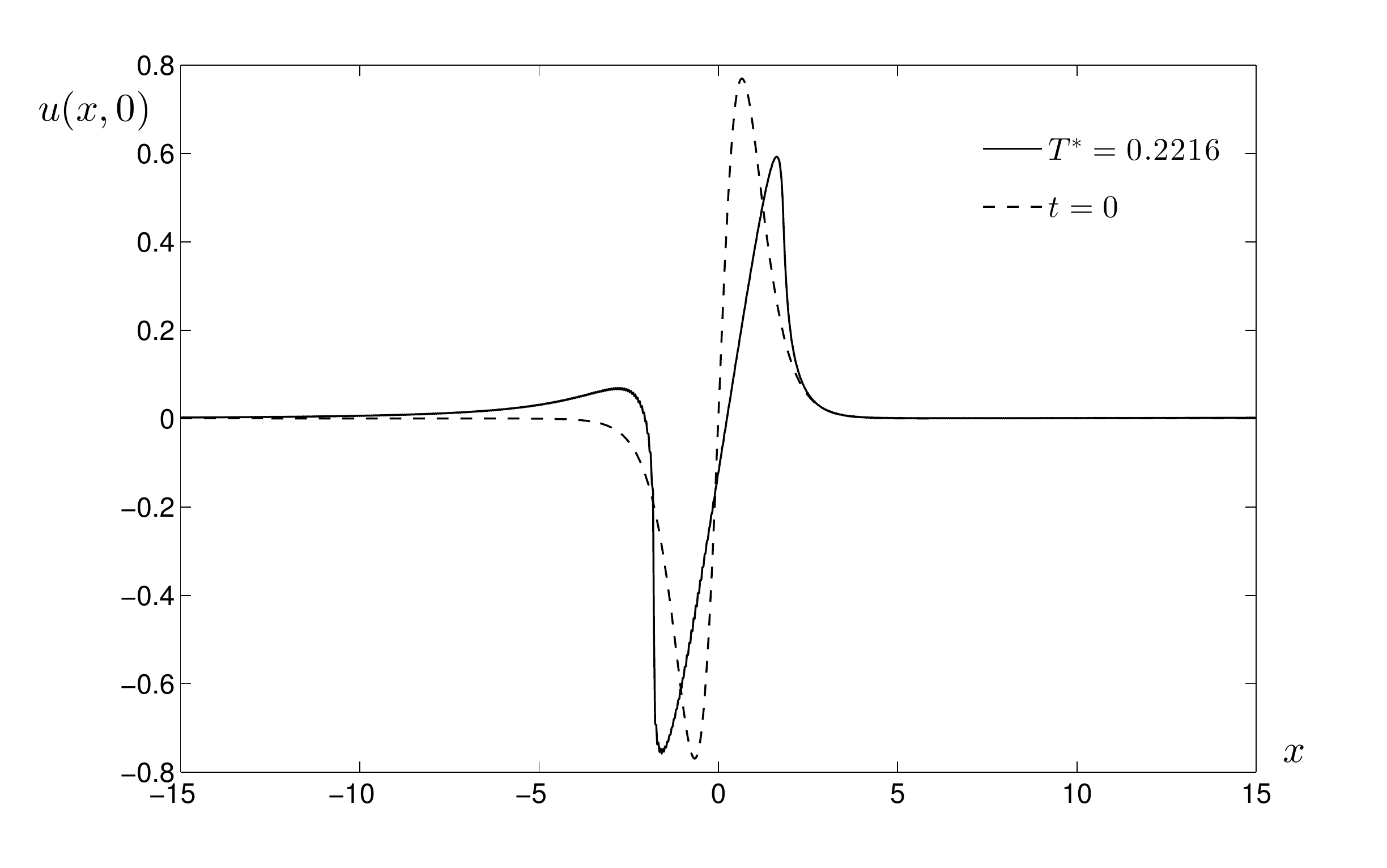}
\caption{The
solution of the KP equation, with $\epsilon=0$, at the location $y=0$ for
initial datum given by \eqref{KP_ini}.}\label{KPfig}
\end{center}
\end{figure}

\begin{figure}
\begin{center}
\includegraphics[height=8.cm,width=11.5cm]{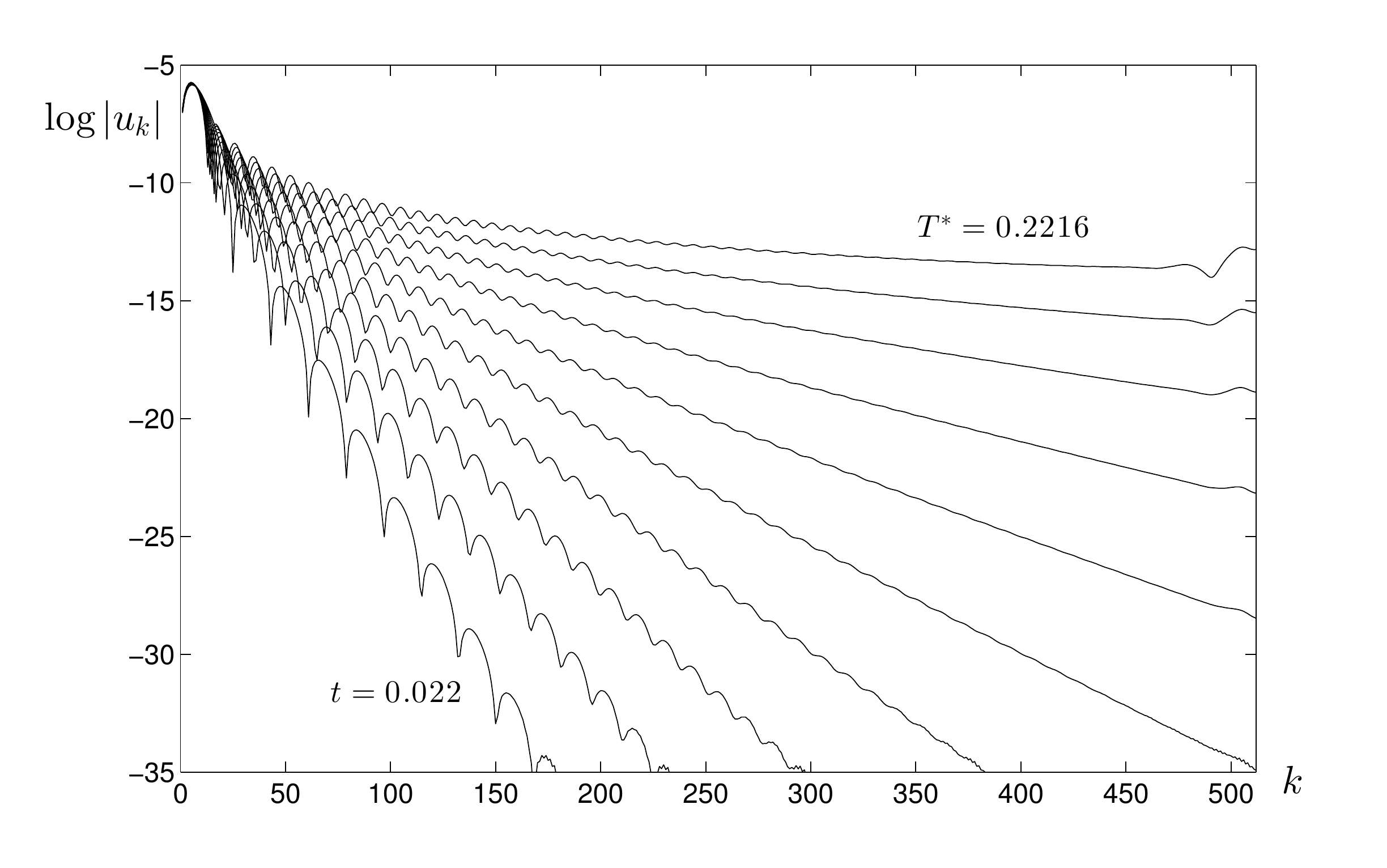}
\caption{The spectrum of the
solution KP equation, with $\epsilon=0$, at the location $y=0$ for
initial datum given by \eqref{KP_ini}.}\label{KPspec}
\end{center}
\end{figure}

\begin{figure}
\begin{center}
\includegraphics[height=8.cm,width=11.5cm]{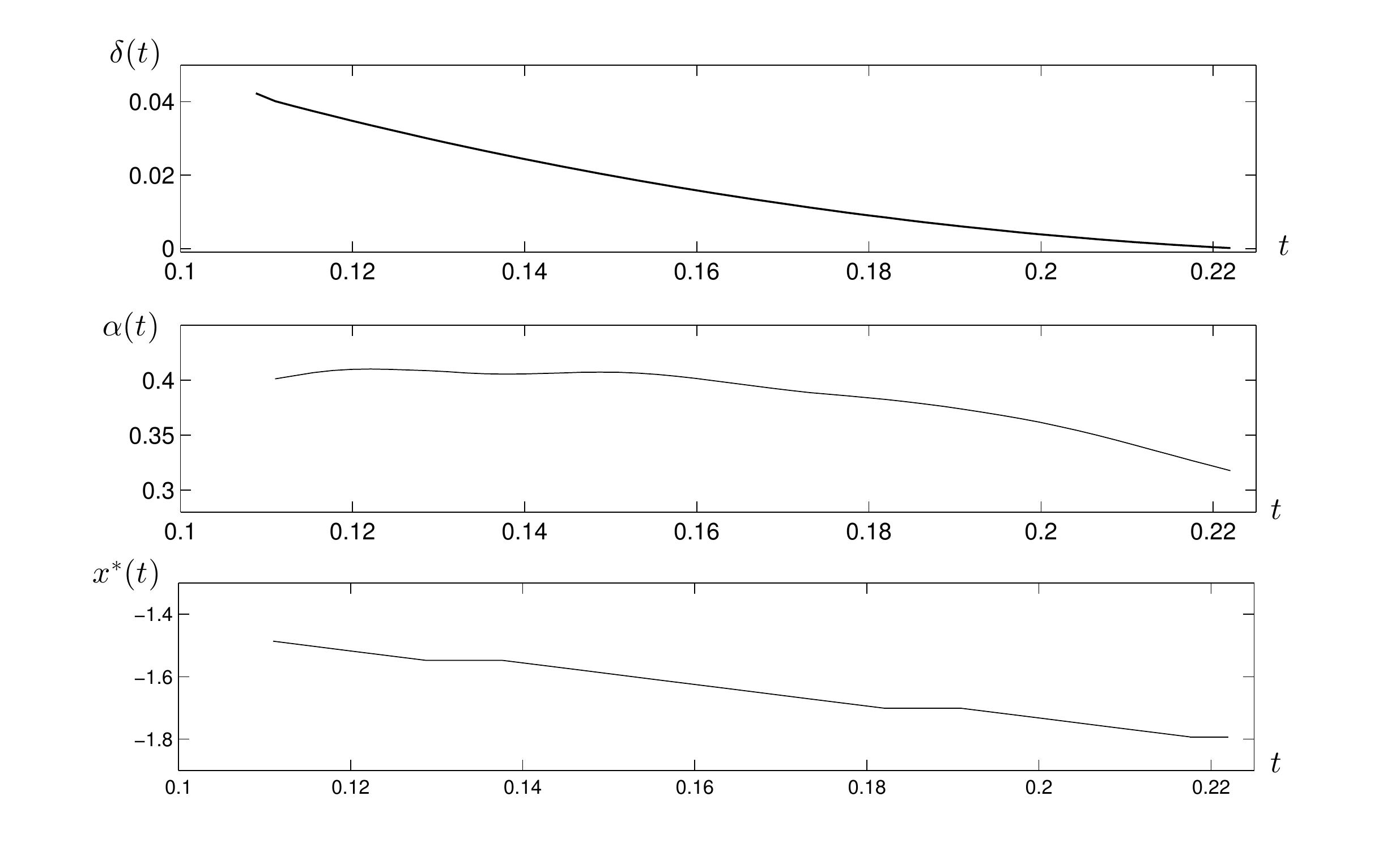}
\caption{The results of the
singularity tracking at $y=0$. One can see that at $t\approx 0.2216$ the strip
of analyticity shrinks to zero as the result of a cubic--root
singularity hitting the real axis. The location of the real coordinate of the singularity
at the singularity time is $x^*\approx -1.79$.}\label{KPfit}
\end{center}
\end{figure}

In Fig.\ref{KPfig} we show the solution, at the singularity time,
at the location $y=0$ and in Fig.\ref{KPspec} it is shown the behavior in time of the spectrum
at location $y=0$.
In Fig.\ref{KPfit} one can see the results of the singularity
tracking method for the initial datum \eqref{KP_ini} at the location
$y=0$. The examination of the oscillatory behavior of the spectrum,
gives the  real location of the singularity;
in Fig.\ref{KPfit} one can see that  $x^*\approx -1.79$.

\begin{figure}
\begin{center}
\includegraphics[height=8.cm,width=11.5cm]{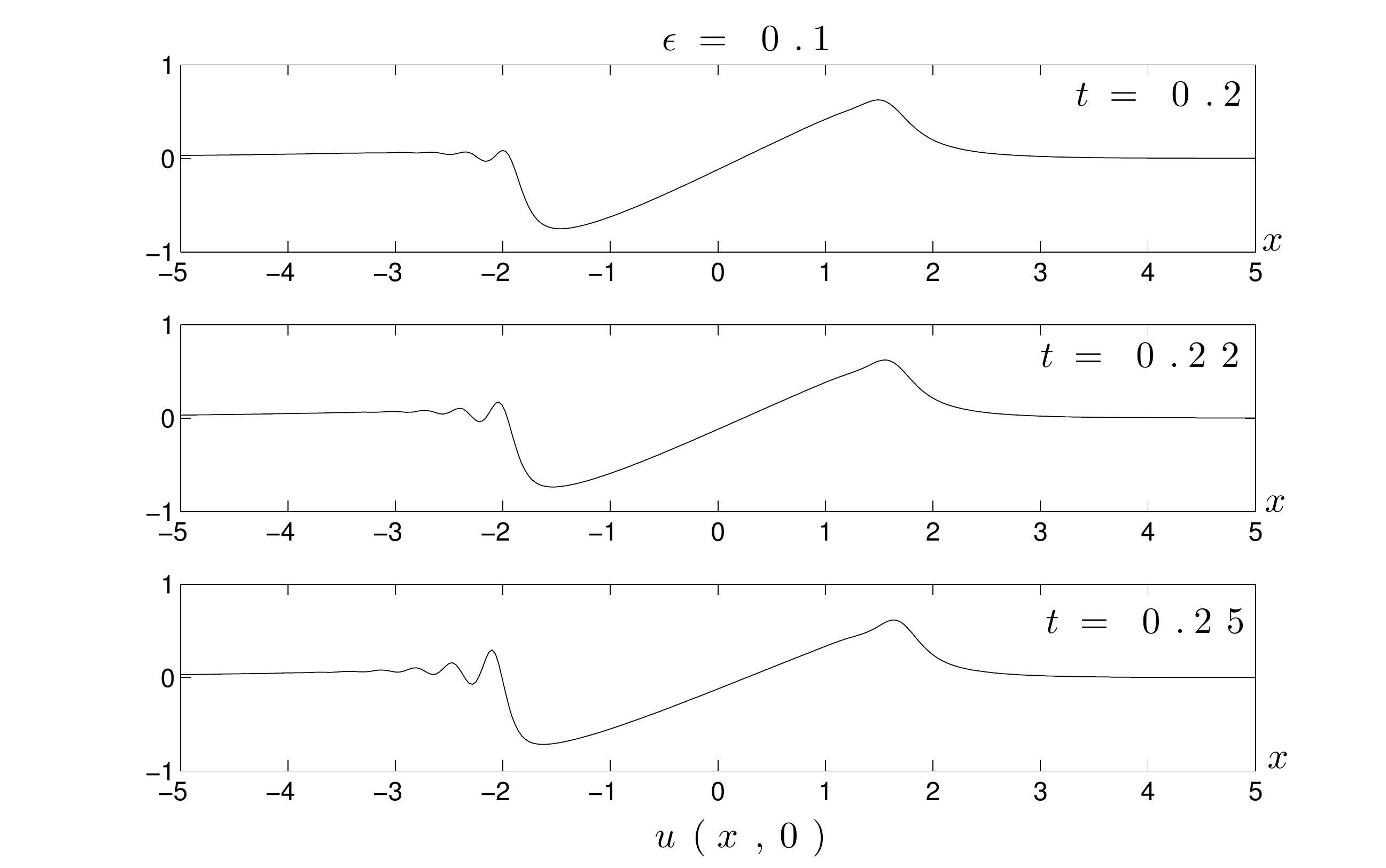}
\caption{The
solution of the KP equation, with $\epsilon=0.1$, at the location $y=0$ for
initial datum given by \eqref{KP_ini} at three different times.}\label{KP01fig}
\end{center}
\end{figure}

We consider now the KP equation with $\epsilon=0.1$, and we want to analyze the behavior of the complex singularities.
As previously observed for the KdV equations, the solution of the KP equation with $\epsilon=0.14$ has a regions of rapid modulated oscillations in the vicinity of $x^*=-1.79$, the shock position of the dispersionless KP solution, as one can see in Fig. \ref{KP01fig}.

\begin{figure}
\begin{center}
\includegraphics[height=8.cm,width=11.5cm]{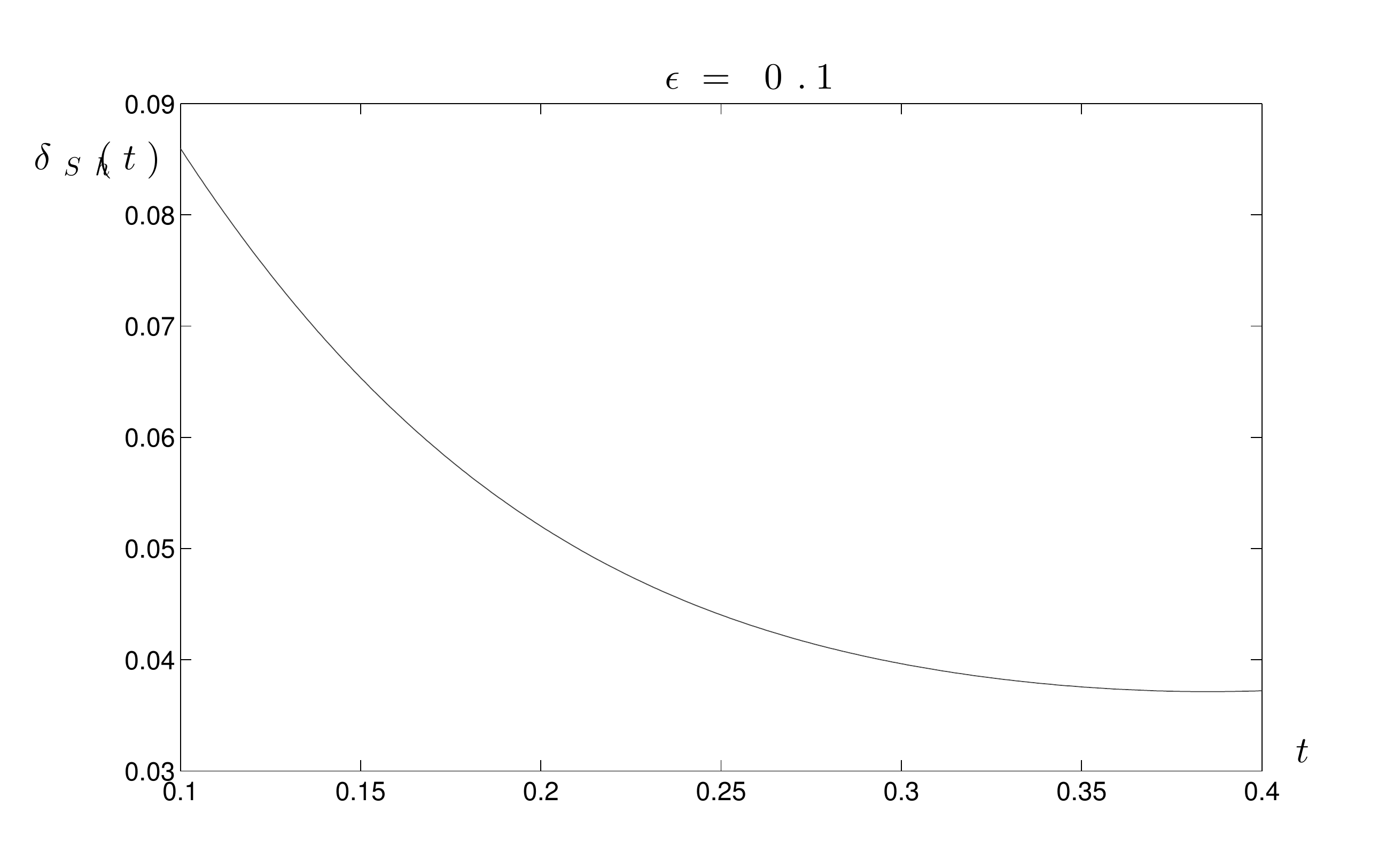}
\caption{The behavior in time of the width of the analyticity strip when $\epsilon=0.1$.}\label{KP01shellFITT}
\end{center}
\end{figure}

The analysis of the spectrum using the shell-summed amplitudes is given in Fig.\ref{KP01shellFITT}.
The behavior in time of the $\delta_{Sh}$ width of the analyticity strip shows that when $\epsilon=0.1$ there is no formation of real singularity.

\begin{figure}
\begin{center}
\includegraphics[height=8.cm,width=11.5cm]{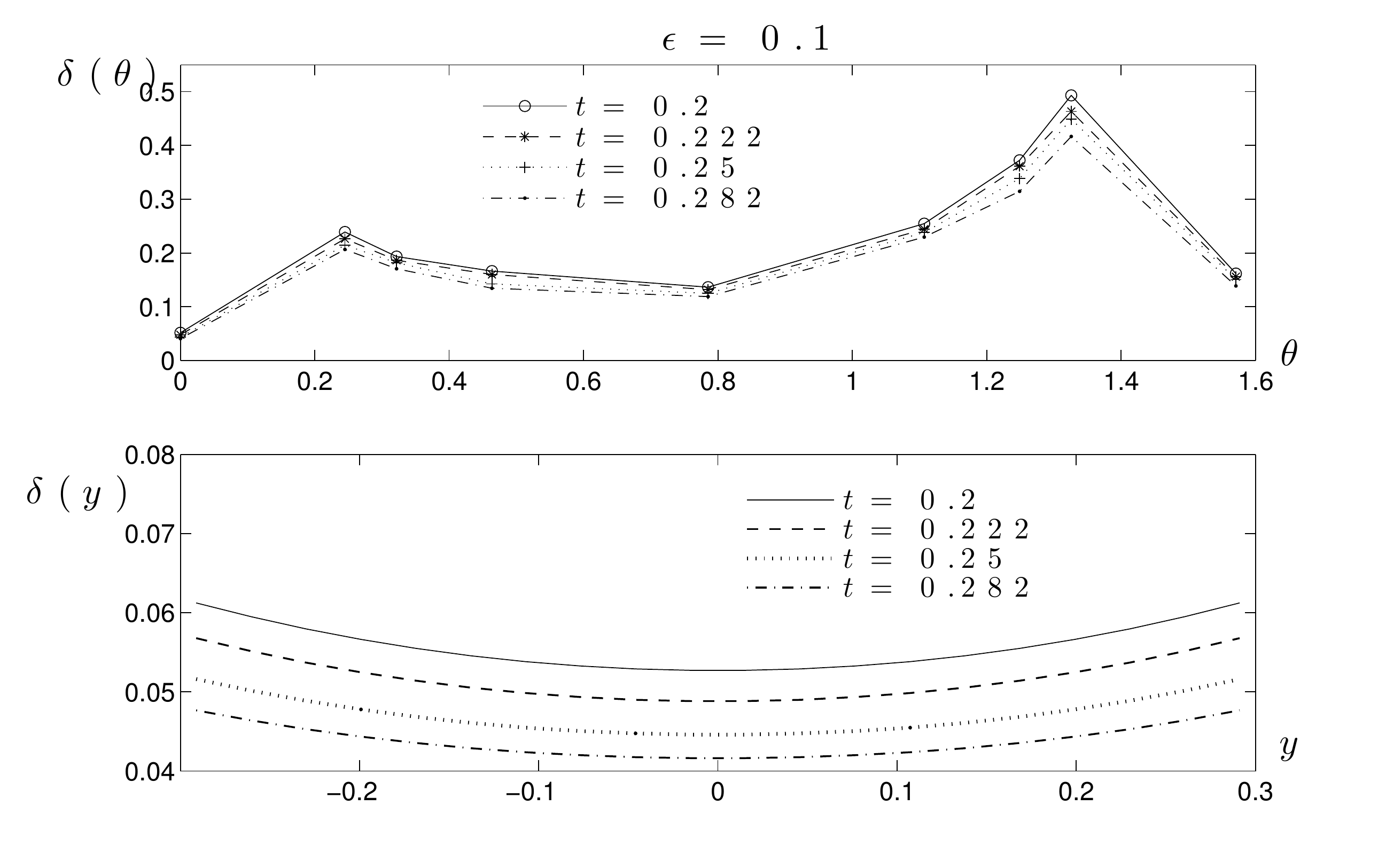}
\caption{The behavior of the $\delta(y)$ and $\delta(\theta)$ of the KP equation with $\epsilon=0.1$ at different times.
The most singular direction is $\theta=0$.
If one estimates the $\delta$ in its dependence on the $y$ variable,
one finds that the location of the complex singularity is at $y\approx 0$.}\label{KP01deltaY}
\end{center}
\end{figure}

In Fig.\ref{KP01deltaY} we show the angular dependence of $\delta$ at different times and
one can see that the most singular direction is $\theta=0$.
This result allows to treat the normal variable as a parameter.
At the bottom of the same Figure we show the dependence of $\delta$ on  $y$,
and one can see that  $\delta(y)$ attains its  minimum at $y=0$.

\begin{figure}
\begin{center}
\includegraphics[height=8.cm,width=11.5cm]{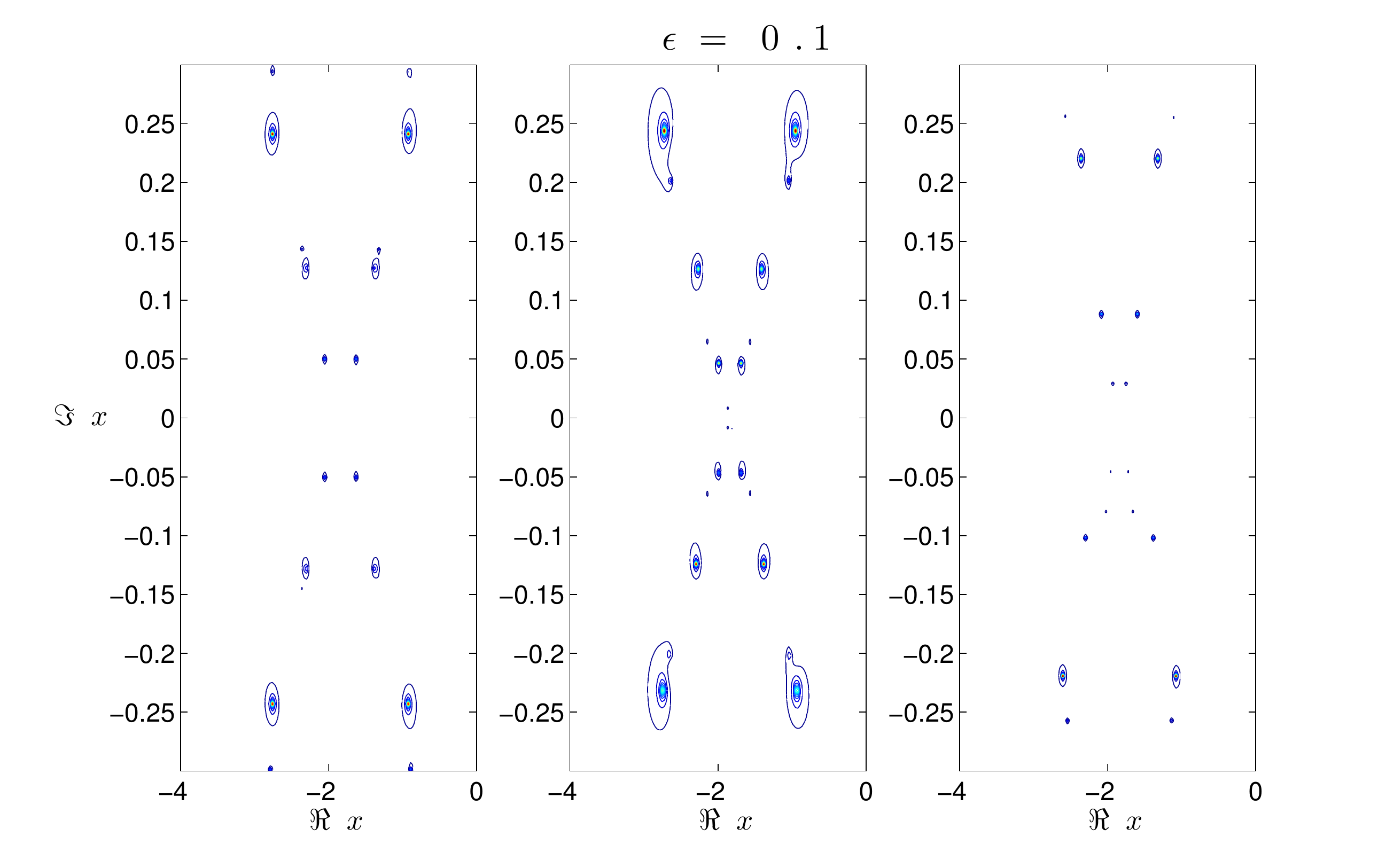}
\caption{Poles at different times for the
solution of the KP equation, with $\epsilon=0.1$, at the location $y=0$. The Pad\'e analysis results.}\label{KP01pade}
\end{center}
\end{figure}

We compute the Pad\'e approximants for the solution of the KP equation, with $\epsilon=0.1$, at the location $y=0$ for
initial datum given by \eqref{KP_ini} at three different times (see Fig.\ref{KP01pade}).
We observe a behavior very similar to  what observed for the dispersive
Burgers equation and the KdV equation.
We conjecture that the region of modulated oscillations in the vicinity of the shocks in the
dispersionless solution of a nonlinear dispersive equation can be explained with the presence of
coalescing complex singularities located on a curve (maybe a straight line) which
approaches the real axis.

\bibliographystyle{amsplain}
\bibliography{Complex_singularities_and_PDEs}{}

\providecommand{\bysame}{\leavevmode\hbox to3em{\hrulefill}\thinspace}
\providecommand{\MR}{\relax\ifhmode\unskip\space\fi MR }
% \MRhref is called by the amsart/book/proc definition of \MR.
\providecommand{\MRhref}[2]{%
  \href{http://www.ams.org/mathscinet-getitem?mr=#1}{#2}
}
\providecommand{\href}[2]{#2}
\begin{thebibliography}{10}

\bibitem{BCS93}
G.~R. Baker, R.E. Caflisch, and M.~J. Shelley, \emph{Singularity formation
  during {Rayleigh-Taylor} instability}, Journal of Fluid Mechanics
  \textbf{252} (1993), 51--75.

\bibitem{BGM96}
G.A. Baker and P.~Graves-Morris, \emph{{Pad{\'e} Approximants}}, Cambridge
  University Press, United States of America, 1996.

\bibitem{Boy09}
John~P. Boyd, \emph{Large-degree asymptotics and exponential asymptotics for
  {F}ourier, {C}hebyshev and {H}ermite coefficients and {F}ourier transforms},
  J. Engrg. Math. \textbf{63} (2009), no.~2-4, 355--399.

\bibitem{Boy00}
J.P. Boyd, \emph{{Chebyshev and Fourier Spectral Methods}}, DOVER Publications,
  Mineoal,New York 11501, 2000.

\bibitem{CO89}
R.~E. Caflisch and O.~F. Orellana, \emph{Singular solutions and ill-posedness
  for the evolution of vortex sheets}, SIAM J. Math. Anal. \textbf{20} (1989),
  no.~2, 293--307.

\bibitem{CS04}
R.~E. Caflisch and M.~Siegel, \emph{A semi-analytic approach to {E}uler
  singularities}, Methods Appl. Anal. \textbf{11} (2004), no.~3, 423--430.

\bibitem{Caf93}
R.E. Caflisch, \emph{Singularity formation for complex solutions of the {3D}
  incompressible {E}uler equations}, Phisica D \textbf{67} (1993), 1--18.

\bibitem{CHQZ06}
C.~Canuto, M.~Y. Hussaini, A.~Quarteroni, and T.~A. Zang, \emph{Spectral
  methods}, Scientific Computation, Springer-Verlag, Berlin, 2006, Fundamentals
  in single domains.

\bibitem{CKP66}
G.F. Carrier, M.~Krook, and C.E. Pearson, \emph{{Functions of a Complex
  Variable: Theory and Technique}}, McGraw--Hill, New York, 1966.

\bibitem{CB05}
C.~Cichowlas and M.-E. Brachet, \emph{Evolution of complex singularities in
  {Kida-Pelz} and {Taylor-Green} inviscid flows}, Fluid Dyn. Res. \textbf{36}
  (2005), 239--248.

\bibitem{CGS12}
G.M. Coclite, F.~Gargano, and V.~Sciacca, \emph{Analytic solutions and
  singularity formation for the peakon b-family equations}, Acta Appl. Math.
  \textbf{122} (2012), 419--434.

\bibitem{COW83}
S.J. Cowley, \emph{{Computer extension and analytic continuation of Blasius'
  expansion for impulsively flow past a circular cylinder}}, J. Fluid Mech.
  \textbf{135} (1983), 389--405.

\bibitem{CBT99}
S.J. Cowley, G.R. Baker, and S.~Tanveer, \emph{On the formation of {Moore}
  curvature singularities in vortex sheets}, Journal of Fluid Mechanics
  \textbf{378} (1999), 233--267.

\bibitem{DLSS06}
G.~{Della Rocca}, M.C. Lombardo, M.~Sammartino, and V.~Sciacca,
  \emph{Singularity tracking for {C}amassa-{H}olm and {P}randtl's equations},
  Appl. Numer. Math. \textbf{56} (2006), no.~8, 1108--1122.

\bibitem{ELZ97}
Nicholas~M. Ercolani, C.~David Levermore, and Taiyan Zhang, \emph{The behavior
  of the {W}eyl function in the zero-dispersion {K}d{V} limit}, Comm. Math.
  Phys. \textbf{183} (1997), no.~1, 119--143.

\bibitem{EGLS94}
N.M. Ercolani, I.R. Gabitov, C.D. Levermore, and D.~Serre, \emph{Singular
  limits of dispersive waves}, B, vol. 320, NATO ASI, 1994.

\bibitem{FF83}
J.-D. Fournier and U.~Frisch, \emph{L'\'equation de {B}urgers d\'eterministe et
  statistique}, J. M\'ec. Th\'eor. Appl. \textbf{2} (1983), no.~5, 699--750.

\bibitem{FMB03}
U.~Frisch, T.~Matsumoto, and J.~Bec, \emph{Singularities of {E}uler flow? not
  out of the blue!}, J. Stat. Phys. \textbf{113} (2003), 761--781.

\bibitem{GLSS09}
F.~Gargano, M.C. Lombardo, M.~Sammartino, and V.~Sciacca, \emph{{Singularity
  Formation and Separation Phenomena in Boundary Layer Theory}}, Partial
  differential equations and fluid mechanics, London Math. Soc. Lecture Note
  Ser., vol. 364, Cambridge Univ. Press, 2009, pp.~81--120.

\bibitem{GSS09}
F.~Gargano, M.~Sammartino, and V.~Sciacca, \emph{Singularity formation for
  {Prandtl's equations}}, Physica D: Nonlinear Phenomena \textbf{238} (2009),
  no.~19, 1975--1991.

\bibitem{GSS11}
F.~Gargano, M.~Sammartino, and V.~Sciacca, \emph{{High Reynolds number
  Navier-Stokes solutions and boundary layer separation induced by a
  rectilinear vortex}}, Computers \& Fluids \textbf{52} (2011), 73--91.

\bibitem{GSSC14}
F.~Gargano, M.~Sammartino, V.~Sciacca, and K.~W. Cassel, \emph{Analysis of
  complex singularities in high-{R}eynolds-number {N}avier-{S}tokes solutions},
  J. Fluid Mech. \textbf{747} (2014), 381--421. \MR{3200689}

\bibitem{GPS98}
R.E. Goldstein, A.I. Pesci, and M.J. Shelley, \emph{{Instabilities and
  singularities in {Hele--Shaw} flow}}, Physics of Fluids \textbf{10} (1998),
  no.~11, 2701--2723.

\bibitem{GK12}
T.~Grava and C.~Klein, \emph{A numerical study of the small dispersion limit of
  the {K}orteweg-de {V}ries equation and asymptotic solutions}, Phys. D
  \textbf{241} (2012), no.~23-24, 2246--2264. \MR{2998126}

\bibitem{Gu89}
A.~J. Guttmann, \emph{Asymptotic analysis of power-series expansions}, Phase
  transitions and critical phenomena, {V}ol.\ 13, Academic Press, London, 1989,
  pp.~1--234.

\bibitem{Henrici}
P.~Henrici, \emph{Applied and computational complex analysis, vol i, ii \&
  iii}, Wiley-Interscience, 1993.

\bibitem{JLM99}
S.~Jin, C.~D. Levermore, and D.~W. McLaughlin, \emph{The semiclassical limit of
  the defocusing {NLS} hierarchy}, Comm. Pure Appl. Math. \textbf{52} (1999),
  no.~5, 613--654.

\bibitem{KMM03}
S.~Kamvissis, K.~D. T.-R. McLaughlin, and P.~D. Miller, \emph{Semiclassical
  soliton ensembles for the focusing nonlinear {S}chr\"odinger equation},
  Annals of Mathematics Studies, vol. 154, Princeton University Press,
  Princeton, NJ, 2003. \MR{1999840 (2004h:37115)}

\bibitem{Ki86}
Shigeo Kida, \emph{Study of complex singularities by filtered spectral method},
  Journal of the Physical Society of Japan \textbf{55} (1986), no.~5,
  1542--1555.

\bibitem{KR13}
C.~Klein and K.~Roidot, \emph{Numerical study of shock formation in the
  dispersionless kadomtsev–petviashvili equation and dispersive
  regularizations}, Physica D: Nonlinear Phenomena \textbf{265} (2013), 1 --
  25.

\bibitem{KR14}
\bysame, \emph{Numerical study of the semiclassical limit of the
  {Davey-Stewartson II} equations}, arxiv:1401.4745 (2014).

\bibitem{KR15}
\bysame, \emph{Numerical study of the long wavelength limit of the toda
  lattice}, Nonlinearity \textbf{28} (2015), 2993--3025.

\bibitem{Kr86a}
Robert Krasny, \emph{A study of singularity formation in a vortex sheet by the
  point-vortex approximation}, Journal of Fluid Mechanics \textbf{167} (1986),
  65--93.

\bibitem{LL87I}
Peter~D. Lax and C.~David Levermore, \emph{The small dispersion limit of the
  {K}orteweg-de {V}ries equation. {I}}, Comm. Pure Appl. Math. \textbf{36}
  (1983), no.~3, 253--290.

\bibitem{LL83II}
\bysame, \emph{The small dispersion limit of the {K}orteweg-de {V}ries
  equation. {II}}, Comm. Pure Appl. Math. \textbf{36} (1983), no.~5, 571--593.

\bibitem{LL83III}
\bysame, \emph{The small dispersion limit of the {K}orteweg-de {V}ries
  equation. {III}}, Comm. Pure Appl. Math. \textbf{36} (1983), no.~6, 809--829.

\bibitem{MCSV13}
K.~Malakuti, R.~E. Caflisch, M.~Siegel, and A.~Virodov, \emph{Detection of
  complex singularities for a function of several variables}, IMA J Appl Math
  \textbf{78} (2013), no.~4, 714--728.

\bibitem{MBF05}
T.~Matsumoto, J.~Bec, and U.~Frisch, \emph{{The Analytic Structure of 2{D}
  Euler Flow at Short Times}}, Fluid Dyn. Res. \textbf{36} (2005), no.~4-6,
  221--237.

\bibitem{MOO78}
D.~W. Moore, \emph{The equation of motion of a vortex layer of small
  thickness}, Studies in Appl. Math. \textbf{58} (1978), no.~2, 119--140.

\bibitem{MOO79}
\bysame, \emph{The spontaneous appearance of a singularity in the shape of an
  evolving vortex sheet}, Proc. Roy. Soc. London Ser. A \textbf{365} (1979),
  no.~1720, 105--119.

\bibitem{OC02}
A.V. Obabko and K.W. Cassel, \emph{{Navier-Stokes solutions of unsteady
  separation induced by a vortex}}, J. Fluid Mech. \textbf{465} (2002),
  99--130.

\bibitem{OC05}
A.V. Obabko and K.W. Cassel, \emph{On the ejection-induced instability in
  {Navier-Stokes} solutions of unsteady separation}, Philosophical Transactions
  of the Royal Society A: Mathematical, Physical and Engineering Sciences
  \textbf{363} (2005), no.~1830, 1189--1198.

\bibitem{PF07}
W.~Pauls and U.~Frisch, \emph{A {B}orel transform method for locating
  singularities of {T}aylor and {F}ourier series}, J. Stat. Phys. \textbf{127}
  (2007), no.~6, 1095--1119.

\bibitem{PMFB06}
W.~Pauls, T.~Matsumoto, U.~Frisch, and J.~Bec, \emph{{Nature of Complex
  Singularities for the 2D Euler Equation}}, Physica D \textbf{219} (2006),
  no.~1, 40--59.

\bibitem{PS98}
M.~C. Pugh and M.~J. Shelley, \emph{Singularity formation in thin jets with
  surface tension}, Communications on Pure and Applied Mathematics \textbf{51}
  (1998), no.~7, 733--795.

\bibitem{RM15}
K.~Roidot and N.~Mauser, \emph{Numerical study of the transverse stability of
  {NLS} soliton solution in several classes of {NLS}-type equations},
  arxiv:1401.5349 (2015).

\bibitem{SC}
Hermann Schlichting, \emph{{Boundary Layer Theory}}, Translated by J. Kestin.
  4th ed. McGraw-Hill Series in Mechanical Engineering, McGraw-Hill Book Co.,
  Inc., New York, 1960. \MR{MR0122222 (22 \#12948)}

\bibitem{SCE96}
D.~Senouf, R.~Caflisch, and N.~Ercolani, \emph{Pole dynamics and oscillations
  for the complex {B}urgers equation in the small-dispersion limit},
  Nonlinearity \textbf{9} (1996), no.~6, 1671--1702.

\bibitem{Sen97I}
David Senouf, \emph{Dynamics and condensation of complex singularities for
  {B}urgers' equation. {I}}, SIAM J. Math. Anal. \textbf{28} (1997), no.~6,
  1457--1489.

\bibitem{Sen97II}
\bysame, \emph{Dynamics and condensation of complex singularities for
  {B}urgers' equation. {II}}, SIAM J. Math. Anal. \textbf{28} (1997), no.~6,
  1490--1513.

\bibitem{Sh92}
M.J. Shelley, \emph{{A study of singularity formation in vortex--sheet motion
  by a spectrally accurate vortex method}}, J. Fluid. Mech. \textbf{244}
  (1992), 493--526.

\bibitem{SC09}
M.~Siegel and R.E. Caflisch, \emph{Calculation of complex singular solutions to
  the 3{D} incompressible {E}uler equations}, Physica D: Nonlinear Phenomena
  \textbf{238} (2009), no.~23–24, 2368 -- 2379.

\bibitem{SSF83}
C.~Sulem, P-L Sulem, and H.~Frisch, \emph{Tracing complex singularities with
  spectral methods}, J. Comput. Phys. \textbf{50} (1983), no.~1, 138--161.

\bibitem{TVZ04}
Alexander Tovbis, Stephanos Venakides, and Xin Zhou, \emph{On semiclassical
  (zero dispersion limit) solutions of the focusing nonlinear {S}chr\"odinger
  equation}, Comm. Pure Appl. Math. \textbf{57} (2004), no.~7, 877--985.
  \MR{2044068 (2005c:35269)}

\bibitem{vDH09}
J.~van~der Hoeven, \emph{Algorithms for asymptotic extrapolation}, J. Symb.
  Comp. \textbf{44} (2009), no.~8, 1000--1016.

\bibitem{vDS80}
L.L. {van Dommelen} and S.F. Shen, \emph{The spontaneous generation of the
  singularity in a separating laminar boundary layer}, J. Comp. Phys.
  \textbf{38} (1980), 125--140.

\bibitem{Wei03}
J.~A.~C. Weideman, \emph{Computing the dynamics of complex singularities of
  nonlinear {PDE}s}, SIAM J. Appl. Dyn. Syst. \textbf{2} (2003), no.~2,
  171--186 (electronic).

\end{thebibliography}

%\bibliographystyle{plain}

% Non-BibTeX users please use

%\begin{thebibliography}{}

%\end{thebibliography}

\end{document}